\documentclass[twocolumn,aps,prc,showpacs,superscriptaddress,floatfix]{revtex4}
\usepackage{epsfig}
\newenvironment{color}[3]
{
}{
}

\newcommand {\rts}	{\sqrt{s_{_{\rm NN}}}}
\newcommand {\pt}	{p_{T}}
\newcommand {\ptt}	{p_{T}^{(t)}}
\newcommand {\pta}	{p_{T}^{(a)}}
\newcommand {\phis}	{\phi_{s}}
\newcommand {\phit}	{\phi_{t}}
\newcommand {\psiEP}	{\psi_{\rm EP}}
\newcommand {\psiRP}	{\psi_{\rm RP}}
\newcommand {\psiPP}	{\psi_2}
\newcommand {\dPsi}	{\Delta\psi}
\newcommand {\dphi}	{\Delta\phi}
\newcommand {\deta}	{\Delta\eta}
\newcommand {\vt}	{v_2^{(t)}}

\newcommand {\vvvt}	{v_3^{(t)}}
\newcommand {\va}	{v_2^{(a)}}

\newcommand {\vvva}	{v_3^{(a)}}
\newcommand {\vtR}	{v_2^{(t,R)}}

\newcommand {\vvvtR}	{v_3^{(t,R)}}
\newcommand {\vf}	{v_2}
\newcommand {\vv}	{v_4}
\newcommand {\vvPsi}	{v_4\{\psi_2\}}
\newcommand {\vvaPsi}	{v_4^{(a)}\{\psi_2\}}
\newcommand {\vvtPsi}	{v_4^{(t)}\{\psi_2\}}
\newcommand {\vvtRPsi}	{v_4^{(t,R)}\{\psi_2\}}
\newcommand {\flow}[1]	{v_2\{#1\}}
\newcommand {\Flow}[2]	{v_{#1}\{#2\}}
\newcommand {\ff}[2]	{v_{#1}\{#2\}}
\newcommand {\fff}[3]	{v_{#1}^{(#2)}\{#3\}}
\newcommand {\FF}[2]	{V_{#1}\{#2\}}
\newcommand {\etagap}	{\eta_{\rm gap}}
\newcommand {\FFeta}[3]	{V_{#1}\{#2,\etagap\mbox{=}#3\}}
\newcommand {\VVuc}     {V_4\{{\rm uc}\}}

\newcommand {\vuc}     {v_4\{{\rm uc}\}}
\newcommand {\vEP}	{$v_2$\{{\sc ep}\}}
\newcommand {\vMRP}	{$v_2$\{{\sc mrp}\}}

\newcommand {\veta}[2]	{v_{#1}\{2,\etagap\mbox{=}#2\}}

\newcommand {\vpt}[1]   {v_{#1}\{\pt\mbox{-}\pt\}}
\newcommand {\vpteta}[2]{v_{#1}\{\pt\mbox{-}\pt,\etagap\mbox{=}#2\}}
\newcommand {\Vpteta}[2]{V_{#1}\{\pt\mbox{-}\pt,\etagap\mbox{=}#2\}}
\newcommand {\mean}[1]	{\langle #1\rangle}
\newcommand {\zyam}	{{\sc zyam}}
\newcommand {\rms}	{{\sc rms}}
\newcommand {\gev}	{{GeV/$c$}}
\newcommand {\pp}	{{$p$+$p$}}
\newcommand {\dAu}	{{$d$+Au}}

\begin{document}


\title{Measurements of Dihadron Correlations Relative to the Event Plane in Au+Au Collisions at $\rts=200$~GeV}

\affiliation{Argonne National Laboratory, Argonne, Illinois 60439, USA}
\affiliation{Brookhaven National Laboratory, Upton, New York 11973, USA}
\affiliation{University of California, Berkeley, California 94720, USA}
\affiliation{University of California, Davis, California 95616, USA}
\affiliation{University of California, Los Angeles, California 90095, USA}
\affiliation{Universidade Estadual de Campinas, Sao Paulo, Brazil}
\affiliation{University of Illinois at Chicago, Chicago, Illinois 60607, USA}
\affiliation{Creighton University, Omaha, Nebraska 68178, USA}
\affiliation{Czech Technical University in Prague, FNSPE, Prague, 115 19, Czech Republic}
\affiliation{Nuclear Physics Institute AS CR, 250 68 \v{R}e\v{z}/Prague, Czech Republic}
\affiliation{University of Frankfurt, Frankfurt, Germany}
\affiliation{Institute of Physics, Bhubaneswar 751005, India}
\affiliation{Indian Institute of Technology, Mumbai, India}
\affiliation{Indiana University, Bloomington, Indiana 47408, USA}
\affiliation{Alikhanov Institute for Theoretical and Experimental Physics, Moscow, Russia}
\affiliation{University of Jammu, Jammu 180001, India}
\affiliation{Joint Institute for Nuclear Research, Dubna, 141 980, Russia}
\affiliation{Kent State University, Kent, Ohio 44242, USA}
\affiliation{University of Kentucky, Lexington, Kentucky, 40506-0055, USA}
\affiliation{Institute of Modern Physics, Lanzhou, China}
\affiliation{Lawrence Berkeley National Laboratory, Berkeley, California 94720, USA}
\affiliation{Massachusetts Institute of Technology, Cambridge, MA 02139-4307, USA}
\affiliation{Max-Planck-Institut f\"ur Physik, Munich, Germany}
\affiliation{Michigan State University, East Lansing, Michigan 48824, USA}
\affiliation{Moscow Engineering Physics Institute, Moscow Russia}
\affiliation{NIKHEF and Utrecht University, Amsterdam, The Netherlands}
\affiliation{Ohio State University, Columbus, Ohio 43210, USA}
\affiliation{Old Dominion University, Norfolk, VA, 23529, USA}
\affiliation{Panjab University, Chandigarh 160014, India}
\affiliation{Pennsylvania State University, University Park, Pennsylvania 16802, USA}
\affiliation{Institute of High Energy Physics, Protvino, Russia}
\affiliation{Purdue University, West Lafayette, Indiana 47907, USA}
\affiliation{Pusan National University, Pusan, Republic of Korea}
\affiliation{University of Rajasthan, Jaipur 302004, India}
\affiliation{Rice University, Houston, Texas 77251, USA}
\affiliation{Universidade de Sao Paulo, Sao Paulo, Brazil}
\affiliation{University of Science \& Technology of China, Hefei 230026, China}
\affiliation{Shandong University, Jinan, Shandong 250100, China}
\affiliation{Shanghai Institute of Applied Physics, Shanghai 201800, China}
\affiliation{SUBATECH, Nantes, France}
\affiliation{Texas A\&M University, College Station, Texas 77843, USA}
\affiliation{University of Texas, Austin, Texas 78712, USA}
\affiliation{Tsinghua University, Beijing 100084, China}
\affiliation{United States Naval Academy, Annapolis, MD 21402, USA}
\affiliation{Valparaiso University, Valparaiso, Indiana 46383, USA}
\affiliation{Variable Energy Cyclotron Centre, Kolkata 700064, India}
\affiliation{Warsaw University of Technology, Warsaw, Poland}
\affiliation{University of Washington, Seattle, Washington 98195, USA}
\affiliation{Wayne State University, Detroit, Michigan 48201, USA}
\affiliation{Institute of Particle Physics, CCNU (HZNU), Wuhan 430079, China}
\affiliation{Yale University, New Haven, Connecticut 06520, USA}
\affiliation{University of Zagreb, Zagreb, HR-10002, Croatia}

\author{H.~Agakishiev}\affiliation{Joint Institute for Nuclear Research, Dubna, 141 980, Russia}
\author{M.~M.~Aggarwal}\affiliation{Panjab University, Chandigarh 160014, India}
\author{Z.~Ahammed}\affiliation{Lawrence Berkeley National Laboratory, Berkeley, California 94720, USA}
\author{A.~V.~Alakhverdyants}\affiliation{Joint Institute for Nuclear Research, Dubna, 141 980, Russia}
\author{I.~Alekseev~~}\affiliation{Alikhanov Institute for Theoretical and Experimental Physics, Moscow, Russia}
\author{J.~Alford}\affiliation{Kent State University, Kent, Ohio 44242, USA}
\author{B.~D.~Anderson}\affiliation{Kent State University, Kent, Ohio 44242, USA}
\author{C.~D.~Anson}\affiliation{Ohio State University, Columbus, Ohio 43210, USA}
\author{D.~Arkhipkin}\affiliation{Brookhaven National Laboratory, Upton, New York 11973, USA}
\author{G.~S.~Averichev}\affiliation{Joint Institute for Nuclear Research, Dubna, 141 980, Russia}
% ...

\collaboration{STAR Collaboration}\noaffiliation

\begin{abstract}
Dihadron azimuthal correlations containing a high transverse momentum ($\pt$) trigger particle are sensitive to the properties of the nuclear medium created at RHIC through the strong interactions occurring between the traversing parton and the medium, i.e. jet-quenching. Previous measurements revealed a strong modification to dihadron azimuthal correlations in Au+Au collisions with respect to \pp\ and \dAu\ collisions. The modification increases with the collision centrality, suggesting a path-length or energy density dependence to the jet-quenching effect. This paper reports STAR measurements of dihadron azimuthal correlations in mid-central (20-60\%) Au+Au collisions at $\rts=200$~GeV as a function of the trigger particle's azimuthal angle relative to the event plane, $\phis=|\phit-\psiEP|$. The azimuthal correlation is studied as a function of both the trigger and associated particle $\pt$. The subtractions of the combinatorial background and anisotropic flow, assuming Zero Yield At Minimum (\zyam), are described. The correlation results are first discussed with the subtraction of the even harmonic (elliptic and quadrangular) flow backgrounds. The away-side correlation is strongly modified, and the modification varies with $\phis$, with a double-peak structure for out-of-plane trigger particles. 
The near-side ridge (long range pseudo-rapidity $\deta$ correlation) appears to drop with increasing $\phis$ while the jet-like component remains approximately constant. The correlation functions are further studied with subtraction of odd harmonic triangular flow background arising from fluctuations. It is found that the triangular flow, while responsible for the majority of the amplitudes, is not sufficient to explain the $\phis$-dependence of the ridge or the away-side double-peak structure. The dropping ridge with $\phis$ could be attributed to a $\phis$-dependent elliptic anisotropy; however, the physics mechanism of the ridge remains an open question. Even with a $\phis$-dependent elliptic flow, the away-side correlation structure is robust. These results, with extensive systematic studies of the dihadron correlations as a function of $\phis$, trigger and associated particle $\pt$, and the pseudo-rapidity range $\deta$, should provide stringent inputs to help understand the underlying physics mechanisms of jet-medium interactions in high energy nuclear collisions.
\end{abstract}


\maketitle

\section{Introduction}

Collisions at the Relativistic Heavy Ion Collider (RHIC) at Brookhaven National Laboratory have created a medium with properties that resemble a nearly perfect liquid of strongly interacting quarks and gluons~\cite{Arsene:2004fa,Back:2004je,Adams:2005dq,Adcox:2004mh}. This conclusion is based upon two pillars of evidence~\cite{Gyulassy:2004zy}: (i) the strong elliptic flow and (ii) jet-quenching-- suppression of high transverse momentum ($\pt$) single hadron yields and dihadron correlations in heavy-ion collisions relative to elementary \pp\ interactions. While suppression of high $\pt$ single hadron yields has limited sensitivity to the medium core, dihadron correlation measurements provide richer and more valuable information about the properties of the created medium~\cite{Jacobs:2004qv,Zhang:2007ja,Wang:2013qca}. There are several key observations that can be made from dihadron correlations with a high $\pt$ trigger particle. 
(i) The correlated hadron yield at high $\pt$, while not much changed on the near side of the trigger particle (where azimuth difference between correlated and trigger particles $|\dphi|<\pi/2$), is strongly suppressed on the away side (where $|\dphi|>\pi/2$)~\cite{Adler:2002tq}. This lends strong support to the partonic energy loss picture~\cite{Gyulassy:1990ye,Wang:1991xy,Baier:2000mf}. 
(ii) The correlated hadron yields at low $\pt$ are strongly enhanced on both the near and away side~\cite{Adams:2005ph}. In particular, the near-side enhancement is tied to long-range correlations in pseudo-rapidity -- the ridge~\cite{Adams:2005ph,Abelev:2009af,Abelev:2009jv}. 
(iii) The away-side correlation broadens from peripheral to central collisions, and exhibits double peaks for select trigger and associated particle $\pt$ ranges~\cite{Adams:2005ph,Adare:2008ae,Aggarwal:2010rf}. The double-peak structure is peculiar and may provide an opportunity to study the underlying physics mechanisms for partonic energy loss, such as gluon radiation~\cite{Vitev:2005yg,Polosa:2006hb}, Mach-cone shock-wave excitation~\cite{Stoecker:2004qu,CasalderreySolana:2004qm,Ruppert:2005uz,Renk:2005si,Khlebnikov:2010yt,Betz:2010qh,Neufeld:2011yh,Ayala:2012bv,Bouras:2014rea,Tachibana:2015qxa}, or simply the bulk medium response~\cite{Alver:2010gr}. Three-particle jet-like correlation studies indicate that the double-peak emission pattern of correlated hadrons is characteristic of medium triangular flow~\cite{Alver:2010gr} and/or Mach-cone shock-waves~\cite{Abelev:2008ac}. 
(iv) The away-side associated particles are partially equilibrated with the bulk medium in mid-central to central collisions, and a higher degree of equilibration is observed for particles which are more aligned back-to-back with the trigger particles~\cite{Adams:2005ph,Aggarwal:2010rf}. This observation may underscore the connection between the medium's path-length and partonic energy loss.

We study the path-length dependence of partonic energy loss in detail in non-central collisions where the overlap region between the two colliding nuclei is anisotropic: the size in the reaction-plane direction is shorter than that perpendicular to it. The reaction plane (RP) is defined by the beam direction and the line connecting the centers of two colliding nuclei. It can be estimated in non-central collisions by determining the azimuthal angle with the highest particle emission probability, using the fact that the particles have an elliptic emission pattern~\cite{Poskanzer:1998yz}. The estimated angle is called the event plane (EP), to emphasize that it is an experimental estimate of the reaction plane with finite resolution. By selecting the trigger particle direction with respect to the event plane, $\phis=|\phit-\psiEP|$ (where $\phit$ is the trigger particle azimuth and $\psiEP$ is the event-plane azimuth), we aim to select different path-lengths through the medium that the away-side parton traverses, providing differential information unavailable to inclusive jet-correlation measurements. 

Previously, the Solenoidal Tracker at RHIC (STAR) experiment has performed an exploratory measurement of azimuthal correlations at high $\pt$ with trigger particles in-plane ($\phis<\pi/4$) and out-of-plane ($\phis>\pi/4$) using non-central 20-60\% Au+Au collisions~\cite{Adams:2004wz}. The results hinted that the away-side correlation with out-of-plane trigger particles is more strongly suppressed than that with in-plane trigger particles. In this paper, we extend those measurements to finer bins in $\phis$ and to lower associated and trigger $\pt$ ranges~\cite{thesisFengAoqi}. 
We also present inclusive jet-like correlation results from minimum bias \dAu\ collisions as a reference to the Au+Au data.
We further study the ridge as a function of $\phis$, and investigate the systematics of the ridge in an attempt to further identify the underlying physics mechanism for the formation of the ridge.
The highlights of the results have been published in Ref.~\cite{Agakishiev:2014ada}. This paper provides extensive results and analysis details. Similar results on the event-plane dependent dihadron correlations have recently been reported by the PHENIX collaboration~\cite{Adare:2018wjb}.

The paper is organized as follows. In Sec.~\ref{sec:analysis} we describe in detail our data analysis of dihadron correlations relative to the event plane. In Sec.~\ref{sec:syst} we discuss our extensive studies of the systematic uncertainties of our results. In Sec.~\ref{sec:results} we report and discuss our results of dihadron correlations relative to the event plane. We finally conclude in Sec.~\ref{conclusion}. We present all raw and background-subtracted dihadron correlation functions relative to the event plane in Appendix~\ref{app}.


\section{Experiment and Data Analysis\label{sec:analysis}}

The data used in this analysis were taken by the STAR experiment~\cite{Ackermann:2002ad} at RHIC at the nucleon-nucleon center of mass energy of $\rts=200$~GeV. The minimum-bias Au+Au data were from Run IV in 2004 at RHIC. The reference minimum-bias \dAu\ data used for comparison were from Run III in 2003. 
The minimum-bias triggers for Au+Au and \dAu\ collisions were provided by the Central Trigger Barrel~\cite{Bieser:2002ah} and the Zero Degree Calorimeters~\cite{Adler:2003sp}. 

The details of the STAR experiment can be found in Ref.~\cite{Ackermann:2002ad}. The main detector used for this analysis is the Time Projection Chamber (TPC)~\cite{Ackermann:1999kc,Anderson:2003ur}. The TPC is surrounded by a solenoidal magnet providing a nearly uniform magnetic field of 0.5 Tesla along the beam direction. 
Particle tracks are reconstructed in the TPC. The primary event vertex was fit using reconstructed tracks which pass certain quality cuts. Events with a primary vertex within $\pm 30$~cm of the geometric center of the TPC along the beam axis are used in the analysis. With this range of primary vertex position, the TPC has good acceptance within the pseudo-rapidity region of $|\eta|\leq 1.1$. The Au+Au collision centrality is defined according to the measured charged hadron multiplicity in the TPC within $|\eta|<0.5$ (reference multiplicity)~\cite{Adams:2003xp}. We choose the 20-60\% centrality data, where good event-plane resolution is achieved, for our analysis (see later).

In our analysis, only tracks that extrapolate to within 2~cm of the primary vertex are used. Tracks are required to be reconstructed with at least 20 out of a maximum of 45 hits in the TPC. The ratio of the number of hits used in track reconstruction to the number of possible hits is required to be greater than 0.51, to eliminate multiple track segments being reconstructed from a single particle trajectory. The same event and track cuts are applied to particle tracks used for event-plane reconstruction and for the subsequent correlation analysis. Particle tracks within $|\eta|<1$ are used in the correlation analysis. 

High $\pt$ particles are selected as triggers off-line to perform the correlation analyses. We select high $\pt$ trigger particles within the $\ptt$ ranges of 3-4~\gev\ and 4-6~\gev. A total of 4.4 million Au+Au events with centrality ranging from 20-60\% are used in this analysis. From the event sample we find 2.1 million trigger particles with $\pt$ values ranging between 3-4~\gev, and 0.36 million trigger particles with $\pt$ values between 4-6~\gev. Associated particles, i.e.~all particles in the event including those correlated with the trigger particles, are grouped into the $\pta$ ranges of 0.15-0.5, 0.5-1.0, 1.0-1.5, 1.5-2.0, 2.0-3.0 (or 2.0-4.0)~\gev. The low $\pta$ cut-off of 0.15~\gev\ is imposed by the magnetic field strength and the TPC aperture. The azimuthal correlation functions in $\dphi$ (azimuthal angle difference between associated particle and trigger particle) are analyzed separately for trigger particles at different azimuthal angles ($\phis$) relative to the event plane.

The associated particle yields are corrected for single-particle track reconstruction efficiency, which is obtained from embedding simulated tracks into real events~\cite{Abelev:2008ab}. It depends on both centrality and $\pt$. The efficiency is found to be insensitive to $\eta$ and is therefore averaged over $\eta$. The $\phi$-dependent part of the acceptance and track reconstruction efficiency are corrected for both the trigger and associated particle yields. This $\phi$-dependent correction is obtained from the inverse of the single-particle $\phi$ distribution, whose average is normalized to unity.
Correction for the $\phi$-dependent efficiencies for both trigger and associated particles removes the majority of the non-uniformity caused by the TPC sector boundaries. The remaining non-uniformity in $\dphi$ is corrected by using an event-mixing technique, where the trigger particle from one event is paired with associated particles from another event within the same centrality bin~\cite{Adams:2005ph}. The two-particle acceptance in $\deta$ (pseudo-rapidity difference between associated particle and trigger particle), which is approximately triangle-shaped in $\deta$, is not corrected to be consistent with earlier publications~\cite{Adams:2005ph}. 
The correlation function is normalized by the corrected number of trigger particles in its corresponding $\phis$ bin. The centrality and $\pt$ dependent aspects of the trigger particle efficiency cancel out in the normalization. 

Tracks that are spatially near each other can be combined into a single reconstructed track due to merged space points of ionization in the STAR TPC. This track merging results in a pair inefficiency at $\deta\sim0$ and a small but finite $\dphi$ whose value depends on the magnetic field polarity, charge combination and the $\pt$'s of the trigger and associated particles~\cite{Abelev:2009jv}. The track merging effect is most significant in central collisions where the TPC hit occupancy is high. The track merging effect in our centrality range of 20-60\% is negligible. 

\subsection{Event-Plane Reconstruction}

\begin{figure}[hbt]
\centerline{\includegraphics[width=0.4\textwidth]{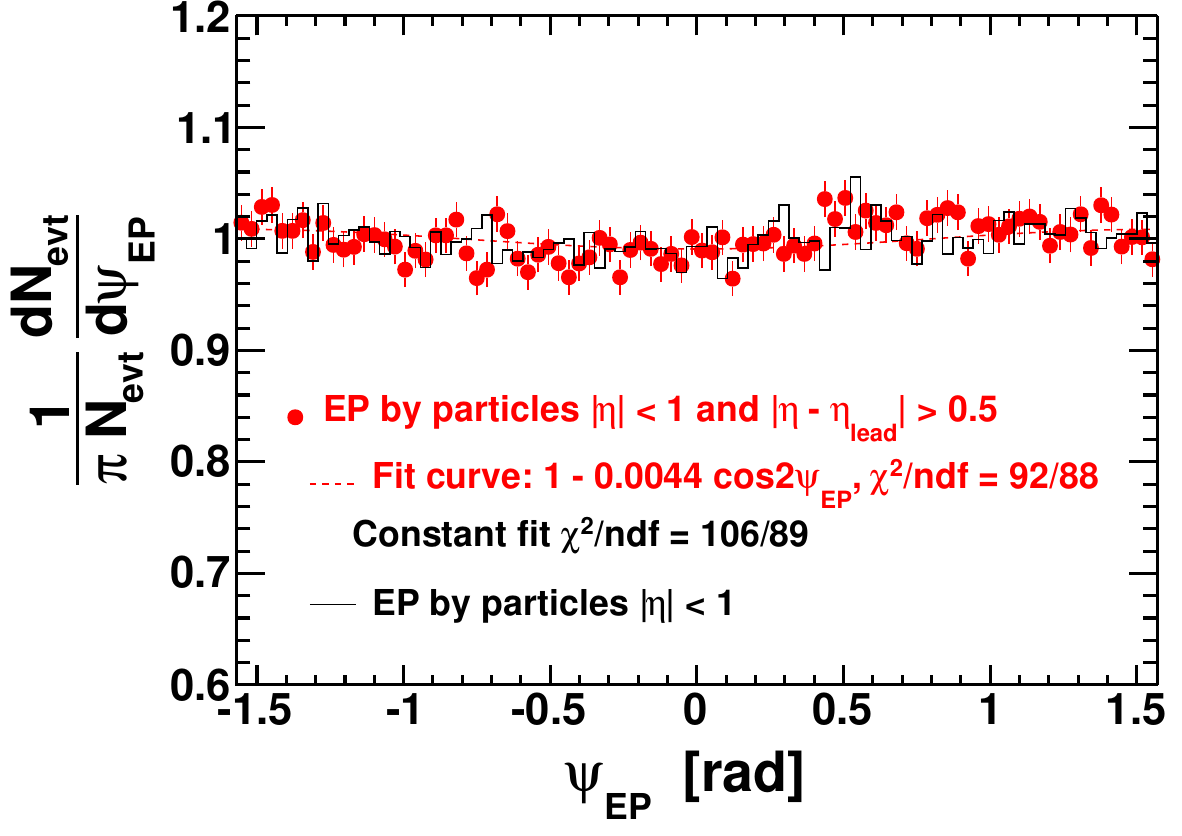}}
\caption{(Color online) Constructed event plane azimuthal angle ($\psiEP$) distributions by the modified reaction-plane ({\sc mrp}) method (points) and the traditional reaction-plane method (histogram). The particles used for constructing the event plane shown in this figure are from $0.15<\pt<1$~GeV/$c$ or $1.5<\pt<2$~GeV/$c$, to be used for correlation analysis for the associated particle $\pta$ bin of $1<\pta<1.5$~GeV/$c$.}
\label{fig:EP}
\end{figure}

We use the second Fourier harmonic in azimuthal angle to determine the event-plane angle $\psiEP$~\cite{Poskanzer:1998yz}, which is not identical to the real reaction-plane angle ($\psiRP$). The event plane is an estimate, with finite resolution, of the second harmonic participant plane (the plane defined by the beam direction and the minor axis of the overlap geometry of participant nucleons)~\cite{Alver:2008zza}. The participant plane angle, $\psiPP$, fluctuates about the reaction plane direction. The particles used to determine the event plane are below $\pt = 2$~\gev. To avoid self-correlations, particles from the $\pt$ bin that is used in the correlation analysis are excluded from event-plane reconstruction. For example, for the associated particle $\pt$ bin of $1.0<\pta<1.5$~\gev, the particles used to calculate the event plane are from $\pt$ ranges of $0.15<\pt<1.0$~\gev\ plus $1.5<\pt<2.0$~\gev. We use the $\pt$-weight method~\cite{Poskanzer:1998yz}, which gives better event-plane resolution due to the stronger anisotropy at larger $\pt$. The slight non-uniform efficiency and acceptance in azimuthal angle were corrected as mentioned previously in the event-plane reconstruction. Figure~\ref{fig:EP} shows examples of the constructed event plane azimuthal angle distributions. As seen from the figure, the constructed event plane $\psiEP$ distribution is approximately uniform. We weight the events by the inverse of the event-plane angle distributions in Fig.~\ref{fig:EP} in our correlation analysis. However, we find negligible difference in our results with and without this event-plane weighting. 

Nonflow correlations, such as di-jets, can influence the determination of the event plane. To reduce this effect, we exclude from EP reconstruction particles within pseudorapidity difference of $|\deta|=|\eta-\eta_{\rm trig}|<0.5$ from the trigger particle. (In other words, we reconstruct an EP for each trigger particle; if there are multiple trigger particles in an event, their event planes are different even though they belong to the same event.)
This method is called the modified reaction-plane ({\sc mrp}) method~\cite{Adams:2004bi}. The traditional reaction-plane method, on the other hand, does not exclude from EP reconstruction those particles in the $\eta$ vicinity of the trigger particle. Remaining possible biases due to correlations between trigger particles and EP particles may be assessed by comparing our results relative to the EP reconstructed from these two different methods with their respective EP resolutions. The results are found to be qualitatively similar, suggesting that any biases may be small. See Appendix~\ref{app:EP} for details.

To extract the near-side jet-like component, we use the difference in azimuthal correlations between those analyzed at small and large $|\deta|$. The {\sc mrp} method, which excludes particles within $|\deta|<0.5$ of the trigger particle in the event, would have different systematic biases in the $\dphi$ correlations at small and large $|\deta|$. Thus, we use the traditional reaction-plane method for the jet-like component. Figure~\ref{fig:EP} shows the $\psiEP$ distributions from the modified reaction-plane method (data points) and the traditional reaction-plane method (histogram). We have checked the correlation between the event plane angles constructed from the traditional method and the {\sc mrp} method, and found they are correlated as expected; the spread between their numerical values is consistent with that caused by EP resolutions.

\begin{figure}[hbt]
\centerline{
\includegraphics[width=0.45\textwidth]{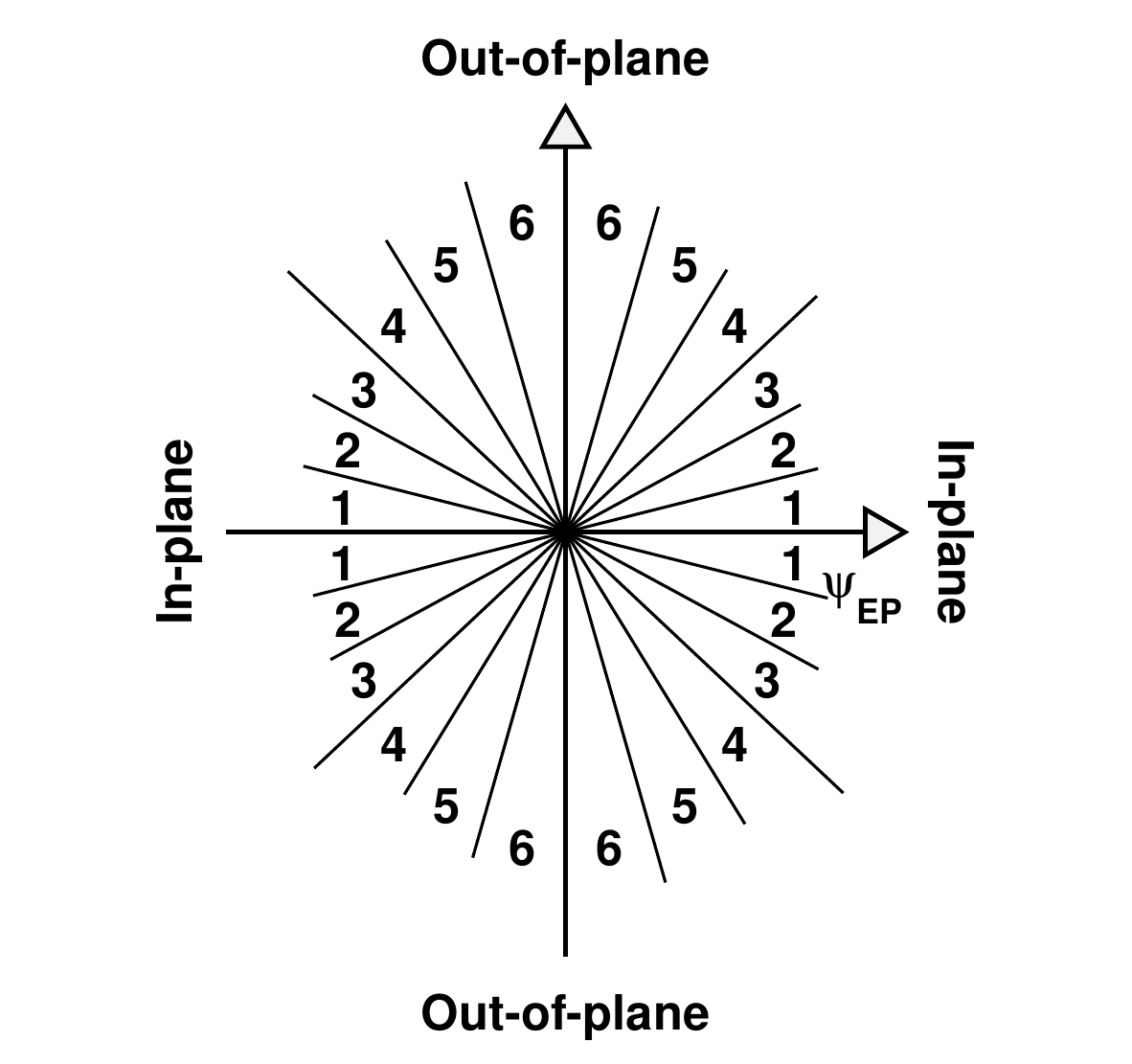}}
\caption{Sketch of six slices in trigger particle azimuthal angle relative to the event plane, $\phis=|\phit-\psiEP|$.}
\label{fig:sketch}
\end{figure}

\begin{figure*}[hbt]
\centerline{\includegraphics[width=1.03\textwidth]{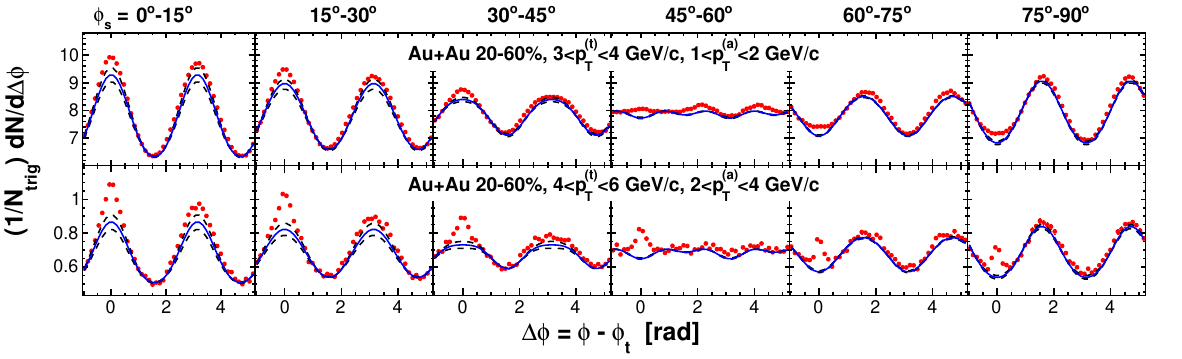}}
\caption{(Color online) Raw dihadron $\dphi$ correlations with trigger particles in six slices of azimuthal angle relative to the event plane, $\phis=|\phit-\psiEP|$. The data are from minimum-bias 20-60\% Au+Au collisions. The trigger and associated particle $\pt$ ranges are $3<\ptt<4$~\gev\ and $1<\pta<2$~\gev\ (upper panel), and $4<\ptt<6$~\gev\ and $2<\pta<4$~\gev\ (lower panel), respectively. Note the lower panels correspond to the kinematic range used in Ref.~\cite{Adler:2002tq}. Both the trigger and associated particles are restricted to within $|\eta|<1$. The triangle two-particle $\deta$ acceptance is not corrected. Statistical errors are smaller than the symbol size. The raw data in the upper panels have been published in Ref.~\cite{Agakishiev:2014ada}. The curves are flow modulated \zyam\ background including $\vf$ and $\vvPsi$ from Eq.~(\ref{eq:bkgd}). The $v_2$ values used are given in Table~\ref{tab:v2} from four-particle $\flow{4}$ and two-particle $\veta{2}{0.7}$ (dashed curves) and the average $\vf$ from the two methods (solid curve). The $\vvPsi$ is taken from the parameterization in Eq.~(\ref{eq:v4}).}
\label{fig:raw}
\end{figure*}

We divide our data into six equal-size slices of trigger particle azimuthal angle relative to the event plane, $\phis$, and analyze azimuthal correlations separately in each slice. Figure~\ref{fig:sketch} shows a schematic view, with the slices labeled numerically 1 to 6, corresponding to $\phis=|\phit-\psiEP|=0$-$\pi/12$, $\pi/12$-$\pi/6$, $\pi/6$-$\pi/4$, $\pi/4$-$\pi/3$, $\pi/3$-$5\pi/12$, and $5\pi/12$-$\pi/2$. We form azimuthal correlations with trigger particles in each slice separately. 
Figure~\ref{fig:raw} shows, as examples, the raw azimuthal correlations in 20-60\% Au+Au collisions for six slices in $\phis$ for trigger and associated particle $\pt$ ranges of $3<\ptt<4$~\gev\ and $1<\pta<2$~\gev\ (upper panel), and $4<\ptt<6$~\gev\ and $2<\pta<4$~\gev\ (lower panel), respectively.
All raw correlation functions are presented in Figs.~\ref{figApp:raw34},~\ref{figApp:raw46},~\ref{figApp:raw34ridge}, and~\ref{figApp:raw46ridge} in Appendix~\ref{app} as a function of trigger $\ptt$, associated $\pta$, and $\phis$.

\subsection{Elliptic and Quadrangular Flow Background~\label{sec:v2}}

The correlation structure sits atop a large combinatorial background. The background has a flow modulation induced by the anisotropies of the trigger particle and the background particles with respect to the participant plane~\cite{Wang:1991qh}. In this analysis we use anisotropic flow parameters measured by two- and multi-particle cumulants~\cite{Poskanzer:1998yz} for the combinatorial background. An alternative approach that has been used to describe dihadron correlation data treats the anisotropic flow modulations as free parameters in a multi-parameter model fit to the dihadron correlation functions in 2-dimensional $\deta$-$\dphi$ space~\cite{Daugherity:2008su,Agakishiev:2011pe}. Results from this alternative approach to the inclusive dihadron correlation data (without a high-$\pt$ trigger or cutting on $\phis$) can be found in Ref.~\cite{Daugherity:2008su,Agakishiev:2011pe}. The multi-parameter fit approach to our $\phis$-dependent high-$\pt$ triggered dihadron correlations is considered in Sec.~\ref{sec:syst:zyam}, but a detailed discussion of the differences in assumptions and conclusions of the two approaches is beyond the scope of this paper.

In this analysis, first, only the $v_2$ and $v_4$ harmonic flow backgrounds are considered. The triangular harmonic flow background and other high-order effects are described in the next subsection, Sec.~\ref{sec:v3bkgd}. Considering only the $v_2$ and $v_4$ harmonics, the flow correlated background is given by~\cite{Bielcikova:2003ku}
\begin{widetext}
\begin{equation}
\frac{dN}{d\dphi}=B\left[1+2\va\vtR\cos(2\dphi)+2\vvaPsi\vvtRPsi\cos(4\dphi)\right],\label{eq:bkgd}
\end{equation}
\end{widetext}
where 
$B$ is the background normalization. In Eq.~(\ref{eq:bkgd}), $\va$ and $\vvaPsi$ are the associated particle's second and fourth harmonics with respect to the second harmonic event plane, $\psiPP$, and $\vtR$ and $\vvtRPsi$ are the average harmonics of the trigger particles, 
$\vtR=\left\langle\cos\left[2\left(\phit-\psiPP\right)\right]\right\rangle^{(R)}$ and
$\vvtRPsi=\left\langle\cos\left[4\left(\phit-\psiPP\right)\right]\right\rangle^{(R)}$, respectively. 
The superscript `$(R)$' indicates that the averages are taken within the $\phit$ region of a slice of width $2c$ at $\phis$: 
$\phis-c<\left|\phit-\psiEP\right|<\phis+c$ 
(where $c=\pi/24$ in our analysis). 
Note that we have used $\phis$ here and in Eq.~(\ref{eq:Tk}) to be the center value of a $|\phit-\psiEP|$ bin, while elsewhere we simply use $\phis=|\phit-\psiEP|$ to indicate a narrow bin in $|\phit-\psiEP|$.
For the $n^{\rm th}$ harmonic anisotropy we have~\cite{Bielcikova:2003ku}
\begin{equation}
v_n^{(t,R)}=\frac{v_n^{(t)}+\delta_{n,{\rm even}}T_n+\sum_{k=2,4,6,...}\left(v_{k+n}^{(t)}+v_{|k-n|}^{(t)}\right)T_k}{1+\sum_{k=2,4,6,...}2v_{k}^{(t)}T_k}\,.\label{eq:vR}
\end{equation}
Here $T_k$ is a short-hand notation for
\begin{equation}
T_k=\cos(k\phis)\frac{\sin(kc)}{kc}\mean{\cos(k\dPsi)}\,,
\label{eq:Tk}
\end{equation}
$\mean{\cos(k\dPsi)}\equiv\mean{\cos k(\psiEP-\psiPP)}$ is the event-plane resolution with respect to the $k^{\rm th}$ harmonic, and $\delta$ is Kronecker's delta. Since the correlation signal we are studying is of the order of a few percent of the background, we need to keep the flow correction in Eq.~(\ref{eq:vR}) up to the order $\vf\vv\sim0.1$\%. Keeping terms for $\vtR$ up to $\vv$ and for $\vvtRPsi$ up to $\vf$, we have
\begin{equation}
\vtR\approx\frac{T_2+(1+T_4)\vt+(T_2+T_6)\vvtPsi}{1+2T_2\vt+2T_4\vvtPsi}\,,\label{eq:v2R}
\end{equation}
and
\begin{eqnarray}
\vvtRPsi&\approx&\frac{T_4+(T_2+T_6)\vt+(1+T_8)\vvtPsi}{1+2T_2\vt+2T_4\vvtPsi}\nonumber\\
&\approx&\frac{T_4+(T_2+T_6)\vt}{1+2T_2\vt}\,.\label{eq:v4R}
\end{eqnarray}
Note the $\vvtPsi$ above is with respect to the second harmonic plane $\psiPP$.
The final flow correction is given by Eqs.~(\ref{eq:bkgd}), (\ref{eq:v2R}), and (\ref{eq:v4R}).

The event-plane resolutions, $\mean{\cos(k\dPsi)}$ ($k=2,4,6$), are obtained from the sub-event method~\cite{Poskanzer:1998yz}. The event is randomly divided into two sub-events $a$ and $b$ with equal multiplicities. The sub-events, excluding the associated particle $\pt$ region, are analyzed to yield event-plane angles which, ideally, should be identical. The difference between the obtained event-plane angles, $\psi_{a}-\psi_{b}$, gives the uncertainty in the event-plane determination of the sub-events~\cite{Poskanzer:1998yz}
\begin{equation}
\mean{\cos(k\dPsi)}_{\mbox{sub-event}}=\sqrt{\mean{\cos k(\psi_{a}-\psi_{b})}}.
\label{eq:subevt}
\end{equation}
The event-plane resolution of the full event can be approximated by~\cite{Poskanzer:1998yz}
\begin{equation}
\mean{\cos(k\dPsi)}\approx\sqrt{2}\mean{\cos(k\dPsi)}_{\mbox{sub-event}}
\label{eq:fulevt}
\end{equation}
in the limit of small event-plane resolution. The factor $\sqrt{2}$ comes in because the multiplicities of the sub-events are smaller than the full event multiplicity by a factor of 2. We use the approximate form of Eq.~(\ref{eq:fulevt}) to assess systematic uncertainties in the event-plane resolutions from different ways of dividing the event into sub-events~(see Sec.~\ref{sec:syst:res}). 

The precise form of the event-plane resolution of the full event is given by~\cite{Poskanzer:1998yz}
\begin{equation}
\mean{\cos(k\dPsi)}=\sqrt{\frac{\pi}{2}}\left(\frac{\chi_{k}}{2}\right)e^{-\frac{\chi_{k}^{2}}{4}}\left[I_{0}\left(\frac{\chi_{k}^{2}}{4}\right)+I_{1}\left(\frac{\chi_{k}^{2}}{4}\right)\right]
\label{eq:res}
\end{equation}
where
\begin{equation}
\chi_{k}(N)=v_{k}\sqrt{2N}\frac{\mean{\pt}}{\sqrt{\mean{\pt^2}}}
\label{eq:chik}
\end{equation}
depends on the harmonic anisotropy magnitude $v_{k}$ and the number of particles $N$ used in event-plane reconstruction. The $\pt$ enters into Eq.~(\ref{eq:chik}) because we weighted each particle by its $\pt$ in constructing the event plane. 
In data analysis we solve for the sub-event $\chi_{k}(N/2)$ by Eq.~(\ref{eq:res}) and the known event-plane resolution of the sub-events from Eq.~(\ref{eq:subevt}), employing an iterative procedure~\cite{Poskanzer:1998yz}. From Eq.~(\ref{eq:chik}) we obtain the full event $\chi_{k}(N)=\sqrt{2}\chi_{k}(N/2)$. We then use Eq.~(\ref{eq:res}) to determine the event-plane resolution of the full event~\cite{Poskanzer:1998yz}. The event-plane resolutions are listed in Table~\ref{tab:v2}. The resolutions depend on the $\pt$ bin because particles in a given $\pt$ bin (to be used for correlation analysis) are excluded from the event-plane reconstruction to avoid self-correlations as aforementioned.

One would naively expect that the event-plane resolution should be different for different trigger particle orientations from the event plane because the influence of di-jets on the event-plane determination should vary: a di-jet aligned with the reaction plane enhances the event-plane reconstruction, resulting in a better resolution, whereas a di-jet perpendicular to the reaction plane reduces the accuracy of the constructed event plane, resulting in a poorer resolution. However, this is a post effect due to the selection based on the relative angle between the trigger particle and the corresponding event plane. The resolutions used in Eq.~(\ref{eq:bkgd}), on the other hand, are those of all triggered events before any selection of the trigger particle orientation is made. We have also verified this with Monte Carlo toy model simulations.

Since only triggered events enter into our correlation measurements, the event-plane resolutions are measured using only these events. The event-plane resolutions from inclusive events (minimum-bias events within the given centrality bin) are found to be within a couple of percent of that from the triggered events (see systematic uncertainty discussion in Sec.~\ref{sec:syst:res}). 

We analyzed the elliptic flow in each of the $\pt$ bins used in our correlation analysis. The obtained elliptic flow parameters are tabulated in Table~\ref{tab:v2} together with their systematic uncertainties. The analysis of the elliptic flow and the assessment of its systematic uncertainty are both described in Sec.~\ref{sec:syst:v2}. We used these $\vf$ parameters for background subtraction. The calculated magnitudes of the elliptic flow modulation, $2\va\vtR$, are listed in Table~\ref{tab:v2v2} together with their systematic uncertainties. The calculated background curves are superimposed in Fig.~\ref{fig:raw}.
As seen from Fig.~\ref{fig:raw} and Table~\ref{tab:v2v2}, the flow modulation is relatively small for $\pi/4<\phis<\pi/3$; this is because $\vtR$ is the average within a given $\phis$ bin, as aforementioned, which is close to zero for $\phis\sim 45^\circ$. It is more so for the $\pi/4<\phis<\pi/3$ bin than for its ``symmetric'' $\pi/6<\phis<\pi/4$ bin because other harmonics also contribute to the average only within the limited $\phis$ bin (cf.~Eq.~(\ref{eq:vR})).

\begin{table*}[hbt]
\caption{Elliptic flow and event-plane resolutions as a function of $\pt$ in 20-60\% minimum-bias Au+Au collisions. The resolutions depend on the $\pt$ bin because particles in a given $\pt$ bin are excluded from the event-plane reconstruction to avoid self-correlations. The errors on $\vf$ are systematic uncertainties given by two-particle $\veta{2}{0.7}$ (with a reference particle $0.15<\pt<2$~\gev) and four-particle $\ff{2}{4}$ (with three reference particles). Systematic uncertainties on the resolutions are negligible.}
\label{tab:v2}
\begin{ruledtabular}
\begin{tabular}{c|cccc}
$\pt$ (\gev) & $\vf$ & $\mean{\cos(2\dPsi)}$ & $\mean{\cos(4\dPsi)}$ & $\mean{\cos(6\dPsi)}$ \\ \hline
0.15 - 0.5 & 0.038 $\pm$ 0.003 & 0.673 & 0.324 & 0.127 \\
0.5 - 1 & 0.082 $\pm$ 0.006 & 0.596 & 0.247 & 0.082 \\
1 - 1.5 & 0.128 $\pm$ 0.010 & 0.637 & 0.286 & 0.104 \\
1.5 - 2 & 0.164 $\pm$ 0.011 & 0.676 & 0.328 & 0.129 \\
2 - 3 & 0.189 $\pm$ 0.012 & 0.704 & 0.360 & 0.150 \\
3 - 4 & 0.194 $\pm$ 0.013 &&& \\
4 - 6 & 0.163 $\pm$ 0.020 &&& \\

\end{tabular}
\end{ruledtabular}
\end{table*}

\begin{table*}[hbt]
\caption{The elliptic flow modulation in the correlation background, $2\va\vtR$, calculated using measurements in Table~\ref{tab:v2}, as a function of $\pta$ (in rows) and $\phis=|\phit-\psiEP|$ (in columns) in minimum-bias 20-60\% Au+Au collisions. Both trigger particle $\pt$ ranges of $3<\ptt<4$~\gev\ and $4<\ptt<6$~\gev\ are listed. Quoted errors are systematic uncertainties. Note the significantly smaller systematic uncertainties out-of-plane than in-plane.}
\label{tab:v2v2}
\begin{ruledtabular}
\begin{tabular}{c|cccccc}
$\pta$ (\gev)	& $0 - \pi/12$ & $\pi/12 - \pi/6$ & $\pi/6 - \pi/4$ & $\pi/4- \pi/3$ & $\pi/3 - 5\pi/12$ & $5\pi/12 - \pi/2$ \\ \hline
 & \multicolumn{6}{c}{$3 < \ptt < 4$~\gev} \\
0.15 - 0.5 & $0.0544\pm0.0046$ & $0.0433\pm0.0039$ & $0.0229\pm0.0025$ & $-0.0028\pm0.0006$ & $-0.0270\pm0.0015$ & $-0.0416\pm0.0028$ \\
0.5 - 1 & $0.1098\pm0.0096$ & $0.0884\pm0.0082$ & $0.0490\pm0.0055$ & $-0.0004\pm0.0018$ & $-0.0466\pm0.0022$ & $-0.0745\pm0.0045$ \\
1 - 1.5 & $0.1793\pm0.0149$ & $0.1435\pm0.0128$ & $0.0776\pm0.0085$ & $-0.0054\pm0.0024$ & $-0.0831\pm0.0042$ & $-0.1301\pm0.0081$ \\
1.5 - 2 & $0.2376\pm0.0178$ & $0.1892\pm0.0152$ & $0.0999\pm0.0100$ & $-0.0128\pm0.0025$ & $-0.1186\pm0.0057$ & $-0.1825\pm0.0105$ \\
2 - 3 & $0.2814\pm0.0194$ & $0.2233\pm0.0166$ & $0.1159\pm0.0108$ & $-0.0199\pm0.0024$ & $-0.1473\pm0.0067$ & $-0.2243\pm0.0121$ \\
 & \multicolumn{6}{c}{$4 < \ptt < 6$~\gev} \\
0.15 - 0.5 & $0.0535\pm0.0047$ & $0.0421\pm0.0041$ & $0.0213\pm0.0028$ & $-0.0045\pm0.0008$ & $-0.0284\pm0.0013$ & $-0.0427\pm0.0026$ \\
0.5 - 1 & $0.1073\pm0.0101$ & $0.0853\pm0.0088$ & $0.0451\pm0.0062$ & $-0.0045\pm0.0025$ & $-0.0502\pm0.0017$ & $-0.0777\pm0.0041$ \\
1 - 1.5 & $0.1758\pm0.0156$ & $0.1390\pm0.0136$ & $0.0717\pm0.0095$ & $-0.0115\pm0.0034$ & $-0.0883\pm0.0035$ & $-0.1344\pm0.0074$ \\
1.5 - 2 & $0.2337\pm0.0186$ & $0.1838\pm0.0162$ & $0.0928\pm0.0113$ & $-0.0201\pm0.0038$ & $-0.1246\pm0.0048$ & $-0.1872\pm0.0097$ \\
2 - 3 & $0.2773\pm0.0202$ & $0.2174\pm0.0177$ & $0.1080\pm0.0123$ & $-0.0280\pm0.0039$ & $-0.1537\pm0.0057$ & $-0.2291\pm0.0113$ \\

\end{tabular}
\end{ruledtabular}
\end{table*}

As mentioned previously, our trigger particle $\pt$ ranges are $3<\ptt<4$~\gev\ and $4<\ptt<6$~\gev. In elementary \pp\ and \dAu\ collisions, the particles in these $\pt$ ranges originate mainly from hard-scatterings and jets. In relativistic heavy ion colllisions, however, a large baryon to meson ratio has been observed in the $\pt$ region around 3~\gev~\cite{Adcox:2001mf,Adams:2006wk}. The reason for the large ratio and the sources of those high $\pt$ particles are still under debate. The coalescence and recombination models~\cite{Hwa:2003bn,Greco:2003mm,Fries:2003kq} can elegantly explain the large baryon to meson ratio from a thermal bath of constituent quarks. On the other hand, the jet-like correlations at small angles relative to trigger particles of $\ptt>3$~\gev, with the long range ridge correlation removed, are measured to be invariant from \pp, \dAu, peripheral to central Au+Au collisions~\cite{Nattrass:2008pn}, and independent of the reaction plane direction in Au+Au collisions, as will be shown in this work. This experimental evidence strongly suggests that those $\ptt>3$~\gev\ particles are mostly of jet origin in Au+Au collisions, just as in \pp\ and \dAu\ collisions. It is possible that recombination may still be at work in our trigger particle $\pt$ ranges, but in such a fashion that the parton(s) prior to recombination have already imprinted angular correlations related to the hard-scatterings~\cite{Fries:2004hd}.

Different sources, such as the recombination~\cite{Hwa:2003bn,Greco:2003mm,Fries:2003kq} and jet fragmentation discussed above, will likely give different anisotropies to those high $\pt$ particles. However, the anisotropy of the trigger particles to be used in the background subtraction in Eqs.~(\ref{eq:bkgd}), (\ref{eq:vR}), (\ref{eq:v2R}), and (\ref{eq:v4R}) should be the experimentally measured net anisotropy~\cite{Ulery:2006iw}, as we have done in this work, irrespective of the different origins. 

\subsection{Triangular and High-Order Harmonic Flow Background~\label{sec:v3bkgd}}

In Eq.~(\ref{eq:bkgd}) we have neglected the odd harmonic terms, such as $2v_1^{(a)}v_1^{(t,R)}\cos(\dphi)$ and $2\vvva\vvvtR\cos(3\dphi)$. Due to symmetry at mid-rapidity, the averages of the odd harmonic coefficients $v_1$, $v_3$, and etc.~vanish. 
However, their fluctuations would yield non-vanishing averages of the products of $v_1^{(a)}v_1^{(t)}$ and $\vvva\vvvt$, thereby contributing to the background in the dihadron correlations. 
If one assumes that the amplitude of the $v_1$ (directed flow) fluctuations is of the same order of magnitude as the maximum $v_1$ in our pseudorapidity range (which was measured to be small~\cite{Abelev:2008jga}), then the $v_1$ fluctuation contribution can be neglected~\cite{Abelev:2009ac,Abelev:2009ad}. In the present work we neglect any direct flow fluctuation effect in our background subtraction. 
More recent developments~\cite{Teaney:2010vd,Jia:2012gu} in the understanding of initial geometry fluctuations, however, suggest that $v_1$ fluctuation effects (sometimes called rapidity-even $v_1$) may not be small as originally thought~\cite{Abelev:2009ac,Abelev:2009ad}. We remark in Sec.~\ref{sec:v3} on the magnitude of the possible $v_1$ fluctuation effects using recent measurements.

Note that the possible effect of statistical global momentum conservation can generate a negative dipole~\cite{Borghini:2000cm} which has the same shape as the $v_1$ fluctuation effect. 
However, the statistical momentum conservation effect is not from $v_1$ fluctuations, but part of the correlation signal, the same as momentum conservation by any other mechanisms, such as dijet production.

It has been shown that the initial fluctuations in the overlap geometry (spatial distribution of participating nucleons) give rise to $v_3$ (triangular flow) fluctuations~\cite{Alver:2010gr,Sorensen:2010zq,Petersen:2010cw}. 
It was found from the Monte Carlo Glauber model~\cite{Miller:2007ri} that the triangularity due to geometry fluctuations can be comparable to the magnitude of the eccentricity, which is connected to the elliptic flow~\cite{Alver:2010gr}. It is thus possible that large triangular flow fluctuations can arise which would give triangular peaks in the flow background~\cite{Alver:2010gr,Sorensen:2010zq,Petersen:2010cw}. This appears to be the case in the AMPT (A Multi-Phase Transport) model and the UrQMD (Ultrarelativistic Quantum Molecular Dynamics) model studied in Ref.~\cite{Alver:2010gr,Xu:2010du,Xu:2011fe} and~\cite{Petersen:2010cw}, respectively. Hydrodynamic calculations with event-by-event geometry fluctuations confirm that sizeable $v_3$ can be generated from initial geometry fluctuations~\cite{Schenke:2010rr,Qiu:2011iv,Schenke:2011bn,Song:2010mg,Schenke:2012wb}. The $v_3$ magnitude is smaller than that of $v_2$ despite the similar initial triangular and elliptic eccentricities of $\epsilon_3$ and $\epsilon_2$, respectively. This is likely due to the larger damping power of shear viscosity on $v_3$ than on $v_2$~\cite{Romatschke:2007mq,Schenke:2010rr,Qiu:2011iv,Schenke:2011bn,Song:2010mg}.

Since the orientation of the triangular overlap shape due to fluctuations is random relative to the event-plane direction~\cite{Lacey:2010av,Teaney:2013dta}, determined by the elliptic anisotropy, the effect of any triangular flow is independent of the event plane. In other words, the triangular flow background would be proportional to $2\vvva\vvvtR\cos(3\dphi)=2\vvva\vvvt\cos(3\dphi)$ independent of $\phis$. With triangular flow, the flow background of Eq.~(\ref{eq:bkgd}) becomes
\begin{widetext}
\begin{equation}
\frac{dN}{d\dphi}=B\left[1+2\va\vtR\cos(2\dphi)+2\vvaPsi\vvtRPsi\cos(4\dphi)+2\vvva\vvvt\cos(3\dphi)\right].\label{eq:bkgd_v3}
\end{equation}
\end{widetext}

We may estimate the effect of triangular flow fluctuations in our correlation measurements. The AMPT and UrQMD models indicate that in the 20-60\% centrality range the triangular flow fluctuation effect is about 10\% of the elliptic flow for our trigger and associated $\pt$ bins, $v_{3}^{2}/v_{2}^{2}\approx0.1$~\cite{Alver:2010gr,Petersen:2010cw,Alver:2010dn}. Event-by-event hydrodynamic calculations yield a similar magnitude of $v_{3}^{2}/v_{2}^{2}$~\cite{Gale:2012rq}. 
Experimental data on inclusive two-particle correlations at $\pt>2$~\gev\ indicate a ratio of the harmonic coefficients also of magnitude $v_{3}^{2}/v_{2}^{2}\approx0.1$ within 20-60\% centrality~\cite{Abelev:2008un}. More recent measurements on triangular anisotropy are consistent with these estimates~\cite{Adare:2011tg,Adamczyk:2013waa}. This suggests that the measured third harmonic term in the inclusive two-particle correlations at low $\pt$ may be dominated by triangular flow fluctuations, just as the second harmonic term is dominated by elliptic flow. As we will show in Sec.~\ref{sec:results:near}, the effect of a triangular flow of this magnitude is sizeable in our dihadron correlation measurements with high $\pt$ trigger particles as well. 
In the main work of our study of high-$\pt$ dihadron correlations relative to the EP, we do not include the possible contributions from $v_3$ anisotropy in the flow background subtraction. In Sec.~\ref{sec:v3} we discuss the effect of the presently measured $v_3$ on our dihadron correlation results.

So far only the $\vv$ contribution correlated with the second harmonic plane $\psiPP$ has been considered as in Eq.~(\ref{eq:bkgd}). This part of $\vv$ is referred to as $\vvPsi$. The other part of $\vv$ arises from fluctuations and is uncorrelated to $\psiEP$. We refer to this part as $\VVuc$. 
The flow background is then given by
\begin{widetext}
\begin{equation}
\frac{dN}{d\dphi}=B\left[1+2\va\vtR\cos(2\dphi)+2\vvaPsi\vvtRPsi\cos(4\dphi)+2\vvva\vvvt\cos(3\dphi)+2\VVuc\cos(4\dphi)\right].\label{eq:bkgd_v4}
\end{equation}
\end{widetext}
Section~\ref{sec:v3} discusses how $\VVuc$ is obtained in the present analysis.

Glauber model~\cite{Miller:2007ri} calculations also show that the quadrangularity, pentagonality, and hexagonality due to geometry fluctuations are equal to the triangularity, all large and comparable to the eccentricity~\cite{Alver:2010dn,Heinz:2013th}. However, it has been suggested that those higher order eccentricities are inefficient in generating sizeable high-order harmonic flow in final state momentum space~\cite{Alver:2010dn}. Experimental data also indicate that the magnitudes and fluctuations of $v_4$ and $v_6$ are small relative to the magnitude of $\vf$~\cite{Adams:2003zg}. 
Although we include $\VVuc$ in our flow background of Eq.~(\ref{eq:bkgd_v4}), the effect of $\VVuc$ is small, as will be discussed in  Sec.~\ref{sec:v3}. It is safe to neglect $v_5^2$ and the higher order anisotropic fluctuation terms in the flow background of Eq.~(\ref{eq:bkgd}).

\subsection{Background Normalization by \zyam\label{sec:analysis:bkgd}}

The flow correlated backgrounds given by Eq.~(\ref{eq:bkgd}), as an example, are shown in Fig.~\ref{fig:raw} as solid curves. The background curves have been normalized assuming that the background-subtracted signal has Zero Yield At Minimum (\zyam)~\cite{Adams:2005ph,Ajitanand:2005jj}. To obtain the \zyam\ normalization factor, we fold the raw correlation function to within the range of $0<\dphi<\pi$ because of the symmetry of the correlation function. We take the ratio of the folded raw correlation to the background curve of Eq.~(\ref{eq:bkgd}), where $B$ is set to unity before taking the ratio. We obtain a continuous range of the size of $\pi/6$ where the average ratio is the smallest. This smallest average ratio is the normalization factor $B$ to be used in the flow background of Eq.~(\ref{eq:bkgd}), which is then subtracted from the raw correlation function to obtain the final correlation signal. 

The background levels can be different for the different $\phis$ slices because of the net effect of the variations in jet-quenching with $\phis$ and the centrality cuts in total charged particle multiplicity in the TPC within $|\eta|<0.5$. Thus, in our correlation analysis, the background level $B$ is treated independently in individual $\phis$ slices.
In the recent proposal~\cite{Sharma:2015qra} to fit the $\phis$-dependent near-side correlations at large $\deta$ (i.e.~the ridge region) by Fourier coefficients and treat them as backgrounds, the background level $B$ is required to be the same in all $\phis$ slices.

Table~\ref{tab:B} lists the obtained background level $B$ as a function of $\phis$ and $\pta$ in 20-60\% Au+Au collisions. Results from both trigger particle $\pt$ ranges of $3<\ptt<4$~\gev\ and $4<\ptt<6$~\gev\ are listed. The background levels listed are not only for the correlation functions with the $|\deta|<2$ region within our acceptance, but also for those in the large $\deta$ region of $|\deta|>0.7$. The latter is used for the ridge studies (see Sec.~\ref{sec:results:near}).
The background level for the lower trigger particle $\ptt$ range is slightly larger. 
This is due to the fact that relatively more events contain multiple jets with the lower trigger particle $\ptt$ and those events are used multiple times in our di-hadron correlation analysis~\cite{Aggarwal:2010rf}.

It is worth emphasizing here that our quantitative results depend on the assumption of the \zyam\ background normalization, and the effects of variations in the \zyam\ normalization within a reasonable range are assessed by systematic uncertainties. However, as will be discussed in Sec.~\ref{sec:syst:zyam}, our qualitative conclusions are not affected by the \zyam\ normalization.

\begin{turnpage}
\begin{table*}[]
\caption{Background level $B$ in flow subtraction by Eq.~(\ref{eq:bkgd}) as a function of $\pta$ (in rows) and $\phis=|\phit-\psiEP|$ (in columns) in minimum-bias 20-60\% Au+Au collisions. Both trigger particle $\pt$ ranges of $3<\ptt<4$~\gev\ and $4<\ptt<6$~\gev\ are listed. The trigger and associated particles are within $|\eta|<1$. Backgrounds are tabulated for the entire $|\deta|<2$ range of our acceptance as well as for the large $\deta$ cut of $|\deta|>0.7$. 
The first error is statistical. The second error is the quadratic sum of the \zyam\ systematic uncertainty and the one-sided systematic uncertainty due to background deviation from \zyam. The former is assessed by varying the $\dphi$ normalization range. The latter is assessed by comparing our \zyam\ background to those obtained from asymmetric correlations of the separate positive and negative $\phit-\psiEP$ regions.}
\label{tab:B}
\begin{ruledtabular}
\begin{tabular}{c|cccccc}
$\pta$ (\gev)	& $0 - \pi/12$ & $\pi/12 - \pi/6$ & $\pi/6 - \pi/4$ & $\pi/4- \pi/3$ & $\pi/3 - 5\pi/12$ & $5\pi/12 - \pi/2$ \\ \hline
%
%
%
%
%
%
 & \multicolumn{6}{c}{$3 < \ptt < 4$~\gev} \\
0.15 - 0.5&$47.41\pm 0.01^{+ 0.06}_{- 0.07}$&$47.37\pm 0.01^{+ 0.07}_{- 0.09}$&$47.28\pm 0.02^{+ 0.02}_{- 0.17}$&$47.22\pm 0.02^{+ 0.06}_{- 0.08}$&$47.16\pm 0.02^{+ 0.02}_{- 0.10}$&$47.04\pm 0.02^{+ 0.04}_{- 0.06}$ \\
0.5 - 1&$22.47\pm 0.01^{+ 0.01}_{- 0.10}$&$22.67\pm 0.01^{+ 0.03}_{- 0.15}$&$22.92\pm 0.01^{+ 0.03}_{- 0.20}$&$23.31\pm 0.01^{+ 0.06}_{- 0.18}$&$23.40\pm 0.01^{+ 0.06}_{- 0.11}$&$23.59\pm 0.01^{+ 0.03}_{- 0.09}$ \\
1 - 1.5&$6.023\pm0.005^{+0.008}_{-0.015}$&$6.072\pm0.005^{+0.020}_{-0.089}$&$6.128\pm0.005^{+0.016}_{-0.106}$&$6.177\pm0.005^{+0.033}_{-0.081}$&$6.128\pm0.005^{+0.022}_{-0.042}$&$6.199\pm0.006^{+0.014}_{-0.028}$ \\
1.5 - 2&$1.683\pm0.002^{+0.005}_{-0.007}$&$1.691\pm0.003^{+0.002}_{-0.034}$&$1.698\pm0.002^{+0.003}_{-0.046}$&$1.700\pm0.003^{+0.010}_{-0.034}$&$1.694\pm0.003^{+0.006}_{-0.036}$&$1.694\pm0.003^{+0.001}_{-0.013}$ \\
2 - 3&$0.655\pm0.002^{+0.004}_{-0.002}$&$0.662\pm0.002^{+0.003}_{-0.017}$&$0.663\pm0.002^{+0.003}_{-0.028}$&$0.660\pm0.002^{+0.002}_{-0.026}$&$0.654\pm0.002^{+0.001}_{-0.014}$&$0.659\pm0.002^{+0.008}_{-0.011}$ \\
%
%
%
%
%
%
%
 & \multicolumn{6}{c}{$4 < \ptt < 6$~\gev} \\
0.15 - 0.5&$46.63\pm 0.04^{+ 0.02}_{- 0.12}$&$46.56\pm 0.04^{+ 0.08}_{- 0.16}$&$46.72\pm 0.04^{+ 0.08}_{- 0.24}$&$46.77\pm 0.04^{+ 0.08}_{- 0.27}$&$46.67\pm 0.05^{+ 0.12}_{- 0.07}$&$46.76\pm 0.05^{+ 0.12}_{- 0.15}$ \\
0.5 - 1&$22.16\pm 0.02^{+ 0.01}_{- 0.07}$&$22.30\pm 0.02^{+ 0.09}_{- 0.22}$&$22.42\pm 0.02^{+ 0.00}_{- 0.32}$&$23.11\pm 0.03^{+ 0.06}_{- 0.19}$&$23.07\pm 0.03^{+ 0.11}_{- 0.09}$&$23.42\pm 0.03^{+ 0.07}_{- 0.17}$ \\
1 - 1.5&$5.947\pm0.012^{+0.003}_{-0.049}$&$5.989\pm0.012^{+0.001}_{-0.084}$&$5.985\pm0.012^{+0.006}_{-0.109}$&$6.113\pm0.013^{+0.040}_{-0.101}$&$6.076\pm0.014^{+0.021}_{-0.061}$&$6.174\pm0.014^{+0.037}_{-0.057}$ \\
1.5 - 2&$1.659\pm0.006^{+0.003}_{-0.041}$&$1.664\pm0.006^{+0.003}_{-0.035}$&$1.673\pm0.006^{+0.017}_{-0.050}$&$1.671\pm0.007^{+0.024}_{-0.051}$&$1.674\pm0.007^{+0.013}_{-0.022}$&$1.712\pm0.007^{+0.014}_{-0.038}$ \\
2 - 3&$0.611\pm0.004^{+0.001}_{-0.006}$&$0.618\pm0.004^{+0.003}_{-0.017}$&$0.613\pm0.004^{+0.001}_{-0.024}$&$0.621\pm0.004^{+0.010}_{-0.028}$&$0.615\pm0.004^{+0.008}_{-0.012}$&$0.615\pm0.005^{+0.004}_{-0.014}$ \\
3 - 4&$0.058\pm0.001^{+0.001}_{-0.002}$&$0.060\pm0.001^{+0.001}_{-0.009}$&$0.061\pm0.001^{+0.002}_{-0.014}$&$0.063\pm0.001^{+0.000}_{-0.016}$&$0.061\pm0.001^{+0.001}_{-0.020}$&$0.062\pm0.001^{+0.001}_{-0.004}$ \\
%
%
%
%
%
%
 & \multicolumn{6}{c}{$3 < \ptt < 4$~\gev, $|\deta|>0.7$} \\
0.15 - 0.5&$19.37\pm 0.01^{+ 0.02}_{- 0.05}$&$19.35\pm 0.01^{+ 0.02}_{- 0.06}$&$19.29\pm 0.01^{+ 0.00}_{- 0.12}$&$19.25\pm 0.01^{+ 0.00}_{- 0.06}$&$19.28\pm 0.01^{+ 0.02}_{- 0.05}$&$19.22\pm 0.01^{+ 0.00}_{- 0.03}$ \\
0.5 - 1&$9.187\pm0.006^{+0.001}_{-0.043}$&$9.268\pm0.006^{+0.022}_{-0.100}$&$9.356\pm0.006^{+0.022}_{-0.100}$&$9.507\pm0.007^{+0.022}_{-0.102}$&$9.548\pm0.007^{+0.009}_{-0.024}$&$9.603\pm0.007^{+0.041}_{-0.028}$ \\
1 - 1.5&$2.452\pm0.003^{+0.004}_{-0.006}$&$2.475\pm0.003^{+0.006}_{-0.050}$&$2.497\pm0.003^{+0.012}_{-0.061}$&$2.512\pm0.003^{+0.014}_{-0.053}$&$2.493\pm0.003^{+0.012}_{-0.025}$&$2.517\pm0.004^{+0.017}_{-0.015}$ \\
1.5 - 2&$0.683\pm0.002^{+0.003}_{-0.005}$&$0.688\pm0.002^{+0.003}_{-0.020}$&$0.691\pm0.002^{+0.003}_{-0.026}$&$0.689\pm0.002^{+0.005}_{-0.017}$&$0.686\pm0.002^{+0.007}_{-0.016}$&$0.685\pm0.002^{+0.005}_{-0.004}$ \\
2 - 3&$0.264\pm0.001^{+0.001}_{-0.004}$&$0.269\pm0.001^{+0.002}_{-0.008}$&$0.269\pm0.001^{+0.002}_{-0.014}$&$0.267\pm0.001^{+0.002}_{-0.015}$&$0.265\pm0.001^{+0.003}_{-0.010}$&$0.264\pm0.001^{+0.004}_{-0.002}$ \\
%
%
%
%
%
%
%
 & \multicolumn{6}{c}{$4 < \ptt < 6$~\gev, $|\deta|>0.7$} \\
0.15 - 0.5&$18.98\pm 0.02^{+ 0.00}_{- 0.05}$&$18.99\pm 0.02^{+ 0.01}_{- 0.10}$&$19.01\pm 0.03^{+ 0.00}_{- 0.11}$&$18.98\pm 0.03^{+ 0.02}_{- 0.16}$&$18.98\pm 0.03^{+ 0.01}_{- 0.01}$&$19.00\pm 0.03^{+ 0.02}_{- 0.07}$ \\
0.5 - 1&$9.014\pm0.015^{+0.003}_{-0.045}$&$9.065\pm0.015^{+0.014}_{-0.116}$&$9.110\pm0.015^{+0.017}_{-0.122}$&$9.374\pm0.017^{+0.035}_{-0.108}$&$9.391\pm0.018^{+0.025}_{-0.077}$&$9.518\pm0.018^{+0.004}_{-0.120}$ \\
1 - 1.5&$2.421\pm0.008^{+0.008}_{-0.012}$&$2.434\pm0.008^{+0.003}_{-0.036}$&$2.435\pm0.008^{+0.009}_{-0.078}$&$2.456\pm0.008^{+0.023}_{-0.045}$&$2.457\pm0.009^{+0.011}_{-0.034}$&$2.501\pm0.009^{+0.008}_{-0.019}$ \\
1.5 - 2&$0.673\pm0.004^{+0.001}_{-0.011}$&$0.669\pm0.004^{+0.003}_{-0.017}$&$0.677\pm0.004^{+0.009}_{-0.032}$&$0.681\pm0.004^{+0.010}_{-0.039}$&$0.677\pm0.005^{+0.007}_{-0.010}$&$0.691\pm0.005^{+0.004}_{-0.018}$ \\
2 - 3&$0.241\pm0.003^{+0.002}_{-0.004}$&$0.250\pm0.003^{+0.004}_{-0.010}$&$0.248\pm0.003^{+0.001}_{-0.014}$&$0.253\pm0.003^{+0.002}_{-0.014}$&$0.245\pm0.003^{+0.004}_{-0.007}$&$0.247\pm0.003^{+0.003}_{-0.004}$ \\
3 - 4&$0.023\pm0.001^{+0.001}_{-0.001}$&$0.024\pm0.001^{+0.000}_{-0.003}$&$0.025\pm0.001^{+0.000}_{-0.004}$&$0.025\pm0.001^{+0.001}_{-0.008}$&$0.023\pm0.001^{+0.001}_{-0.007}$&$0.024\pm0.001^{+0.000}_{-0.002}$ \\

\end{tabular}
\end{ruledtabular}
\end{table*}
\end{turnpage}


\section{Systematic Uncertainties\label{sec:syst}}

Background subtraction is the major source of systematic uncertainty in our results. We first study the dihadron correlations with even harmonic flow subtraction. The even harmonic flow background, as given by Eq.~(\ref{eq:bkgd}), has three important ingredients: the anisotropic flow measurements $\vf$ and $\vv$, the event-plane resolutions, and the background magnitude $B$. We discuss these systematic uncertainties in Sections~\ref{sec:syst:v2}-\ref{sec:syst:zyam}, respectively. They have effects on the dihadron correlation functions presented in Sec.~\ref{sec:results:corr} and the away-side correlation widths and magnitudes presented in Sec.~\ref{sec:results:away}. 

We also report results on near-side jet-like and ridge correlations in Sec.~\ref{sec:results:near}. Uncertainties in $\vf$ and the \zyam\ background normalization contribute to the uncertainties in the ridge correlation results. They do not affect the jet-like correlation results, in which they largely cancel because $\vf$ is approximately independent of pseudo-rapidity within our acceptance. Additional systematic uncertainties arise from the assumption of a uniform ridge in $\deta$, which affects both the ridge and jet-like results. These additional systematic uncertainties are discussed in Sec.~\ref{sec:syst:ridge}.

We also study the dihadron correlations with background subtraction, including odd harmonic flow. Various systematics are discussed together with the correlation results in Sec.~\ref{sec:v3}.

\subsection{Systematic Uncertainty due to Anisotropic Flow\label{sec:syst:v2}}

The anisotropic flow (mainly elliptic flow) background which is to be subtracted from the dihadron correlation is the anisotropy caused by particle correlations to the participant plane~\cite{Wang:2009af,Wang:2008gp}. There are several measurements of elliptic flow; many of them are affected to various degrees by nonflow contributions that are caused by particle correlations unrelated to the reaction plane (or participant plane), such as resonance decays and jet-correlations. One technique, called the event-plane method, is to construct the event plane from all charged particles except those of interest and then calculate \vEP$=\mean{\cos2(\phi-\psiEP)}/\mean{\cos2\dPsi}$ for the particles of interest, where $\mean{\cos2\dPsi}$ is the event-plane resolution~\cite{Poskanzer:1998yz}. This method is affected by nonflow contributions in both sets of particles, those of interest and those used to construct the event plane. The \vEP\ already contains flow fluctuation effects which should be included in the jet-correlation background. 

Another method, called the two-particle method, is to calculate $\flow{2}=\sqrt{\mean{\cos2\dphi}}$ using all particle pairs of interest~\cite{Poskanzer:1998yz}. This method is affected by nonflow only in the particles of interest used for correlation studies. This flow parameter also contains flow fluctuation effects. The two-particle cumulant method can also be applied between the particle of interest and a reference particle. The anisotropy of the particle of interest is then the ratio of the two-particle cumulant to the anisotropy of the reference particles, which can be in turn obtained from the two-particle cumulant between reference particle pairs. (More details are given in Sec.~\ref{sec:v3}.) This method of mixed pair cumulant is intrinsically similar to the event-plane method.

The third method, called the four-particle method, is to obtain $\flow{4}$ from the four-particle cumulant~\cite{Adler:2002pu}. 
This method is less affected by nonflow from particle clustering because the nonflow arising from two particle correlations is eliminated, and the nonflow from three particle correlations does not contribute. This method is subject to nonflow from higher orders (four-particle correlation and above) but those contributions are suppressed by high orders of multiplicity~\cite{Adler:2002pu}. The flow fluctuation will give a negative contribution to $\flow{4}$~\cite{Adler:2002pu}.

The fourth method is to decompose the low $\pt$ two-particle correlation (the so-called untriggered correlation, without the requirement of a trigger particle) into a near-angle Gaussian, a dipole, and a quadrupole, and infer $\vf$\{{\sc 2d}\} from the fitted quadrupole~\cite{Kettler:2009zz}. The method attempts to geometrically separate the reaction-plane correlated $\vf$ from other (i.e.~nonflow) correlations (small-angle correlations and large-angle dipole). 
However, the method assumes a particular functional form for those nonflow correlations, whereas the goal of this paper is to study the magnitude and shape of those nonflow (jet) correlations, defined to be the data minus harmonic (flow) backgrounds. 

The measured $\flow{2}$ and \vMRP\ are similar and they both significantly overestimate elliptic flow due to large contributions from nonflow and fluctuations. While the flow fluctuation effect should be included in our background subtraction, nonflow should be excluded. The major component of nonflow is the measured small-angle two-particle correlation~\cite{Daugherity:2008su,Agakishiev:2011pe,Abdelwahab:2014sge}. To suppress nonflow, a pseudo-rapidity $\eta$-gap ($\etagap$) is often applied between the particle pair in the $\ff{n}{2}$ measurement, and in the \vEP\ measurement, between the particle of interest and the particles used in EP reconstruction. In this analysis, we apply $\etagap=0.7$ to obtain the two-particle cumulant elliptic flow, $\veta{2}{0.7}$. However, the away-side two-particle correlations, presumably due to jet-like correlations, cannot be eliminated~\cite{Xu:2012ue,Wang:2013qca,Wang:2014sfa}. This is because the inter-jet correlation in $\eta$ is broad (nearly uniform in the STAR TPC acceptance) due to the unconstrained underlying parton kinematics in the longitudinal direction.

We use $\ff{2}{2}$ as our upper systematic bound for $v_2$. The $\ff{2}{2}$ is measured in 10\%-size centrality bins. Two-particle cumulants between the particle of interest and a reference particle, $\FF{n}{\pt\mbox{-ref}}$, and between two reference particles, $\FF{n}{\mbox{ref-ref}}$, are calculated. The particle of interest is from a particular $\pt$ bin, while the reference particle is from $0.15<\pt<2$~\gev. To reduce nonflow one particle is taken from $\eta<-0.35$ and the other from $\eta>0.35$, with an $\etagap=0.7$ in-between. The $v_n$ are referred to as $\veta{n}{0.7}$ or simply as $\ff{n}{2}$. The cumulants are calculated by the Q-cumulant method and divided by the corresponding number of pairs in each event. The cumulants are averaged over the event sample with a unit weight (not weighted by the number of pairs). The anisotropy of the particle of interest is simply given by
\begin{equation}
\ff{n}{2}(\pt)=\frac{\FFeta{n}{\pt\mbox{-ref}}{0.7}}{\sqrt{\FFeta{n}{\mbox{ref-ref}}{0.7}}}\,.\label{eq:vn}
\end{equation}
The $\ff{n}{2}$ of the four individual centralities are averaged by weighting each centrality by the number of particles of interest. 

The measured $\flow{4}$ likely underestimates elliptic flow because the flow fluctuation effect in $\flow{4}$ is negative~\cite{Adler:2002pu}. We note that $\flow{4}$ may still contain some nonflow effects. However, the agreement between $\flow{4}$ and the elliptic flow measurement using the Lee-Yang-Zero method suggests that such nonflow effects are small~\cite{Abelev:2008ae}. We therefore use $\flow{4}$ as our lower bound of $\vf$ systematic uncertainty, the same as in Refs.~\cite{Adams:2005ph,Aggarwal:2010rf}. 
The $\ff{2}{4}$ is obtained as follows. Two four-particle cumulants are calculated. One is for quadralets of one particle of interest and three reference particles, referred to as $\FF{2}{\pt\mbox{-ref}^3}$. The other is for quadralets of four reference particles, referred to as $\FF{2}{\mbox{ref}^4}$. Since nonflow is negligible in $\ff{n}{4}$, no $\etagap$ is applied; all four particles are from the entire region of $|\eta|<1$. Similar to $\ff{2}{2}$, the Q-cumulant method is used to calculate $\ff{2}{4}$. Self-correlations are properly removed. 
The four-particle anisotropy of the particle of interest is given by
\begin{equation}
\ff{2}{4}(\pt)=\FF{2}{\pt\mbox{-ref}^3}/(\FF{2}{\mbox{ref}^4})^{3/4}\,.\label{eq:v24}
\end{equation}
Again the $\ff{2}{4}(\pt)$ of the four individual centralities are averaged by weighting each centrality by the number of particles of interest. 

As the default $\vf$, we use the average, 
\begin{equation}
v_2=(\ff{2}{2}+\ff{2}{4})/2\,.\label{eq:v2}
\end{equation}
We use the range bracketed by $\ff{2}{2}$ and $\ff{2}{4}$ as our systematic uncertainty on $\vf$. 
Table~\ref{tab:v2} lists the default $\vf$ values together with systematic uncertainties for different $\pt$ bins in 20-60\% Au+Au collisions. 

We parameterized the $\vv$ measurement~\cite{Adams:2004bi} as 
\begin{equation}
\vvPsi=1.15v_2^2\,,\label{eq:v4}
\end{equation}
and used this parameterization for both trigger and associated particles in our flow correction~\cite{Abelev:2008ac}. The uncertainties in $v_2$ are propagated to $\vv$. Note that the $\vv$ fluctuation effects related to the second harmonic event plane, which should be included in our flow background, are already included in the $\vv$ measurement which was carried out with respect to the second harmonic event plane~\cite{Adams:2004bi,Adams:2003zg}. Fluctuations in $\vv$ related to the fourth harmonic event plane could be potentially not small~\cite{Adams:2003zg} and are not included in the available measurement of $\vv$. However, these fluctuation effects come into our two-particle correlation background as $\vv^{2}$ (not through the cross-term of $\vf\vv$) and are therefore negligible for our centrality range. Nevertheless, in Sec.~\ref{sec:v3}, we also include this fluctuation effect in flow subtraction.

The flow backgrounds are shown by the solid curves in Fig.~\ref{fig:raw}. The systematic uncertainties due to anisotropic flow parameters are shown by the dashed curves. The normalization of each background curve is adjusted by \zyam\ to match the raw correlation function such that the background-subtracted correlation is zero at the minimum (see Sec.~\ref{sec:analysis:bkgd}). As seen from the figures, the dashed curves are not symmetric about the solid curve. This is mainly due to the \zyam\ normalization, as the normalization region is around $\dphi\approx\pm1$, not at $\pm\pi/2$.

The coefficient $\va\vtR$ in Eq.~(\ref{eq:bkgd}) determines the size of the modulation in the flow background. These coefficients are tabulated in Table~\ref{tab:v2v2}. For in-plane trigger particles, $\vtR$ is positive as given by Eq.~(\ref{eq:vR}) or~(\ref{eq:v2R}). The correlated elliptic flow uncertainties in $\va$ and $\vtR$ gives a large uncertainty in $\va\vtR$. For out-of-plane trigger particles, however, $\vtR$ is negative. The correlated uncertainties in $\va$ and $\vtR$ tend to cancel each other, resulting in a small uncertainty in $\va\vtR$. This is apparent in the systematic uncertainties listed in Table~\ref{tab:v2v2}. It is shown in the systematic uncertainty background curves in Fig.~\ref{fig:raw}, where the uncertainty for in-plane correlations is large, while for out-of-plane correlations it is small.

\subsection{Systematic Uncertainty due to Event-Plane Resolution\label{sec:syst:res}}

The event-plane resolutions enter into the flow background modulation together with the anisotropic flow parameters, via $v_{n}\mean{\cos(k\dPsi)}$. Terms with $k=n$ are not affected by uncertainties in the event-plane resolutions, because $v_{n}\mean{\cos(n\dPsi)}$ are the measured anisotropic flow parameters. The event-plane resolutions $\mean{\cos(k\dPsi)}$ of different $k$'s are likely correlated, hence the uncertainty in $v_{n}\mean{\cos(k\dPsi)}$ for $k\neq n$ due to uncertainties in the resolutions may be greatly reduced. To be conservative, we assume the uncertainties in the event-plane resolutions to be uncorrelated in our estimation of their effects on our correlation results.

The systematic uncertainty of the event-plane resolution was determined by repeating the sub-event method, but splitting the particles by charge instead of by random determination, as done in the default case.
It is also assessed by comparing the event-plane resolution from triggered events only (default) to inclusive events, and by applying a weighting of the number of trigger particles (default) and not applying this weighting.
In addition, differences in event-plane resolutions were assessed with (default) and without event-plane flattening by weighting of the inverse of $\phi$-dependent efficiencies. The event-plane resolution uncertainties thus estimated are typically less than 1\% for $\mean{\cos(2\dPsi)}$, and less than 2-3\% for $\mean{\cos(4\dPsi)}$ and $\mean{\cos(6\dPsi)}$. 

The effects of the estimated event-plane resolution uncertainties on the final background-subtracted correlation functions are significantly smaller than those caused by the uncertainties on anisotropic flow, and are therefore neglected.

\subsection{Effect of Finite Centrality Bin Width}

For the data reported in this paper, the entire 20-60\% Au+Au centrality range is treated as a single centrality bin in which the event-plane resolutions and elliptic flow are obtained and the azimuthal correlation is analyzed. Alternatively, the analysis was repeated in each of the four 10\%-size centrality bins using the corresponding event-plane resolutions and the elliptic flow measurements. Those correlation results were added together, weighted by the number of trigger particles in each centrality bin. The recombined results are consistent with using a single 20-60\% centrality bin, well within the systematic uncertainties due to those in flow subtraction and \zyam\ normalization. This is because the measured elliptic flow $\vf$ is fairly constant over the entire 20-60\% centrality range, so that $\mean{\vt\va}\approx\mean{\vt}\mean{\va}$. 
The event-plane resolutions vary with centrality mainly due to the multiplicity change. However, the event-plane resolutions enter into the flow background of Eq.~(\ref{eq:bkgd}) linearly, and because the high $\pt$ trigger particle multiplicity scales almost linearly with the total multiplicity, the effect of the centrality-varying event-plane resolution is minimal in the flow correction calculated from the single 20-60\% centrality bin or summed from multiple narrower centrality bins.

\subsection{Systematic Uncertainty due to \zyam\ Background Normalization\label{sec:syst:B}}

Naively one would expect the background level $B$ in Eq.~(\ref{eq:bkgd}) to be the same for all $\phis$ slices because the underlying background should not depend on the signal (or orientation of the trigger particle). However, there could be biases in the event samples with trigger particles at different $\phis$ such that they contain slightly different underlying background multiplicities due to the possible difference in jet-like correlated multiplicities at different $\phis$ and the overall constraints caused by centrality cuts on the reference multiplicity. In our analysis we use different $B$ values for different $\phis$ slices, each independently obtained using \zyam\ on the correlation function of the corresponding slice. 

One source of systematic uncertainty on $B$ is due to the limited range in $\dphi$ where the background-subtracted correlations appear to have a minimum `plateau'. This part of the systematic uncertainty is assessed by varying the size of the normalization range in $\dphi$ between $\pi/12$ and $\pi/4$ (the default range is $\pi/6$), similar to Ref.~\cite{Adams:2005ph}. 

The \zyam\ assumption likely gives an upper limit to the underlying background level. One could make an improved assessment of the background level with more stringent requirements, such as using three-particle correlation \zyam~\cite{Abelev:2008ac}. However, the analysis of three-particle correlation within a limited $\phis$ range of the trigger particle is difficult.

In this paper, we assess this part of the systematic uncertainty on $B$ by comparing to the \zyam\ backgrounds obtained separately from correlation functions at positive $\phit-\psiEP$ and negative $\phit-\psiEP$. 
Those \zyam\ backgrounds are always lower than our default $B$ from \zyam\ of the combined correlation function of positive and negative $\phit-\psiEP$. 
This is because the separately analyzed correlation functions are asymmetric about $\dphi=0$ and $\dphi=\pi$, and the \zyam\ is determined by only one side of the correlation function~\cite{Wang:2008hg,Konzer:2009xp}, whereas in our combined correlation functions reported here, the two sides of the separately analyzed asymmetric correlation functions are averaged. We treat the difference between the \zyam\ background from this paper and that obtained from the asymmetric correlation functions as an additional, one-sided systematic uncertainty on $B$.

We may also study the background level by fitting the \zyam-background-subtracted correlation functions with a combination of Gaussians and a free parameter for an offset from zero. Specifically, we fit the correlation data to three Gaussians (a near-side Gaussian at $\dphi=0$ and two away-side Gaussians symmetric about $\dphi=\pi$), and four Gaussians (adding a fourth Gaussian at $\dphi=\pi$ with the same width as the near-side Gaussian). Some of the fits yielded unphysical offsets because of the limited constraint of the correlation data on the fit model. For the other fits, the fitted offsets are comparable to the systematic uncertainty obtained from the comparisons to the asymmetric correlation functions discussed above. 
The Gaussian fits to the correlation functions without the offset will be discussed in Sec.~\ref{sec:results:peaks}.

The different sources of systematic uncertainties on $B$ are added in quadrature. The total systematic uncertainty is listed in Table~\ref{tab:B} together with the statistical uncertainty. We take the quadratic sum of the statistical and systematic uncertainties as the total uncertainty for $B$ on our correlation results.

\subsection{Is the Away-Side Double-Peak an Artifact of \zyam?\label{sec:syst:zyam}}

As will be shown in Sec.~\ref{sec:results:corr}, the $v_2$ and $v_4$ background-subtracted correlation functions on the away side are single-peaked at $\dphi=\pi$ for triggered particles in-plane, but double-peaked for trigger particles out-of-plane beyond the flow systematic uncertainties. Since the subtracted background is flow-modulated, the natural question is whether the away-side double-peak structure is due to an unrealistic systematic uncertainty. To address this question, it is worth noting that the flow background modulation changes phase when the trigger particle moves from in-plane to out-of-plane, as shown in Fig.~\ref{fig:raw}. A smaller elliptic flow would make the in-plane correlation more peaked at $\dphi=0$ and $\pi$ and the out-of-plane correlation more dipped at $\pi$ (hence more double peaked on the away side). On the other hand, a larger elliptic flow would make the out-of-plane away-side correlation less double-peaked. One would need a $\sim$15\% larger $\va\vtR$ than in Table~\ref{tab:v2v2}, significantly beyond the systematic uncertainty from the anisotropy measurements, to eliminate the away-side double-peak for the out-of-plane $\phis$ slice. However, this large $\va\vtR$ would result in double-peaked away-side correlations for some of the other $\phis$ slices.

The background magnitude affects the absolute magnitude of the flow modulation subtracted from the raw data in obtaining the correlation signal. Since the background normalization is determined by the \zyam\ description, the question arises whether the away-side double-peak for the out-of-plane $\phis$ slices is an artifact of a significantly smaller background level than \zyam\ beyond the \zyam\ normalization systematic uncertainty. The answer is negative because the flow background is the lowest at $\dphi=\pi$ for out-of-plane trigger particles. Allowing a non-zero flow-modulated ``pedestal'' into the correlation signal will exaggerate the double-peak feature, i.e., the dip at $\dphi=\pi$ will be even deeper than the double peaks. In other words, if the true background is lower than \zyam, then the away-side correlation functions for out-of-plane trigger particles will be more double-peaked. Only when the background is larger than \zyam\ would the dihadron correlation signal become single-peaked; however, as a result the signal strength would become negative.

\begin{figure*}[hbt]
\centerline{\includegraphics[width=1.03\textwidth]{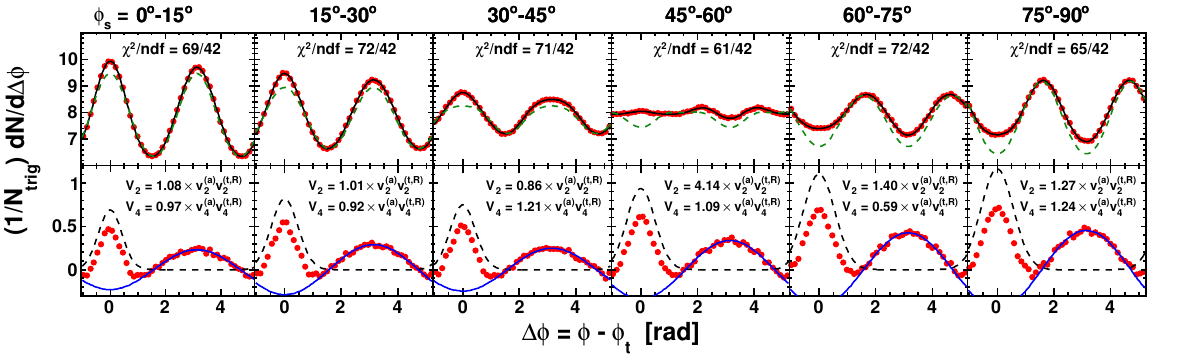}}
\caption{(Color online) Fit by Eq.~(\ref{eq:estruct}) to the raw correlation data in the upper panels of Fig.~\ref{fig:raw}. (Upper panels) The solid curves are the fit results and the dashed curves are the fitted flow background. (Lower panels) The correlation functions after subtracting the fitted flow background. The text in each plot gives the fitted $V_2$ and $V_4$ results relative to the measured $\va\vtR$ and $\vvaPsi\vvtRPsi$, respectively. The fitted same-side Gaussian and negative dipole are depicted individually in the dashed and solid curves, respectively.}
\label{fig:estruct}
\end{figure*}

In summary, to eliminate the away-side double-peak, one needs either a larger anisotropic flow than measured while fixing the background normalization by \zyam, or a larger background normalization than \zyam\ while fixing the anisotropic flow as measured. To investigate further the interplay between background normalization and anisotropic flow and its effect on the dihadron correlation signal, we performed a study of free fits to the raw correlation data, treating the anisotropic flow and the background magnitude as free parameters. In order to do so, one needs a prescription for the correlation signal functional form. It has been shown that the sum of a near-side Gaussian, a negative dipole, and a quadrupole (reflecting elliptic flow) can adequately describe the two-particle azimuthal correlation at low $\pt$ without the requirement of a high $\pt$ trigger particle~\cite{Daugherity:2008su,Agakishiev:2011pe}. Thus, we fit our raw correlation data by
\begin{eqnarray}
\frac{dN}{d\dphi}&=&B\left(1+2V_{2}\cos2\dphi+2V_{4}\cos4\dphi\right)+\nonumber\\
&&A_{\rm ns}\exp\left(-\frac{(\dphi)^{2}}{2\sigma_{\rm ns}^2}\right)-A_{\rm dipole}\cos\dphi,\label{eq:estruct}
\end{eqnarray}
treating the flow modulations $V_{2}$ and $V_{4}$, the near-side Gaussian parameters $A_{\rm ns}$ and $\sigma_{\rm ns}$, and the negative dipole magnitude $A_{\rm dipole}$ as free parameters. Figure~\ref{fig:estruct} (upper panels) shows the fits by Eq.~(\ref{eq:estruct}) to the raw correlation functions in six $\phis$ slices for $3<\ptt<4$~\gev\ and $1<\pta<2$~\gev. The fits are shown by the solid curves. The dashed curves show the fitted flow backgrounds. The lower panels of Fig.~\ref{fig:estruct} show the correlation functions after subtracting the fitted flow backgrounds. The fitted near-side Gaussian and the negative dipole are depicted individually. 

As seen from the $\chi^{2}/${\sc ndf} written in each upper panel, the fits by Eq.~(\ref{eq:estruct}) are generally good. This is also true for the other $\ptt$ and $\pta$ bins. However, the fitted flow modulations (written in the lower panels) are significantly larger than the measured ones for the out-of-plane $\phis$ slices, much beyond their systematic uncertainties quoted in Table~\ref{tab:v2v2}. In other words, in order to eliminate the away-side double-peak, an anisotropic flow that is much larger than that measured by the two-particle cumulant method is required, consistent with our earlier observation. Moreover, the deviations of the fitted flow modulations from the measured ones vary from slice to slice (non-monotonically), which should not be the case if the measured flow parameters that we used were simply in error. Qualitatively the same features are observed for the other $\ptt$ and $\pta$ bins. 
These free fit results suggest that the near-side Gaussian and the negative dipole in the fit model of Eq.~(\ref{eq:estruct}) likely do not correspond to the nonflow dihadron correlation signal sought after in this analysis with a high $\pt$ trigger particle.

We have also used other single-peaked functional forms, e.g.~a near-side Gaussian and an away-side Gaussian, in our fit. Similar conclusions were reached. The away-side double-peak for the out-of-plane trigger particles cannot be eliminated without using a flow subtraction much larger than experimentally determined, either with or without \zyam. Thus, we conclude that the away-side double-peak structure is not an artifact of the \zyam\ flow subtraction procedure used in this analysis.

\subsection{Systematic Uncertainties on Jet-Like and Ridge Correlations\label{sec:syst:ridge}}

\begin{figure}[hbt]
\centerline{\includegraphics[width=0.35\textwidth]{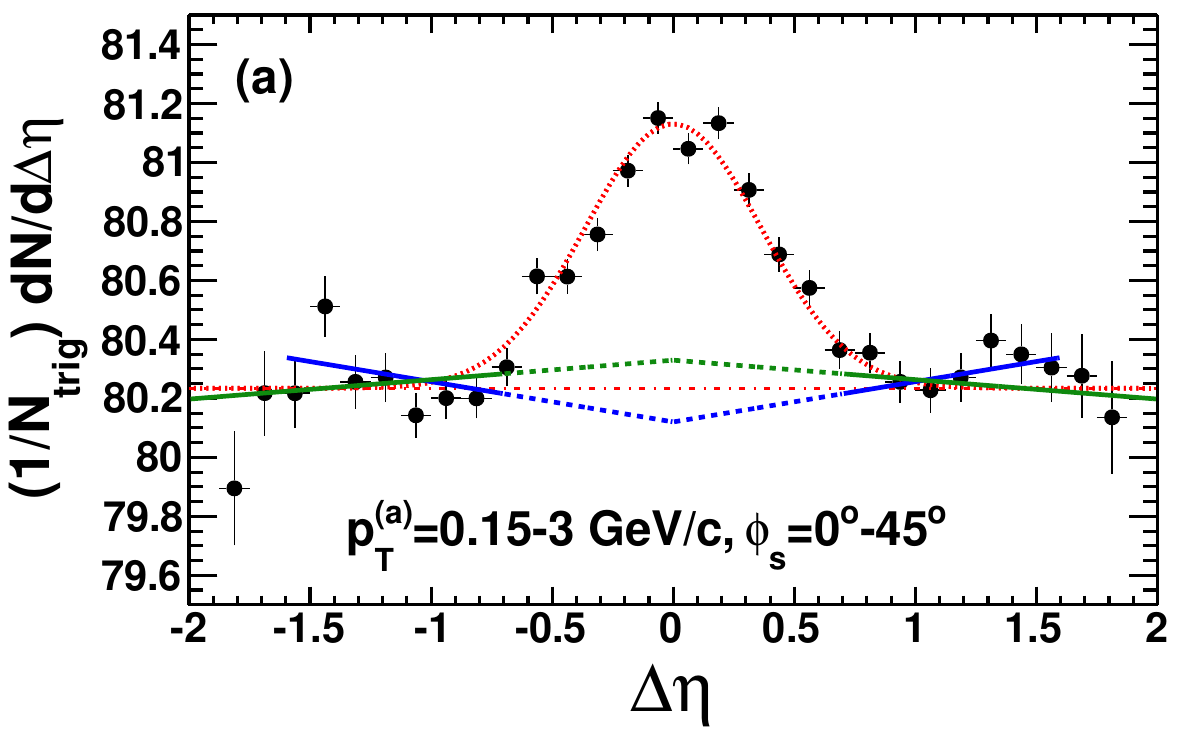}}
\centerline{\includegraphics[width=0.35\textwidth]{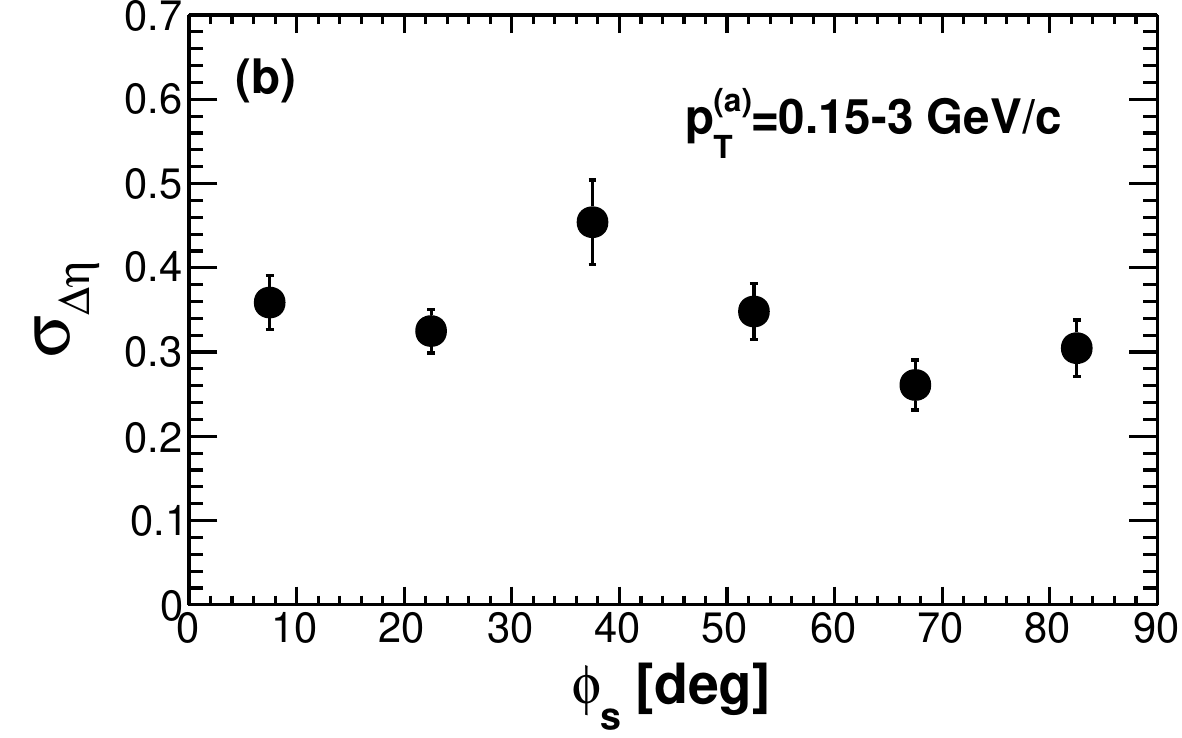}}
\centerline{\includegraphics[width=0.35\textwidth]{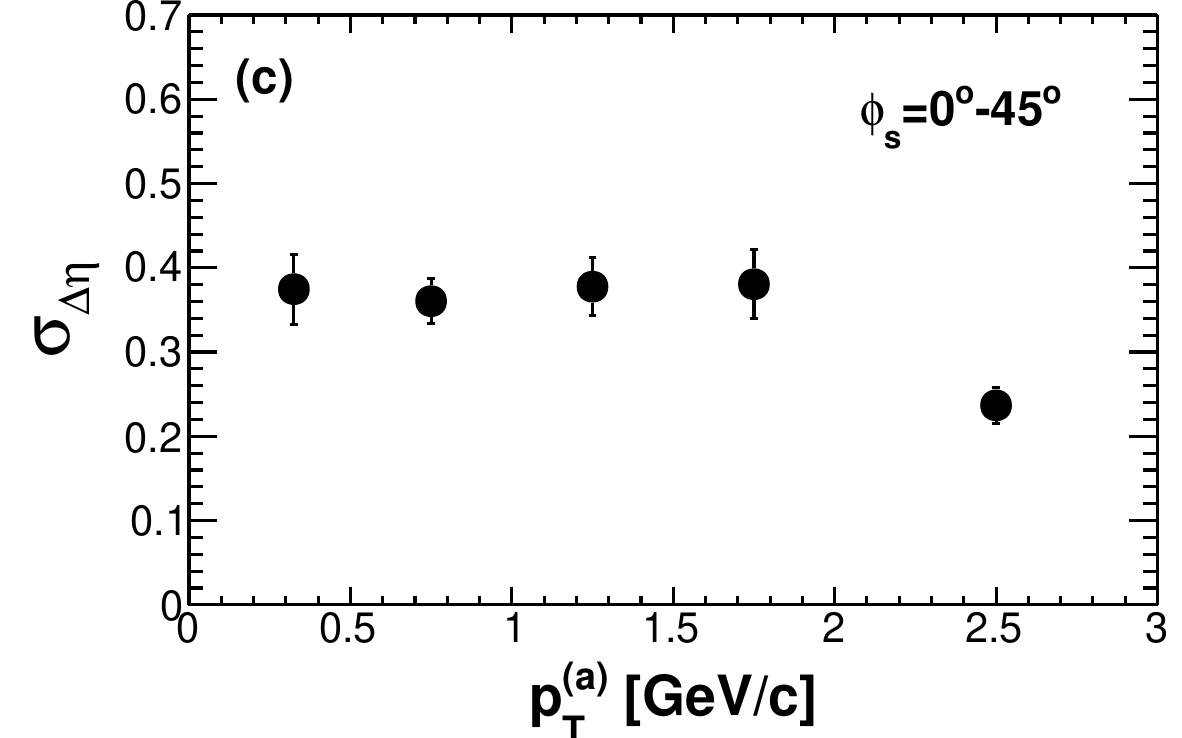}}
\caption{(Color online) (a) Raw $\deta$ correlation of near-side associated hadrons ($|\dphi|<1$) integrated over $0<\phis=|\phit-\psiEP|<\pi/4$ and $0.15<\pta<3$~\gev, corrected by the two-particle $\deta$ acceptance. The dotted curve is a single Gaussian fit and the dot-dashed horizontal line is the fit pedestal; the solid lines are linear fits to the regions $0.7<|\deta|<1.6$ and $0.7<|\deta|<2.0$, respectively, and the dashed lines are their extrapolations. (b) Gaussian fit $\sigma$ to near-side $\deta$ correlation in $0.15<\pta<3$~\gev\ as a function of $\phis$. (c) Gaussian fit $\sigma$ to near-side $\deta$ correlation integrated over $0<\phis<\pi/4$ as a function of $\pta$. The data are from minimum-bias 20-60\% Au+Au collisions. The trigger particle $\pt$ range is $3<\ptt<4$~\gev. Error bars are statistical.}
\label{fig:deta}
\end{figure}

To obtain the jet-like component, we take the difference of the correlation functions from $|\deta|<0.7$ and $|\deta|>0.7$ (properly weighted by the relative two-particle $\deta$ acceptance). The assumption in this procedure is that the ridge is uniform in $\deta$ (after taking into account the trivial two-particle $\deta$ acceptance) and is therefore subtracted away in the difference~\cite{Abelev:2009af}. Measurements at low $\pt$ without a trigger particle indicate that the ridge is broad but drops with increasing $\deta$~\cite{Adams:2004pa}. If this is true for trigger particle correlations as studied here, our ``jet" measurement contains a residual ridge contribution. To estimate this effect, we study $\deta$ correlation functions for near-side associated particles ($|\dphi|<1$). An example is shown in Fig.~\ref{fig:deta} for $3<\ptt<4$~\gev\ and $0.15<\pta<3$~\gev\ in the 20-60\% centrality bin. The in-plane direction ($0<\phi<\pi/4$) is used because, as will be shown later, the ridge resides mainly in the in-plane direction. We compare the ridge contributions to the $|\deta|<0.7$ region as extrapolated from a constant ridge fit and from a linear fit~\cite{Abelev:2009af}, both done in the large $\deta$ range of $|\deta|>0.7$. Because of possible edge effects in the $\deta$ acceptance, we also limit our fit range within $0.7<|\deta|<1.6$. We assign the difference, $\pm15\%$, as the systematic uncertainty on the jet-like component yield due to the assumption of a uniform ridge. 

In this paper, we consider all correlated particles at $|\deta|>0.7$ and $|\dphi|<1$ to be part of the ridge. The ridge yield we report in this paper is defined to be the integral of the correlated particle yield over $0.7<|\deta|<2.0$ (and $|\dphi|<1$). Thus, the assumption of the ridge shape does not affect the ridge yield.

We have assumed that the jet-like component is contained within $|\deta|<0.7$, and assigned the entire correlated yield in $|\deta|>0.7$ as ridge. This introduces uncertainty in the ridge yield as well as in the jet-like yield. Moreover, the fraction of the jet-like component that leaks out of the $\deta$ cut is subtracted in obtaining the jet-like part, thus the effect of the leakage is doubled in the extracted jet-like component. To study this effect, we fit the $\deta$ correlation function (such as that shown in the upper panel of Fig.~\ref{fig:deta}) to a Gaussian with centroid at $\deta=0$ and a constant pedestal (i.e.~a uniform ridge). The Gaussian width is shown in Fig.~\ref{fig:deta}(b) as a function of $\phis$ for $1<\pta<2$~\gev\ and in Fig.~\ref{fig:deta}(c) as a function of $\pta$ for integrated $\phis$. The Gaussian width does not significantly depend on $\phis$ or $\pta$. We estimate the effect of the leakage of the jet-like component to be about 10\% of the jet-like yield, assigned as a single-sided negative uncertainty on the ridge yield, and a single-sided positive uncertainty, twice as large, on the jet-like yield. The physics of the correlation widths will be discussed in Sec.~\ref{sec:results:peaks}.

The systematic uncertainty on the jet-like yield due to flow uncertainty is small because the large uncertainties due to $\vf$ are cancelled, assuming $\vf$ is constant over $\deta$. This should be a good assumption because the PHOBOS experiment found that $\vf$ was constant within the $\eta$ acceptance of the STAR TPC (dropping only towards larger $|\eta|$)~\cite{Back:2004zg,Back:2004mh}. 

\begin{figure}[hbt]
\centerline{\includegraphics[width=0.4\textwidth]{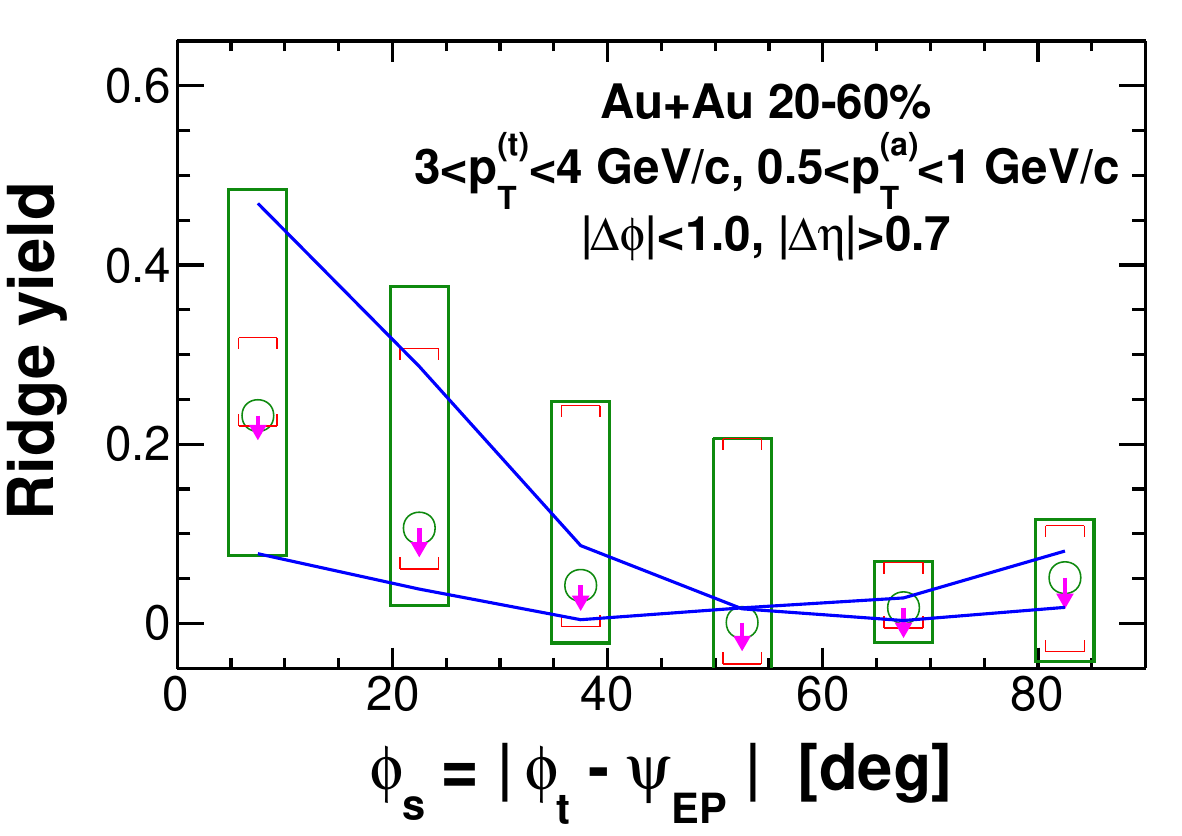}}
\caption{(Color online) Illustration of the different systematic uncertainties on the ridge yield (defined in Sec.~\ref{sec:results:near}) within $|\dphi|<1$ and $|\deta|>0.7$ as a function of $\phis=|\phit-\psiEP|$. The data are from minimum-bias 20-60\% Au+Au collisions. The trigger and associated particle $\pt$ ranges are $3<\ptt<4$~\gev\ and $0.5<\pta<1$~\gev, respectively. Statistical errors are smaller than the symbol size. The various systematic uncertainties are due to: (i) flow subtraction by Eq.~(\ref{eq:bkgd}) (shown by the solid curves), (ii) background normalization uncertainty (shown in brackets), assessed by varying \zyam\ background normalization range and by comparing to \zyam\ from asymmetric correlations separately for positive and negative $\phit-\psiEP$, and (iii) leakage from the jet-like component into the $|\deta|>0.7$ region (shown by the arrows). The total systematic uncertainties are shown by the boxes.}
\label{fig:syst}
\end{figure}

\begin{figure*}[hbt]
\centerline{\includegraphics[width=1.03\textwidth]{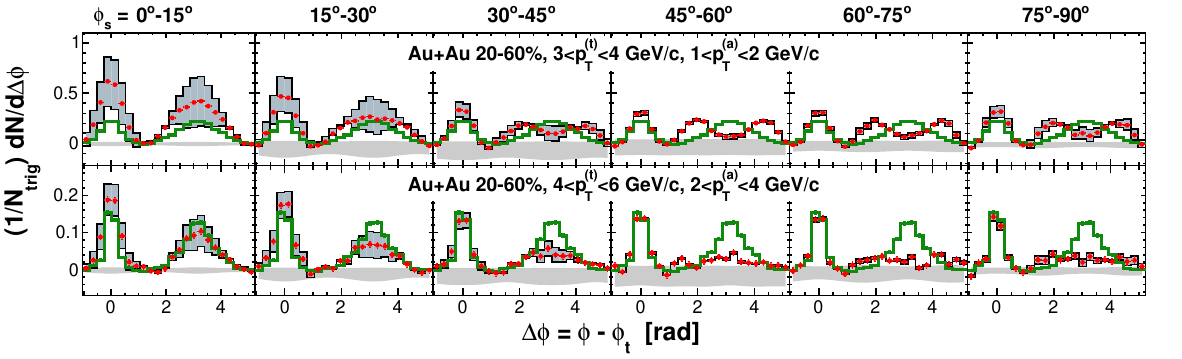}}
\caption{(Color online) Background-subtracted dihadron correlations with trigger particle in six slices of azimuthal angle from the event plane, $\phis=|\phit-\psiEP|$. The trigger and associated particle $\pt$ ranges are $3<\ptt<4$~\gev\ and $1<\pta<2$~\gev\ (upper panel), and $4<\ptt<6$~\gev\ and $2<\pta<4$~\gev\ (lower panel), respectively. Note that the bottom row corresponds to the kinematic range used in Ref.~\cite{Adler:2002tq}. Both the trigger and associated particles are restricted to be within $|\eta|<1$. The triangle two-particle $\deta$ acceptance is not corrected. The data points are from minimum-bias 20-60\% Au+Au collisions. Flow background is subtracted by Eq.~(\ref{eq:bkgd}) using measurements in Table~\ref{tab:v2} and the parameterization in Eq.~(\ref{eq:v4}). Systematic uncertainties are shown in the thin histograms embracing the shaded area due to flow subtraction and in the horizontal shaded band around zero due to \zyam\ background normalization. Statistical errors are smaller than the symbol size. For comparison, the inclusive dihadron correlations from \dAu\ collisions are superimposed as the thick (green) histograms (only statistical errors are depicted).}
\label{fig:corr}
\end{figure*}

Figure~\ref{fig:syst} illustrates the various systematic uncertainties on the extracted ridge yield. (i) The systematic uncertainties due to flow subtraction are shown by the solid curves. The uncertainty is dominant at small $\phis$; the flow uncertainty at large $\phis$ is small. (ii) The systematic uncertainty due to background normalization uncertainty is shown in brackets, as assessed by varying background normalization range and by comparing to background normalizations of asymmetric correlation functions at positive and negative $\phit-\psiEP$ separately. (iii) There is an additional systematic uncertainty in the extracted ridge yield because the jet-like correlation can be broader than 0.7 in $\deta$ and the jet-like yield beyond $|\deta|>0.7$ is contained in the extracted ridge yield. This part of the systematic uncertainty is shown by the arrows. The total systematic uncertainties are obtained by the quadratic sum of the individual sources and shown by the boxes.

\section{Results and Discussion\label{sec:results}}

We first present in Sections~\ref{sec:results:corr}--\ref{sec:results:peaks} our dihadron correlation results with subtraction of the $v_2$ and $v_4$ harmonic flow backgrounds of Eq.~(\ref{eq:bkgd}). We then present in Section~\ref{sec:v3} the corresponding dihadron correlation results with additional background subtraction, including $v_3$ harmonic and other high-order effects by Eqs.~(\ref{eq:bkgd_v3} and (\ref{eq:bkgd_v4}).

\subsection{Correlation Functions\label{sec:results:corr}}

Figure~\ref{fig:corr} shows the $v_2$ and $v_4$ background-subtracted dihadron azimuthal correlations in 20-60\% Au+Au collisions as a function of the trigger particle orientation relative to the event plane, $\phis$. 
The subtracted flow background is given by Eq.~(\ref{eq:bkgd}) using measurements in Table~\ref{tab:v2} and the parameterization of $\vvPsi$ by Eq.~(\ref{eq:v4}). 
The thin histograms embracing the shaded area indicate the systematic uncertainties due to anisotropic flow. The horizontal shaded band around zero indicates the systematic uncertainties due to \zyam\ background normalization. The slight modulations of the edges of the band are because of the anisotropic flow in the combintorial background. For comparison, the minimum-bias \dAu\ inclusive dihadron correlation (without differentiating with respect to an ``event plane'') is superimposed in each panel in Fig.~\ref{fig:corr}. The trigger and associated particle $\pt$ ranges are $3<\ptt<4$~\gev\ and $1<\pta<2$~\gev\ (upper panel), and $4<\ptt<6$~\gev\ and $2<\pta<4$~\gev\ (lower panel), respectively.
These kinematic ranges correspond to those for the raw correlations shown in Fig.~\ref{fig:raw}. The background-subtracted correlations for all trigger and associated particle $\pt$ ranges are presented in Appendix~\ref{app} in Figs.~\ref{figApp:corr34} and~\ref{figApp:corr46}.

As seen in Fig.~\ref{fig:corr}, the near-side peaks in Au+Au collisions are evident for all trigger particle orientations. The single-peak shape of the near-side correlation remains relatively unchanged from in-plane ($\phis\sim0$) to out-of-plane ($\phis\sim\pi/2$). However, the amplitude of the near-side peak decreases with $\phis$, becoming similar to that from \dAu\ collisions at large $\phis$. Our previous studies have shown that the near-side correlation, while not much modified at high $\pt$, is enhanced in Au+Au collisions relative to \pp\ and \dAu\ collisions at low to modest $\pt$~\cite{Adams:2005ph,Abelev:2009af,Aggarwal:2010rf}. The present results show that the near-side enhancement is mostly present for trigger particles oriented in-plane and the modification for trigger particles oriented at $\phis\sim\pi/2$ is minimal for this centrality bin. 

Unlike the near side, the away-side correlation structure evolves when trigger particles move from in-plane to out-of-plane for the 20-60\% centrality bin. The away side has a single peak when the trigger particles are oriented close to the event plane. Only when the trigger particle direction is far away from the event plane does the double-peak structure emerge on the away side. In addition, the away-side modification increases with increasing associated particle $\pta$. Our previous studies showed that the away-side correlation structure is significantly modified in central Au+Au collisions, and the modification is the largest in the intermediate $\pt$ range~\cite{Adams:2005ph,Aggarwal:2010rf}. The present result indicates that the away-side modification has a strong dependence on the trigger particle direction relative to the event plane. The strongest away-side modification is found for trigger particles perpendicular to the event plane (see Fig.~\ref{fig:corr}). However, the systematic uncertainty due to flow subtraction is presently large; when the upper systematic bound of $v_2$ is used, the change from in-plane to out-of-plane is less dramatic. The results nevertheless suggest that the medium path-length dependence in $\phis$ plays an important role, and should provide useful input to theoretical modeling of partonic energy loss in the nuclear medium.

The lower panel of Fig.~\ref{fig:corr} shows the high $\pt$ associated particle results. The ``disappearance" of the away-side correlation at high associated particle $\pt$, first observed for this kinematic range in the inclusive dihadron correlations in Ref.~\cite{Adler:2002tq}, has a clear dependence on the trigger particle orientation. When the trigger particles move from $\phis\sim0$ to $\pi/2$, the path-length increases, and the away-side peak(s) become diminished.

\begin{figure}[hbt]
\centerline{\includegraphics[width=0.4\textwidth]{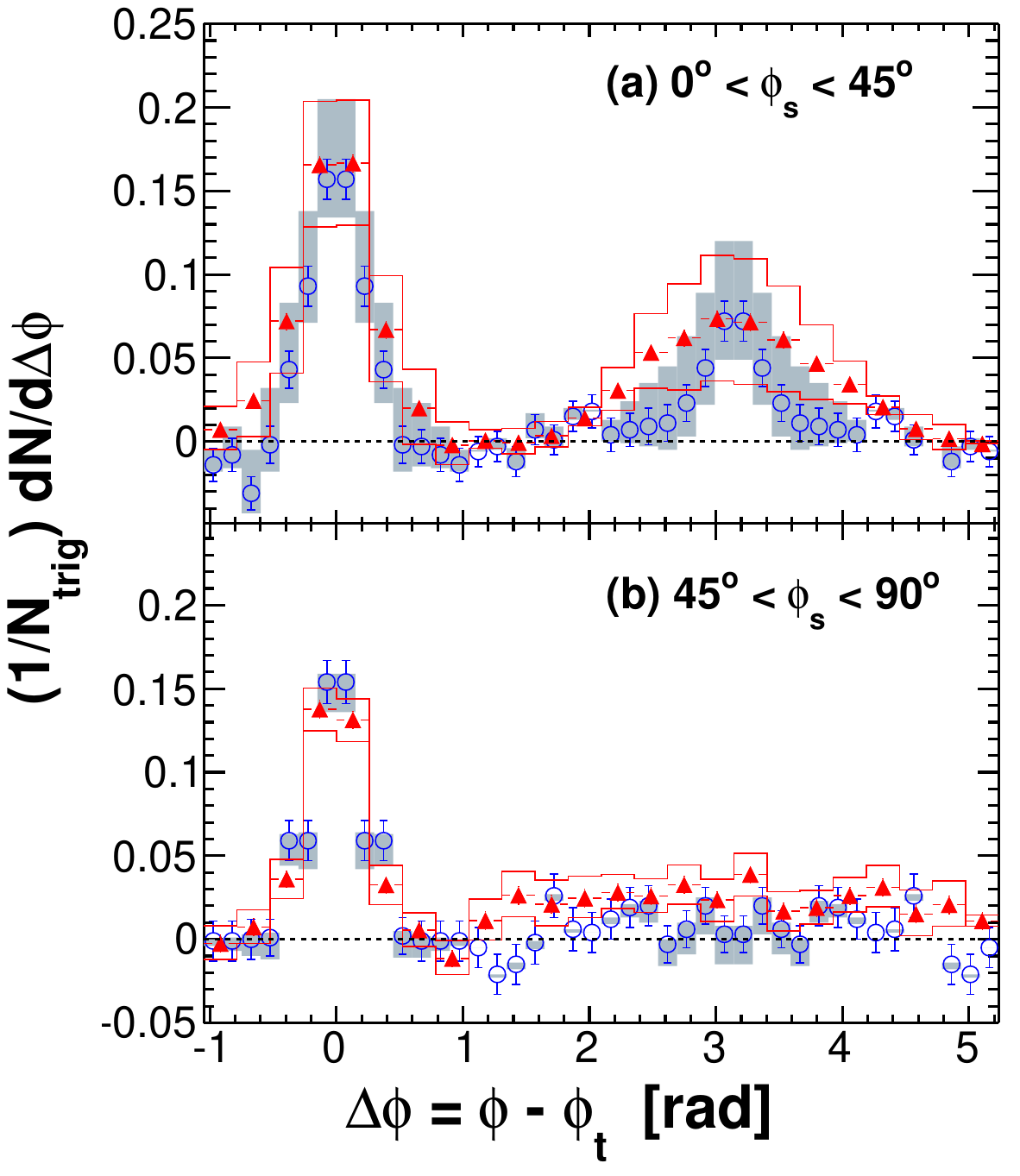}}
\caption{(Color online) Comparison of dihadron correlation results from this work (triangles) with those from Ref.~\cite{Adams:2004wz} (circles) for (a) in-plane ($\phis<\pi/4$) and (b) out-of-plane ($\phis>\pi/4$) trigger particles. Data are from 20-60\% Au+Au collisions. The trigger and associated particle $\pt$ ranges are $4<\ptt<6$~\gev\ and $2<\pta<4$~\gev\, respectively. Both the trigger and associated particles are restricted within $|\eta|<1$. The triangle two-particle $\deta$ acceptance is not corrected. Error bars are statistical. Systematic uncertainties on background subtraction by Eq.~(\ref{eq:bkgd}) (including those due to anisotropic flow $v_2$ and due to background normalization from different \zyam\ normalization ranges) are shown in histograms for results from this work and in shaded areas for results from Ref.~\cite{Adams:2004wz}. For fair comparison, the systematic uncertainty due to background deviations from \zyam\ is not included.}
\label{fig:Kirill}
\end{figure}

STAR has previously published dihadron correlations for in-plane ($\phis<\pi/4$) and out-of-plane ($\phis>\pi/4$) trigger particles~\cite{Adams:2004wz}. We sum up our correlation results from slices 1-3 and 4-6 to obtain the in-plane and out-of-plane correlations, respectively. We have also analyzed the data treating $\phis<\pi/4$ as a single slice to obtain the in-plane correlation and $\phis>\pi/4$ for the out-of-plane correlation. The resultant correlation functions are consistent with those reported here that were summed from individual slices. 
Figure~\ref{fig:Kirill} compares results from this work to those in Ref.~\cite{Adams:2004wz}. The histograms show systematic uncertainties of the results from this work, while the shaded boxes show those of the results from Ref.~\cite{Adams:2004wz}. The analysis reported here differs from that in Ref.~\cite{Adams:2004wz} in two ways: (i) In the average $\vf=(\veta{2}{0.7}+\flow{4})/2$ used in this analysis the two-particle cumulant flow was obtained with a $\etagap=0.7$, whereas in the average used in Ref.~\cite{Adams:2004wz} all particle pairs were included in the two-particle cumulant flow, which contains a more significant nonflow effect; (ii) The flow correlation is corrected up to $\vv$ in this analysis, while correction only up to $\vf$ was done in Ref.~\cite{Adams:2004wz}. 

\begin{figure}[hbt]
\centerline{\includegraphics[width=0.5\textwidth]{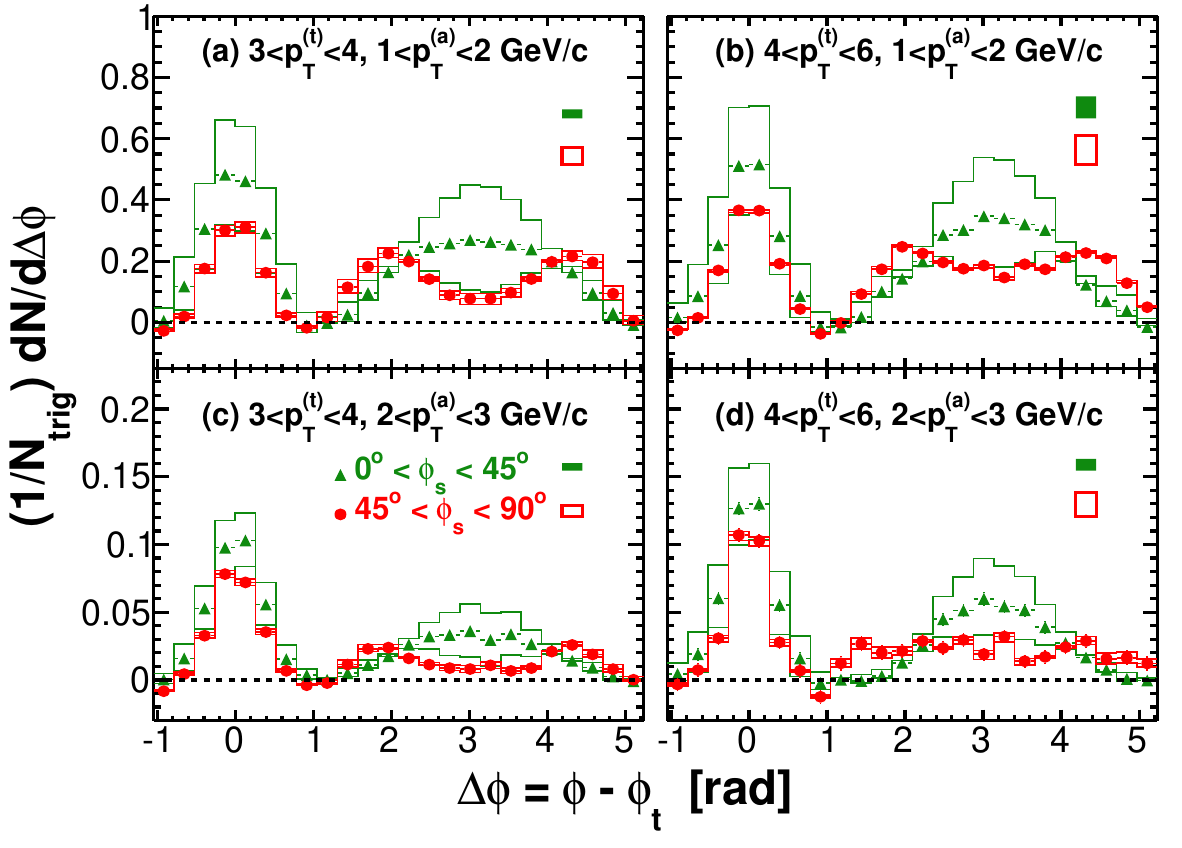}}
\caption{(Color online) Background-subtracted dihadron correlations with trigger particles in-plane ($\phis<\pi/4$) and out-of-plane ($\phis>\pi/4$) in 20-60\% Au+Au collisions. The results are for $3<\ptt<4$~\gev\ (left panels) and $4<\ptt<6$~\gev\ (right panels), and $1<\pta<2$~\gev\ (upper panels) and $2<\pta<3$~\gev\ (lower panels). Both the trigger and associated particles are restricted within $|\eta|<1$. The triangle two-particle $\deta$ acceptance is not corrected. Flow background is subtracted by Eq.~(\ref{eq:bkgd}) using measurements in Table~\ref{tab:v2} and the parameterization in Eq.~(\ref{eq:v4}). Error bars are statistical. Systematic uncertainties due to those on anisotropic flow $v_2$ are shown in histograms; those due to \zyam\ background normalization are indicated by the vertical sizes of the filled and hollow boxes in the legends for in-plane and out-of-plane trigger particles, respectively.}
\label{fig:highpt}
\end{figure}

Figure~\ref{fig:highpt} shows the in-plane and out-of-plane correlation results for two trigger $\ptt$ ranges and two associated particle $\pta$ ranges for the 20-60\% Au+Au collisions. The histograms show the systematic uncertainties due to flow subtraction; those due to \zyam\ background normalization are shown as boxes in the legends. With admittedly large systematic uncertainties, a difference seems evident between in-plane and out-of-plane dihadron correlations for both trigger $\ptt$ ranges and both associated particle $\pta$ bins. 
The near-side correlated yield appears larger for in-plane than out-of-plane triggers. As will be discussed in Sections~\ref{sec:results:near} and~\ref{sec:results:connection}, the difference is due to the larger ridge contribution in-plane than out-of-plane, and the jet-like contributions are similar for in-plane and out-of-plane. A more significant difference appears on the away side between in-plane and out-of-plane correlations. For in-plane trigger particles, the away-side correlations appear to peak at $\dphi=\pi$. For out-of-plane trigger particles, the away-side correlations are double-peaked. The double-peak structure is stronger for the lower trigger particle $\ptt$ range. The away-side structure is studied in more detail in Sec.~\ref{sec:results:away} below.

\subsection{Discussion on the Away-Side Results\label{sec:results:away}}

\begin{figure}[hbt]
\centerline{\includegraphics[width=0.4\textwidth]{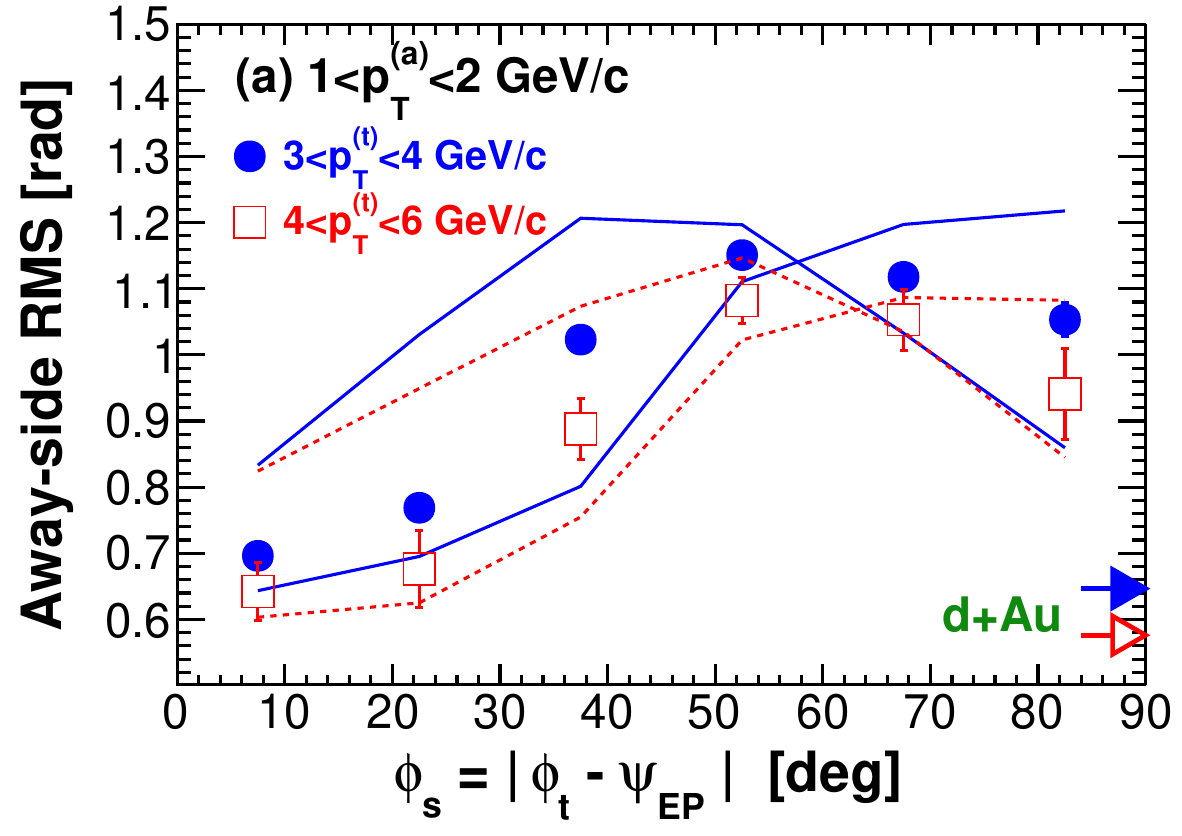}}
\centerline{\includegraphics[width=0.4\textwidth]{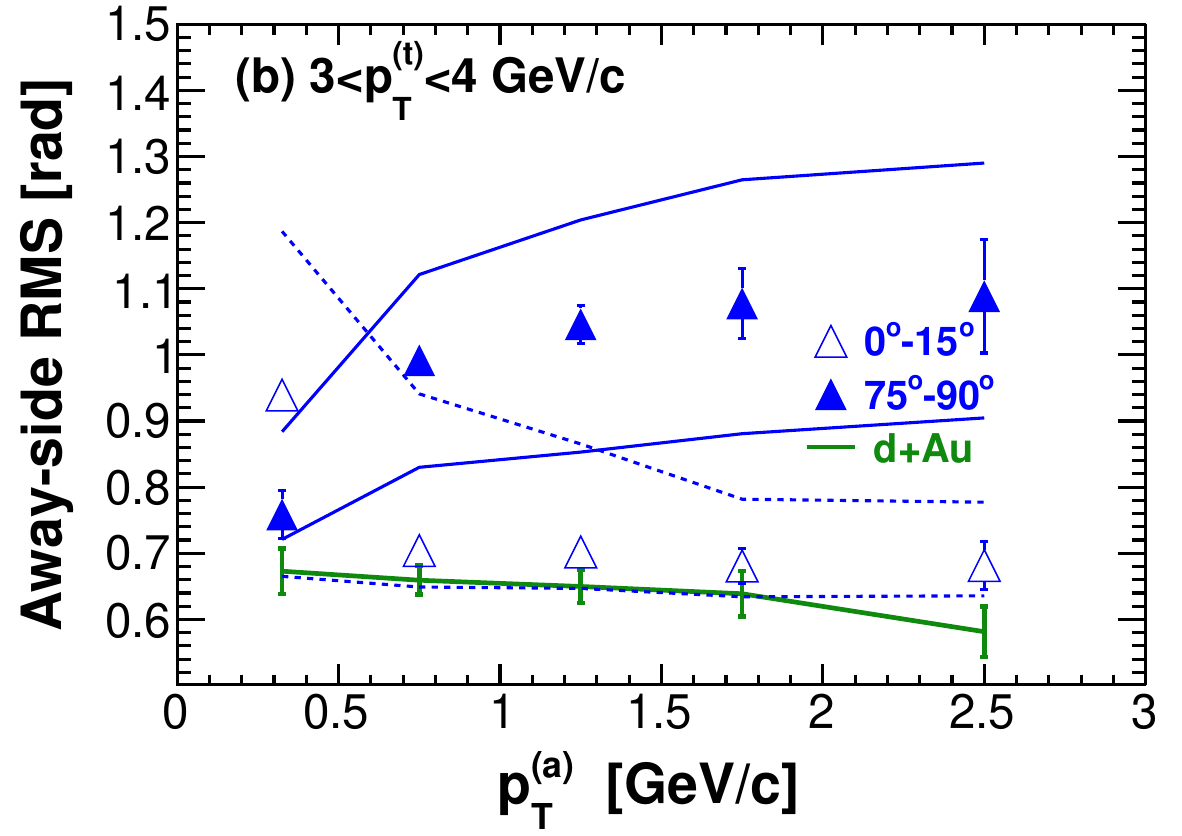}}
\caption{(Color online) (a) The away-side \rms\ of the dihadron correlation function versus the trigger particle azimuth relative to the event plane, $\phis=|\phit-\psiEP|$, in 20-60\% Au+Au collisions for $1<\pta<2$~\gev. Two trigger $\ptt$ selections are shown: $3<\ptt<4$~GeV/$c$ (solid circles) and $4<\ptt<6$~GeV/$c$ (hollow squares). (b) The away-side \rms\ for slice 1 (hollow triangles) and slice 6 (solid triangles) versus the associated particle $\pta$ in 20-60\% Au+Au collisions. The trigger particle $\pt$ range is $3<\ptt<4$~\gev. Error bars are statistical. The curves indicate systematic uncertainties due to flow subtraction, solid curves for the solid data points and dashed curves for the hollow data points. The systematic uncertainty due to \zyam\ background normalization is not shown. The corresponding \dAu\ results are indicated by the arrows (solid arrow for $3<\ptt<4$~GeV/$c$ and hollow arrow for $4<\ptt<6$~GeV/$c$) in the upper panel and by the lower solid line connecting error bars in the lower panel.}
\label{fig:RMS}
\end{figure}

In order to quantify the modification in the away-side correlation structure, we calculate the width of the distribution by
\begin{equation}
\mbox{\rms} = \left(\frac{\displaystyle{\int_1^{2\pi-1}}\frac{dN}{d\dphi}(\dphi-\pi)^2d\dphi}{\displaystyle{\int_1^{2\pi-1}}\frac{dN}{d\dphi}d\dphi}\right)^{1/2}\,.
\end{equation}
We show in Fig.~\ref{fig:RMS}(a) the \rms\ of the away-side correlation as a function of the trigger particle orientation $\phis$ for 20-60\% Au+Au collisions. The associated particle $\pt$ range is $1<\pta<2$~\gev. Two trigger particle $\pt$ ranges are shown: $3<\ptt<4$~\gev\ and $4<\ptt<6$~\gev. The central value of the \rms\ increases with increasing $\phis$ by approximately a factor of 1.5 from in-plane to out-of-plane. The distribution becomes more double-peaked as $\phis$ increases. No difference is observed between the two trigger $\ptt$ selections. Only when the upper bound of elliptic flow is used for background subtraction does the away-side \rms\ difference between $\phis=0$ and $\pi$ diminish, but the change of \rms\ with $\phis$ becomes nonmonotonic.

For comparison, the \dAu\ results are indicated by the arrows in Fig.~\ref{fig:RMS}(a). As seen, the \rms\ in 20-60\% Au+Au collisions from slices 1 and 2 is not much larger than in \dAu. This may be consistent with the path-length effect. However, we note that the correlation amplitudes in Au+Au collisions for the in-plane slices are larger than in \dAu\ collisions, as discussed below. This suggests that the away-side single peak in Au+Au and \dAu\ collisions may come from different physics mechanisms. As will be discussed in Sec.~\ref{sec:results:near}, the near-side correlation for in-plane trigger particles has a large contribution from the ridge, and it is likely that there is an accompanying back-to-back ridge on the away side.

Figure~\ref{fig:RMS}(b) shows the \rms\ as a function of the associated particle $\pta$ for slices 1 and 6 in 20-60\% centrality. The \rms\ remains constant for slice 1, and is not much broader than the \dAu\ result for all measured $\pta$ bins. The \rms\ for slice 6 increases with $\pta$ and then seems to saturate. The double-peak structure is strongest when the trigger particle is perpendicular to the reaction plane and the associated particle is not soft. 
Results for other slices vary smoothly between slices 1 and 6. The features for trigger particles of $4<\ptt<6$~\gev\ are qualitatively the same.

\begin{figure}[hbt]
\centerline{\includegraphics[width=0.4\textwidth]{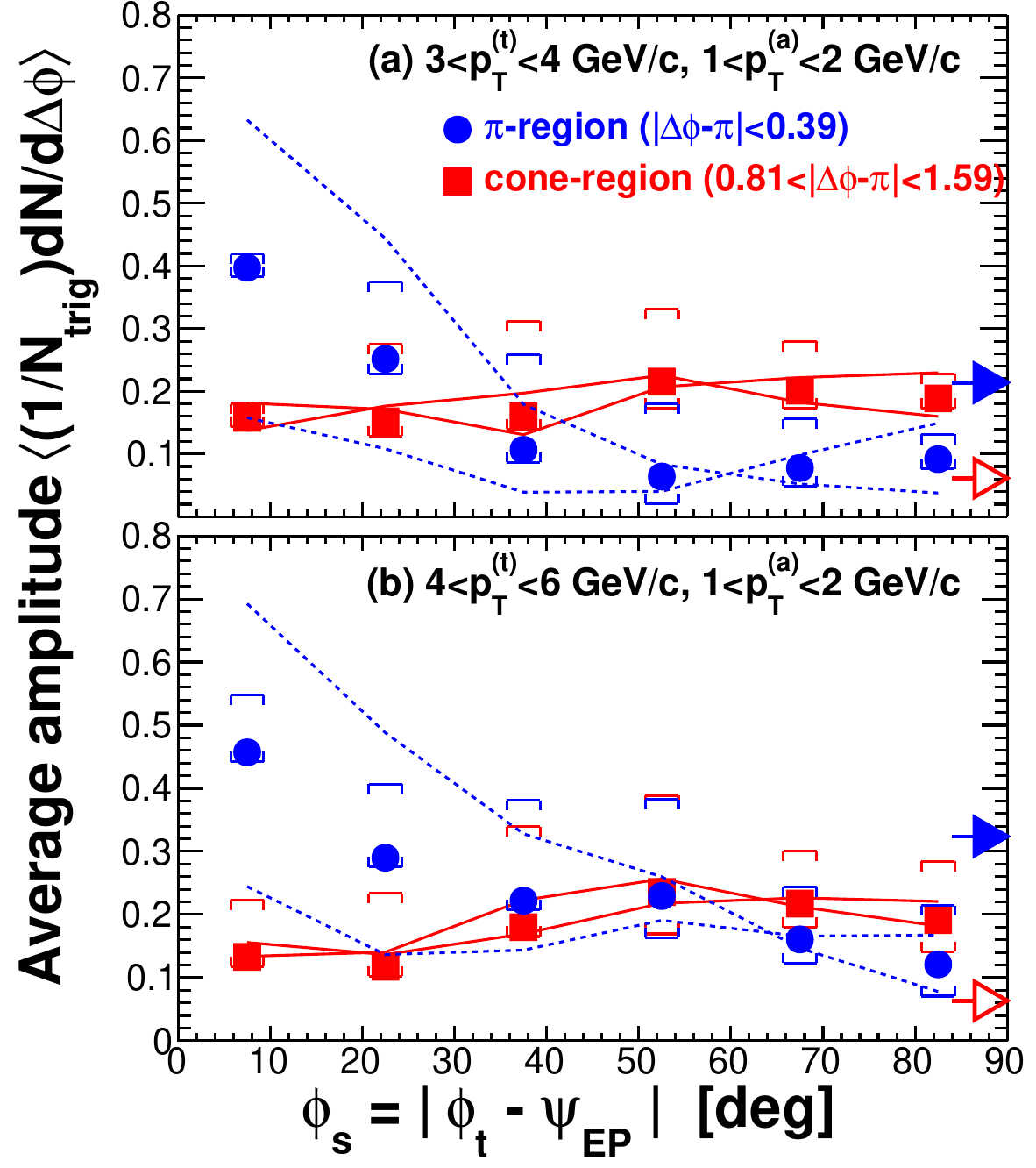}}
\caption{(Color online) The away-side dihadron correlation amplitudes in the $\pi$-region ($|\dphi-\pi|<0.39$) and the double-peak region ($0.81<|\dphi-\pi|<1.59$) as a function of the trigger particle azimuth relative to the event plane, $\phis=|\phit-\psiEP|$, in 20-60\% Au+Au collisions. Statistical errors are smaller than the symbol size. The curves indicate systematic uncertainties due to flow subtraction, and the brackets indicate those due to \zyam\ background normalization. Both trigger $\pt$ ranges are shown: (a) $3<\ptt<4$~\gev\ and (b) $4<\ptt<6$~\gev. The associated particle $\pt$ range is $1<\pta<2$~\gev. The \dAu\ results in the $\pi$- and double-peak regions are indicated by the solid and hollow arrows, respectively.}
\label{fig:ampl}
\end{figure}

The away-side double-peak structure observed in the inclusive dihadron correlation (i.e.~without differentiating trigger particle azimuthal angles relative to the reaction plane)~\cite{Adams:2005ph} has stimulated much interest~\cite{Stoecker:2004qu,CasalderreySolana:2004qm,Ruppert:2005uz,Renk:2005si,Khlebnikov:2010yt,Betz:2010qh,Neufeld:2011yh,Ayala:2012bv,Tachibana:2015qxa}. The three-particle jet-like correlation studies indicate that the double-peak correlated hadrons are emitted event-by-event, not an effect of averaging of single peaks in individual events~\cite{Abelev:2008ac}. 
To study the double-peak structure in more detail, we show in Fig.~\ref{fig:ampl} the average correlation amplitude on the away side in the $\pi$-region ($|\dphi-\pi|<0.39$) and in the double-peak region ($0.81<|\dphi-\pi|<1.59$) as a function of $\phis$ in 20-60\% Au+Au collisions. Two trigger $\ptt$ selections are shown; no significant difference is observed. 
With default elliptic flow subtraction, the amplitudes in the $\pi$-region drop with increasing $\phis$, from a value larger than that in \dAu\ (as indicated by the arrows) to a value significantly smaller than that in \dAu. If the upper systematic bound of the elliptic flow is subtracted, the $\pi$-region amplitude seems to vary nonmonotonically with $\phis$. On the other hand, the double-peak region amplitude seems rather constant with $\phis$ in Au+Au collisions, and is significantly stronger than that in \dAu\ collisions for both trigger particle $\ptt$ selections. This suggests that the double-peak emission of correlated hadrons may also be present underneath the single away-side peak for in-plane trigger particles. See the discussion in Sec.~\ref{sec:results:connection}.

\begin{figure}[hbt]
\centerline{\includegraphics[width=0.4\textwidth]{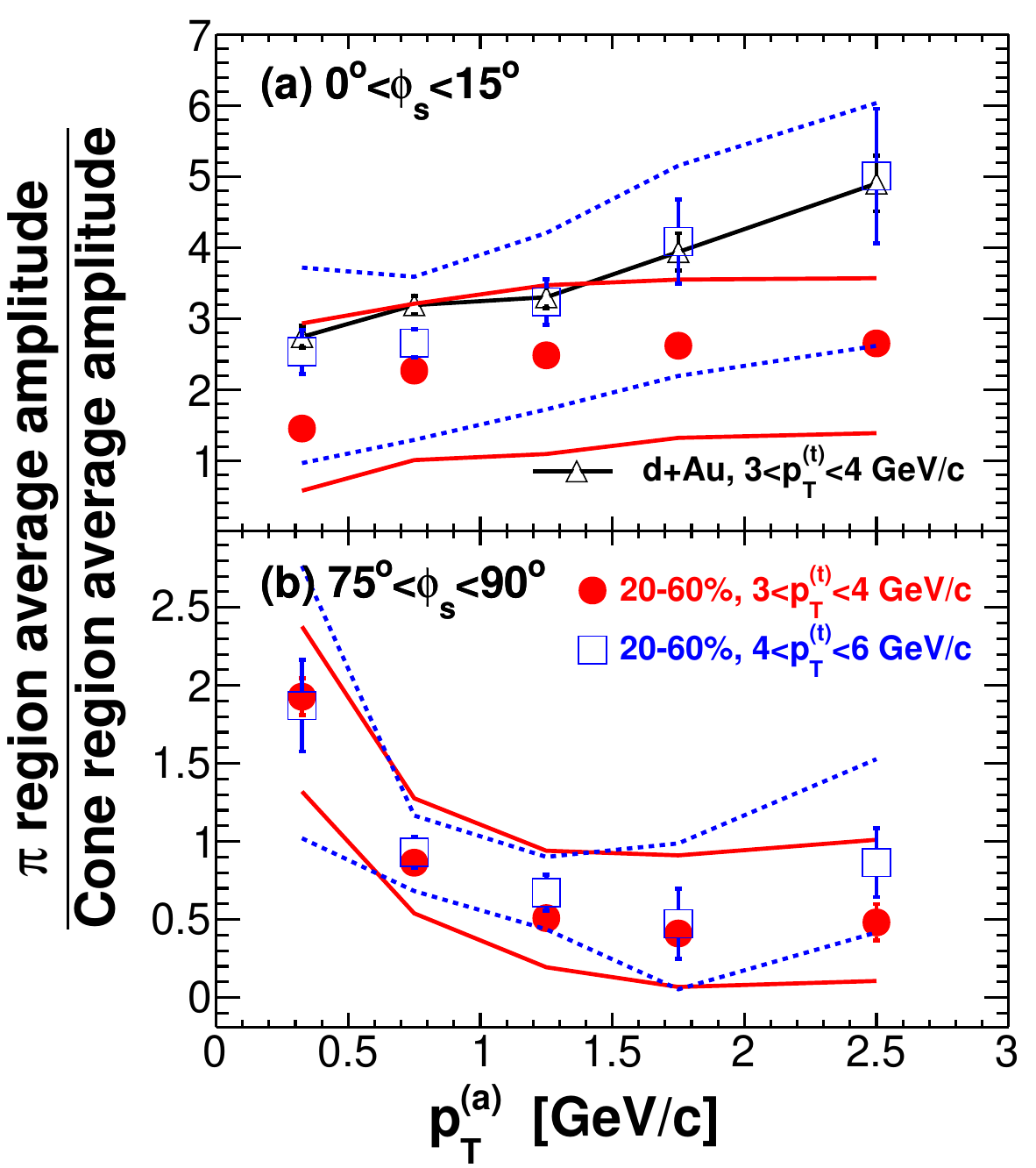}}
\caption{(Color online) Associated particle $\pta$ dependence of the ratio of away-side dihadron correlation amplitude in the $\pi$-region ($|\dphi-\pi|<0.39$) to that in the double-peak region ($0.81<|\dphi-\pi|<1.59$). Two $\phis=|\phit-\psiEP|$ slices are shown: (a) $0<\phis<\pi/12$ and (b) $5\pi/12<\phis<\pi/2$. The data are from minimum-bias 20-60\% Au+Au collisions. Both trigger $\pt$ ranges of $3<\ptt<4$~\gev\ and $4<\ptt<6$~\gev\ are shown. Error bars are statistical. The curves indicate systematic uncertainties due to flow subtraction. The systematic uncertainty due to \zyam\ background normalization is not shown. The \dAu\ results are indicated by the open triangles connected by the line in (a), where the error bars are statistical.}
\label{fig:awayRatio}
\end{figure}

Comparison of the relative amplitudes in the $\pi$-region and the double-peak region shown in Fig.~\ref{fig:ampl} again reveals the degree of the double-peak structure. In order to study the $\pt$ dependence of the relative amplitudes, Fig.~\ref{fig:awayRatio}(a) and (b) show the amplitude ratios of $\pi$-region to double-peak region in slices 1 and 6, respectively. The amplitude ratio in slice 1 increases with $\pta$ for the higher $\ptt$ trigger particles. The trend is not much different from that observed in \dAu\ collisions (shown by the black line). The increasing trend suggests that for in-plane trigger particles the away-side correlation is dominated by physics mechanisms other than double-peak emission, such as punch-though jets and/or back-to-back ridge. The increasing trend may also be present for the lower $\ptt$ triggers, but the systematic uncertainty in this analysis prevents a firm conclusion. On the other hand, for slice 6 the amplitude ratio decreases with $\pta$. The away-side jet-like correlation at $\dphi=\pi$ is essentially diminished; what remains are hadrons that form the double-peak structure. It is also worth noting that the away-side amplitude ratio for out-of-plane trigger particles (lower panel of Fig.~\ref{fig:awayRatio}) is significantly smaller than for in-plane trigger particles (upper panel of Fig.~\ref{fig:awayRatio}). This is again the consequence of the significant away-side broadening from in-plane to out-of-plane. 

\subsection{Discussion on the Near-Side Results\label{sec:results:near}}

\begin{figure*}[hbt]
\centerline{\includegraphics[width=1.03\textwidth]{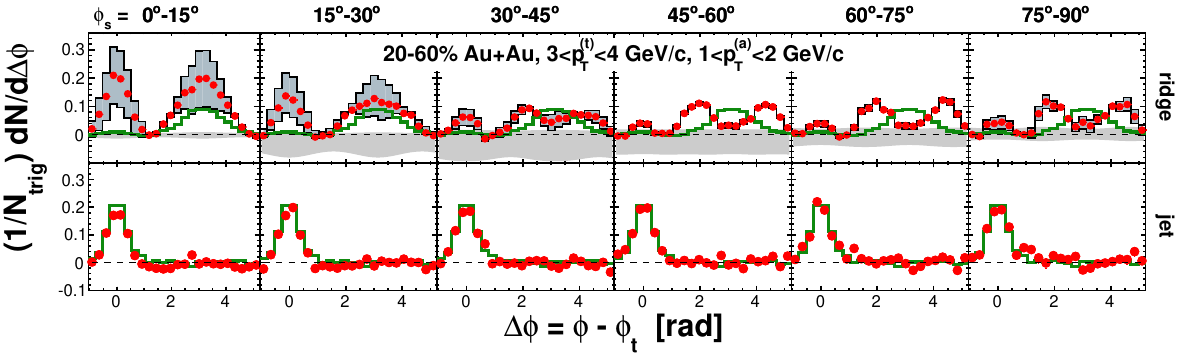}}
\caption{(Color online) Upper panels: background-subtracted dihadron correlations with trigger particles in six slices of azimuthal angle relative to the event plane, $\phis=|\phit-\psiEP|$, with a cut on the trigger-associated pseudo-rapidity difference of $|\deta|>0.7$. The triangle two-particle $\deta$ acceptance is not corrected. Statistical errors are smaller than the symbol size.  Flow background is subtracted by Eq.~(\ref{eq:bkgd}) using measurements in Table~\ref{tab:v2} and the parameterization in Eq.~(\ref{eq:v4}). Systematic uncertainties are shown in the black histograms due to flow subtraction and in the horizontal shaded band around zero due to \zyam\ background normalization. The near-side correlation is due to the ridge. Lower panels: The difference between raw dihadron correlations in $|\deta|<0.7$ and $|\deta|>0.7$, after multiplying a coefficient onto the latter such that the resultant difference is zero on average on the away side in the range $|\dphi-\pi|<1$. This correlation represents the jet-like component of the dihadron correlations. Statistical errors are smaller than the symbol size. Systematic uncertainties are small. The results are for $3<\ptt<4$~\gev\ and $1<\pta<2$~\gev\ in 20-60\% Au+Au collisions. In both panels the corresponding inclusive dihadron correlations from \dAu\ collisions (thick histograms) are superimposed for comparison.}
\label{fig:jetRidge}
\end{figure*}

Recall that in Fig.~\ref{fig:corr}, we observe a significant change in the near-side peak amplitude. The near-side amplitude drops with increasing $\phis$. For the 20-60\% centrality, the amplitude at large $\phis$ is not much different from the \dAu\ result, perhaps indicating minimal medium modification. On the other hand, the amplitude at small $\phis$ appears significantly larger than in \dAu, suggesting a large medium effect. This might be counterintuitive at first glance. Due to jet-quenching, the near-side jets predominately emerge outward from the surface of the medium, so variation in the medium thickness traversed by the near-side jets between in-plane and out-of-plane directions is not naively expected.

It has been shown by the inclusive dihadron correlation that the near-side correlation strength is enhanced in Au+Au with respect to \pp\ and \dAu\ collisions~\cite{Adams:2005ph,Aggarwal:2010rf}, and the enhancement is mainly due to the large contribution from the ridge~\cite{Adams:2005ph,Abelev:2009af}. In order to investigate the underlying physics mechanism for the near-side structure change with trigger particle orientation, we separate contributions from the ridge and the jet-like component by analyzing the correlation data in two different $\deta$ regions~\cite{Abelev:2009af}: $|\deta|>0.7$ where the ridge is the dominant contributor, and $|\deta|<0.7$ where both the ridge and jet-like correlations contribute. Figure~\ref{fig:jetRidge} (upper panel) shows the $v_2$ and $v_4$ background-subtracted dihadron correlation function from $|\deta|>0.7$ for trigger and associated particle $\pt$ ranges of $3<\ptt<4$~\gev\ and $1<\pta<2$~\gev\ in 20-60\% Au+Au collisions. (The $|\deta|>0.7$ correlation functions for all kinematic ranges are presented in Appendix~\ref{app} in Figs.~\ref{figApp:corr34ridge} and~\ref{figApp:corr46ridge}.) The near-side correlation for $|\deta|>0.7$ is due to the ridge because the jet-like contribution is mostly confined within $|\deta|<0.7$. The ridge correlation shows a significant drop with increasing $\phis$. The ridge contribution is close to zero for trigger particles perpendicular to the reaction plane in the 20-60\% centrality bin. 

The near-side ridge correlation at large $\deta$, after two-particle $\deta$ acceptance correction, was found to be nearly uniform in $\deta$~\cite{Abelev:2009af}. If the ridge is uniform over the entire measured $\deta$ range, then the ridge can readily be subtracted by taking the difference between the raw (not background-subtracted) correlations from the small and large $\deta$ regions as
\begin{widetext}
\begin{equation}
\frac{dN_{\rm jet-like}}{d\dphi}=\int_{-0.7}^{0.7}\frac{d^{2}N_{\rm raw}}{d\dphi d\deta}d\deta-\mathcal{A}\left(\int_{-2.0}^{-0.7}\frac{d^{2}N_{\rm raw}}{d\dphi d\deta}d\deta+\int_{0.7}^{2.0}\frac{d^{2}N_{\rm raw}}{d\dphi d\deta}d\deta\right)\,.\label{eq:jet}
\end{equation}
\end{widetext}
The coefficient $\mathcal{A}$ accounts for the $\deta$ acceptance difference between $|\deta|<0.7$ and $|\deta|>0.7$, and can easily be obtained from the acceptance ratio of the two $\deta$ regions. It can also be obtained by requiring the away side of the resultant average correlation magnitude to be zero because the away-side correlation (after $\deta$ acceptance correction) is also uniform within the measured $\deta$ range in the TPC~\cite{Adams:2005ph}. We use the latter method to obtain $\mathcal{A}$ such that the resultant away-side average correlation signal within $|\dphi-\pi|<1$ is zero, namely
\begin{widetext}
\begin{equation}
\int_{\pi-1}^{\pi+1}d\dphi\int_{-0.7}^{0.7}\frac{d^{2}N_{\rm raw}}{d\dphi d\deta}d\deta-\mathcal{A}\int_{\pi-1}^{\pi+1}d\dphi\left(\int_{-2.0}^{-0.7}\frac{d^{2}N_{\rm raw}}{d\dphi d\deta}d\deta+\int_{0.7}^{2.0}\frac{d^{2}N_{\rm raw}}{d\dphi d\deta}d\deta\right)=0\,.
\end{equation}
\end{widetext}
The obtained coefficient is approximately $\mathcal{A}\approx1.45$. The resultant difference by Eq.~(\ref{eq:jet}) represents the dihadron correlation of the near-side jet-like component under the assumption that the near-side ridge is uniform in $\deta$ within our measured range. The $\dphi$ correlation of the jet-like component obtained by Eq.~(\ref{eq:jet}) is free of large systematic uncertainties because the anisotropic flow, approximately independent of $\eta$, is largely cancelled in the difference. 

The obtained $\dphi$ correlation of the jet-like component is shown in the lower panel of Fig.~\ref{fig:jetRidge}. The corresponding \dAu\ result is superimposed on the figure. The $\dphi$ correlation of the jet-like component is approximately independent of the trigger particle orientation, in contrast to the ridge component shown in the upper panel of Fig.~\ref{fig:jetRidge}. The near-side jet-like correlations are consistent between \dAu\ and Au+Au collisions. The $\dphi$ correlation functions of the jet-like component for all trigger and associated particle $\pt$ ranges are presented in Appendix~\ref{app} in Figs.~\ref{figApp:corr34jet} and~\ref{figApp:corr46jet}.

\begin{figure}[hbt]
\centerline{\includegraphics[width=0.4\textwidth]{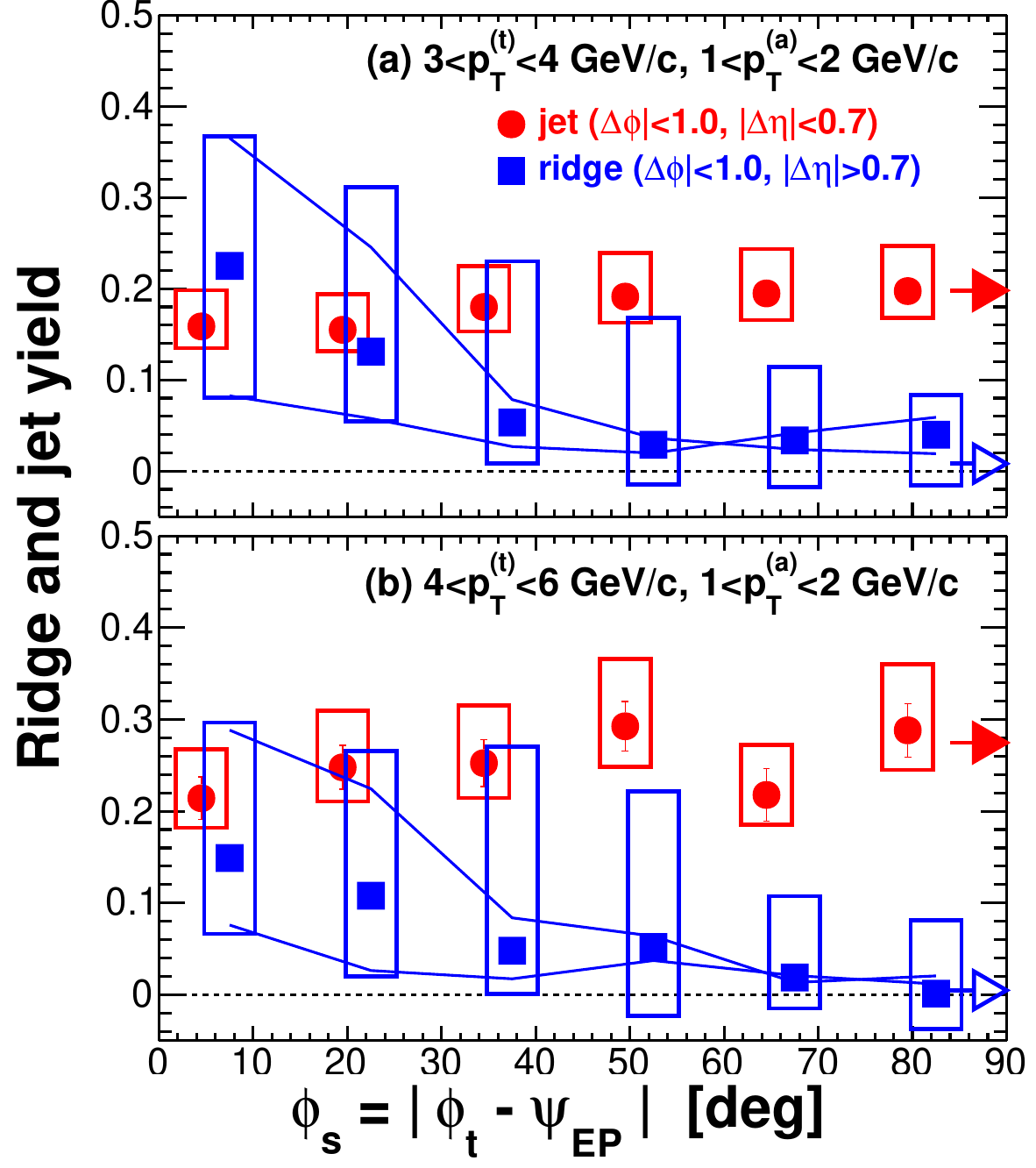}}
\caption{(Color online) The near-side jet-like and ridge yields as functions of the trigger particle azimuth relative to the event plane, $\phis=|\phit-\psiEP|$. The results are from 20-60\% Au+Au collisions. Two trigger $\ptt$ ranges are shown: (a) $3<\ptt<4$~\gev\ and (b) $4<\ptt<6$~\gev. The associated particle $\pta$ range is $1<\pta<2$~\gev. The jet-like yield is from $|\dphi|<1$ and $|\deta|<0.7$ and the ridge yield is from $|\dphi|<1$ and $|\deta|>0.7$. Error bars are statistical. The systematic uncertainties are shown by the boxes. For the ridge yield they include those from anisotropic flow (indicated by the curves) and \zyam\ background normalization. The systematic uncertainties on the jet-like component are due to leakage of jet-like correlations out to $|\deta|>0.7$ and the assumption that the ridge is uniform in $\deta$. The \dAu\ results in the jet and ridge regions are indicated by the filled and hollow arrows, respectively.}
\label{fig:jetRidgeYield}
\end{figure}

To quantify the near-side modification, we study the ridge and jet-like yields as a function of $\phis$. We extract the ridge yield in $|\deta|>0.7$ and $|\dphi|<1$ from the \zyam\ background-subtracted correlations, such as those in the upper panel of Fig.~\ref{fig:jetRidge}, by
\begin{widetext}
\begin{equation}
\mbox{Ridge yield}=\frac{1}{N_{\rm trig}}\int_{-1}^{1}d\dphi\left(\int_{-2.0}^{-0.7}\frac{d^{2}N}{d\dphi d\deta}d\deta+\int_{0.7}^{2.0}\frac{d^{2}N}{d\dphi d\deta}d\deta\right)\,.
\end{equation}
\end{widetext}
We extract the jet-like yield in $|\deta|<0.7$ and $|\dphi|<1$ from the correlations of the jet-like component, such as those in the lower panel of Fig.~\ref{fig:jetRidge}, by
\begin{equation}
\mbox{Jet-like yield}=\frac{1}{N_{\rm trig}}\int_{-1}^{1}d\dphi\int_{-0.7}^{0.7}\frac{d^{2}N}{d\dphi d\deta}d\deta\,.
\end{equation}
Note the $\deta$ acceptance is not corrected in the $\dphi$ correlations of the ridge or the jet-like component; hence, neither are the extracted corresponding yields.
The extracted ridge and jet-like yields are shown in Fig.~\ref{fig:jetRidgeYield} as functions of $\phis$ in the 20-60\% centrality bin. The boxes indicate the total systematic uncertainty; the individual sources of systematic uncertainties and their correlations have been discussed earlier in Sec.~\ref{sec:syst}.
As seen from Fig.~\ref{fig:jetRidgeYield}, the jet-like yield is approximately independent of $\phis$ in Au+Au collisions, and consistent with the \dAu\ data. The ridge yield in Au+Au collisions at small $\phis$ (in-plane) is significant, but it decreases quickly with increasing $\phis$. The ridge yield at large $\phis$ (out-of-plane) is consistent with zero. The ridge is dominated by events where trigger particles are within approximately $\pi/4$ of the event plane.

\begin{figure}[hbt]
\centerline{\includegraphics[width=0.4\textwidth]{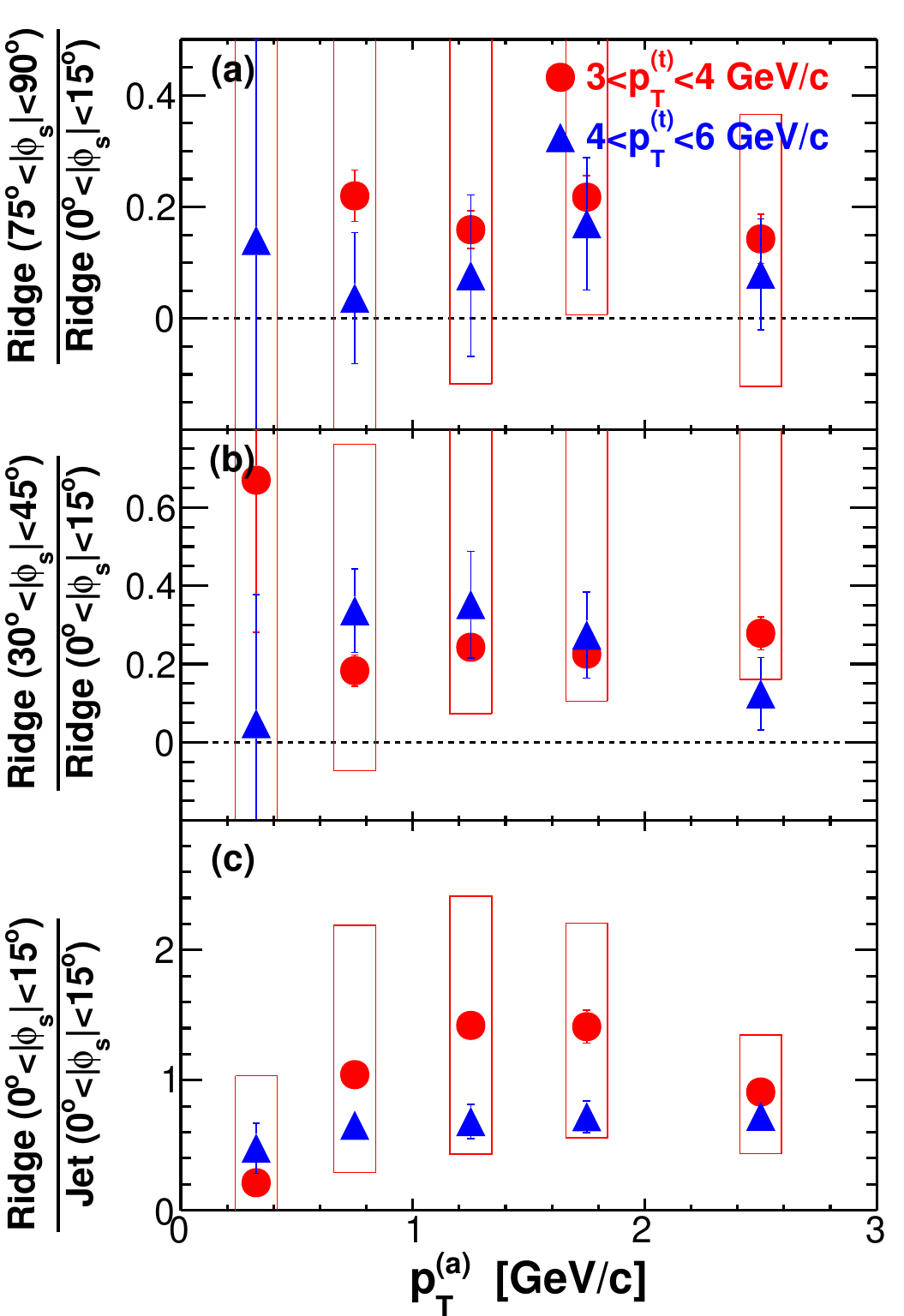}}
\caption{(Color online) (a) Ratio of the ridge yield from $5\pi/12<\phis<\pi/2$ to that from $0<\phis<\pi/12$. (b) Ratio of the ridge yield from $\pi/6<\phis<\pi/4$ to that from $0<\phis<\pi/12$. (c) Ratio of the ridge yield to the jet-like yield from $0<\phis<\pi/12$. The ridge yield is from $|\dphi|<1$ and $|\deta|>0.7$ and the jet-like yield is from $|\dphi|<1$ and $|\deta|<0.7$. Data are from 20-60\% Au+Au collisions. Both trigger $\ptt$ ranges of $3<\ptt<4$~\gev\ and $4<\ptt<6$~\gev\ are shown. Error bars are statistical. Boxes indicate systematic uncertainties on the $3<\ptt<4$~\gev\ data; those for $4<\ptt<6$~\gev\ are similar.}
\label{fig:ridgeRatio}
\end{figure}

The trend of decreasing ridge amplitude with increasing $\phis$ is seen in all measured $\pta$ bins. To quantify this, we show in Fig.~\ref{fig:ridgeRatio}(a) and (b) the $\pta$ dependence of the ratio of ridge yield in $5\pi/12<\phis<\pi/2$ and $\pi/6<\phis<\pi/4$, respectively, to that in $0<\phis<\pi/12$. Both trigger particle $\ptt$ selections are shown. The systematic uncertainties, shown for $3<\ptt<4$~\gev\ in the boxes, have taken into account correlations among the different sources of systematic uncertainties. Within the systematic uncertainty there is no observable difference between the two $\ptt$ selections. The ridge ratios from different $\phis$ slices appear to be independent of $\pta$. The ridge decreases with $\phis$ universally for all $\pta$. The ridge yield out-of-plane is consistent with zero at all associated particle $\pta$ for both the trigger particle $\ptt$ selections.

Motivated by the preliminary version of our data, Chiu and Hwa~\cite{Chiu:2008ht} suggested that alignment between jet propagation and medium flow direction, likely to be found for in-plane trigger particles, may be responsible for the ridge; radiated gluons (within a small angle of the parton direction) become thermalized with the medium and combine with medium partons to form the ridge when they are aligned in the same direction. 
This model, called the Correlated Emission Model (CEM), predicts a measurable asymmetry in the $\dphi$ correlation of the near-side ridge for trigger particles on a single side of the event plane~\cite{Chiu:2008ht}. 
We note that this correlated emission of ridge particles with the medium flow direction may be rather general, not necessarily restricted to the recombination of radiated and medium gluons. For instance, it is possible that initial fluctuations of color flux tubes together with the stronger in-plane transverse flow can produce similar effects~\cite{Voloshin:2003ud,Shuryak:2007fu,Dumitru:2008wn,Dusling:2009ar,Gavin:2008ev,Takahashi:2009na}. We discuss this color flux tube fluctuation model further in Sec.~\ref{sec:results:connection}.

There is strong experimental evidence suggesting that the jet-like component and the ridge are produced by different physics mechanisms~\cite{Abelev:2009jv,Nattrass:2008pn}, thus their $\pt$ dependences are expected to be different. To quantitatively study this, we show in Fig.~\ref{fig:ridgeRatio}(c) the ratio of the ridge yield to the jet-like yield for $0<\phis<\pi/12$. Again, the systematic uncertainties shown in boxes have already taken into account correlations among different sources of systematics. 
Within the systematic uncertainties, the ridge over jet-like component ratio appears to be constant over the measured $\pta$. This may suggest, contrary to the other findings, that the ridge and the jet-like component may be of the same origin. However, it is possible that differences in the $\pta$ spectra of the jet-like and the ridge component are small for our trigger $\ptt$ ranges compared to our systematic uncertainties. The $\pt$ spectra of the jet-like component and the ridge will be further discussed below.

\subsection{Connections between Near- and Away-Side\label{sec:results:connection}}

We have observed that the away-side amplitude in the $\pi$-region decreases strongly with increasing $\phis$, as does the near-side ridge amplitude. We have also observed that both the away-side amplitude in the double-peak region and the near-side jet-like amplitude remain approximately constant with $\phis$. This raises the question whether the near-side and the away-side are connected, or stem from the same physics origin, even though high $\pt$ trigger particles strongly bias the near-side towards surface emission. In order to gain further insights, we study the near- and away-side correlation properties together as a function of $\phis$ and $\pta$.

\begin{figure}[hbt]
\centerline{\includegraphics[width=0.45\textwidth]{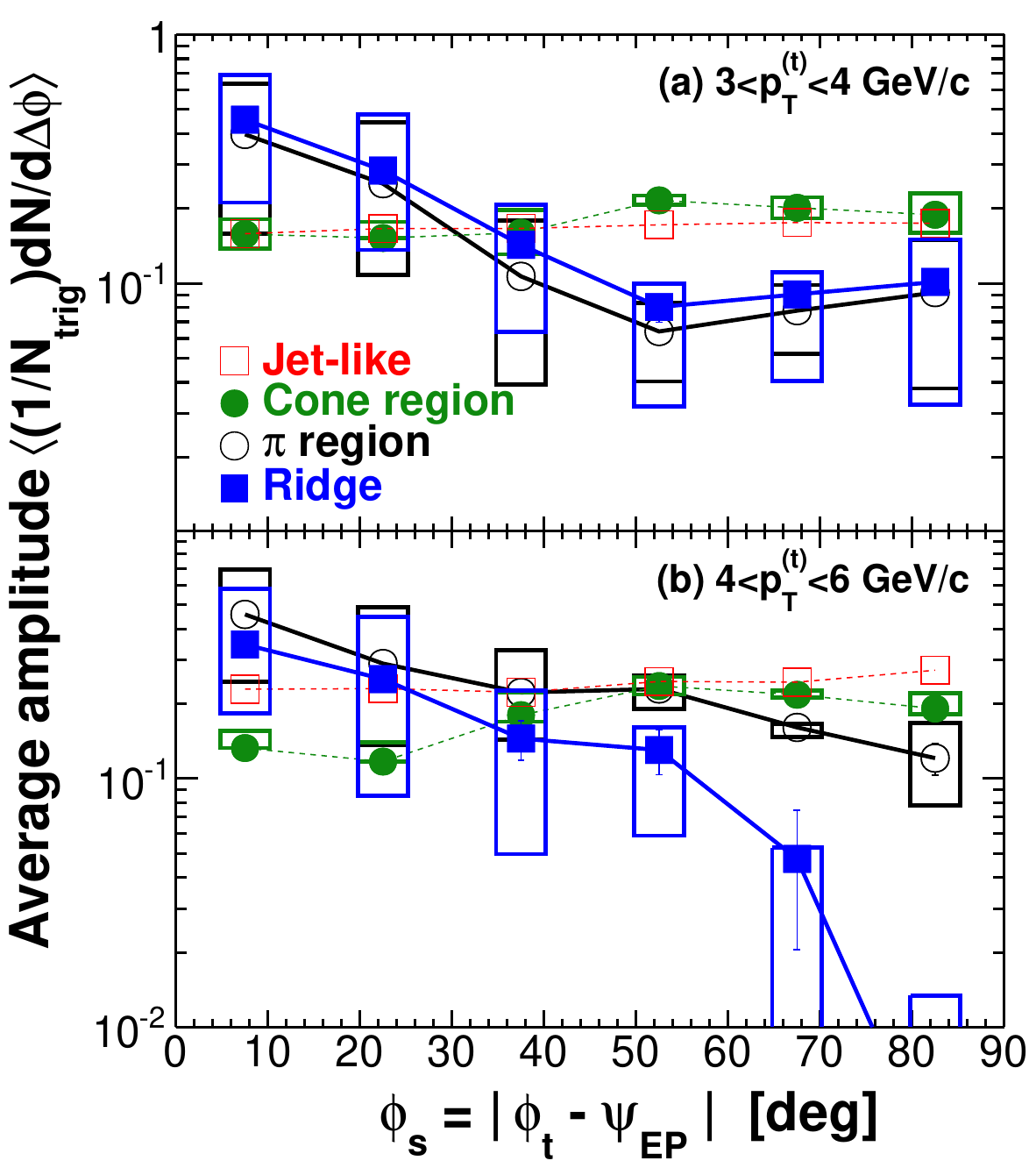}}
\caption{(Color online) Average correlation amplitude as a function of $\phis=|\phit-\psiEP|$ for the near-side jet-like component ($|\dphi|<1$, $|\deta|<0.7$), the double-peak region ($|\dphi-\pi\pm1.2|<0.39$, $|\eta|<1$), the $\pi$-region ($|\dphi-\pi|<0.39$, $|\eta|<1$), and the acceptance-scaled near-side ridge ($|\dphi|<1$, $|\deta|>0.7$). Data are from 20-60\% Au+Au collisions. Both trigger $\ptt$ selections are shown: (a) $3<\ptt<4$~\gev\ and (b) $4<\ptt<6$~\gev. The associated particle $\pt$ range is $1<\pta<2$~\gev\ for both panels. Error bars are statistical. Boxes indicate systematic uncertainties due to anisotropic flow. The systematic uncertainties due to \zyam\ background normalization, common to the $\pi$-region, double-peak region and ridge amplitudes, are not shown.}
\label{fig:ampl_phis}
\end{figure}

Figure~\ref{fig:ampl_phis} shows the average correlation amplitudes of the away-side $\pi$-region and double-peak region and the near-side ridge and jet-like component. The averages are taken within the same window size of $\pm 0.39$. The ridge amplitude is scaled by a factor of $1+\mathcal{A}\approx2.45$, which is approximately the acceptance factor to scale $|\deta|>0.7$ to the entire $\deta$ range assuming a uniform ridge. The jet-like amplitude and the double-peak region amplitude have a similar dependence on $\phis$. 
The similarity suggests that the near-side jet-like component and the away-side double-peak region might be closely related. 

Figure~\ref{fig:ampl_phis} also shows that the ridge amplitude and the away-side $\pi$-region amplitude have a similar dependence on $\phis$. The magnitudes are also similar between the ridge and the $\pi$-region. This is especially true for the lower $\ptt$ range. On the other hand, the jet-like and double-peak region amplitudes have a rather different dependence on $\phis$ than the ridge and $\pi$-region amplitudes. This suggests that the near-side ridge and the away-side $\pi$-region may be connected. Furthermore, they seem not to be connected to the jet-like component or to the component in the double-peak region.

\begin{figure*}[hbt]
\centerline{\includegraphics[width=0.7\textwidth]{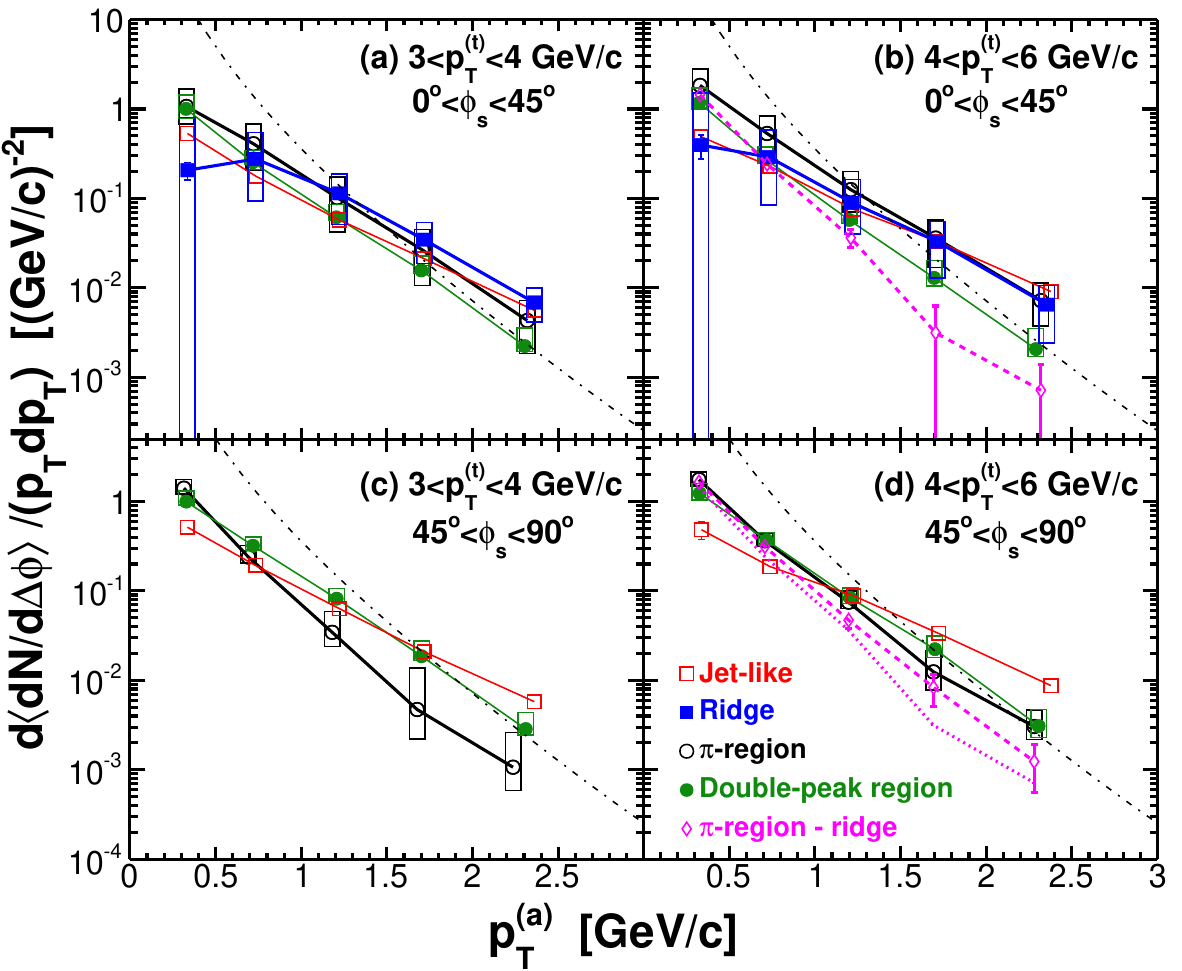}}
\caption{(Color online) Correlated particle $\pt$ spectra in different $\dphi$ regions: the near-side jet-like component ($|\dphi|<1$, $|\deta|<0.7$), the double-peak region ($|\dphi-\pi\pm1.2|<0.39$, $|\eta|<1$), the $\pi$-region ($|\dphi-\pi|<0.39$, $|\eta|<1$), and the near-side ridge ($|\dphi|<1$, $|\deta|>0.7$). The ridge amplitude is scaled by the two-particle $\deta$ acceptance ratio of approximately 2.45. The data are from minimum-bias 20-60\% Au+Au collisions. Error bars are statistical. Systematic uncertainties (including those on anisotropic flow and on \zyam\ background normalization) are shown as boxes for the double-peak region, the $\pi$-region, and the ridge spectra. Two trigger $\ptt$ ranges and two $\phis$ regions are shown: (a) $3<\ptt<4$~\gev\ and $0<\phis<\pi/4$, (b) $4<\ptt<6$~\gev\ and $0<\phis<\pi/4$, (c) $3<\ptt<4$~\gev\ and $\pi/4<\phis<\pi/2$, and (d) $4<\ptt<6$~\gev\ and $\pi/4<\phis<\pi/2$. The dot-dashed curve is the inclusive charged hadron spectrum with an arbitrary normalization. All other lines are to guide the eye. The dotted curve in (d) replicates the dashed curve in (b).}
\label{fig:ampl_pt}
\end{figure*}

\begin{table*}[hbt]
\caption{Inverse slope parameter $T$ (MeV/$c$) from an exponential fit to the associated particle $\pt$ spectra of correlated amplitudes in different $\dphi$ regions: $dN/(\pt d\pt)\propto\exp(-\pt/T)$. Systematic uncertainties for the jet-like spectra and the ($\pi$-region $-$ ridge) spectra are small. Statistical and systematic uncertainties in the inclusive charged hadron spectrum are both negligible.}
\label{tab:slope}
\begin{ruledtabular}
\begin{tabular}{c|llll}
& \multicolumn{2}{c}{$0<\phis<\pi/4$} & \multicolumn{2}{c}{$\pi/4<\phis<\pi/2$}\\
& $3<p_{T}^{(t)}<4$ GeV/$c$ & $4<p_{T}^{(t)}<6$ GeV/$c$ & $3<p_{T}^{(t)}<4$ GeV/$c$ & $4<p_{T}^{(t)}<6$ GeV/$c$\\\hline
Jet-like & $465 \pm 6 {\rm (stat)}$ & $518 \pm 13 {\rm (stat)}$ & $460 \pm 6 {\rm (stat)}$ & $518 \pm 14 {\rm (stat)}$\\
Double-peak region & $331 \pm 2 {\rm (stat)} ^{+21}_{-21} {\rm (syst)}$ & $307 \pm 5 {\rm (stat)} ^{+27}_{-21} {\rm (syst)}$ & $342 \pm 2 {\rm (stat)} ^{+12}_{-3} {\rm (syst)}$ & $335 \pm 4 {\rm (stat)} ^{+12}_{-5} {\rm (syst)}$\\
$\pi$-region & $359 \pm 2 {\rm (stat)} ^{+7}_{-31} {\rm (syst)}$ & $360 \pm 4 {\rm (stat)} ^{+5}_{-26} {\rm (syst)}$ & $231 \pm 3 {\rm (stat)} ^{+47}_{-17} {\rm (syst)}$ & $291 \pm 6 {\rm (stat)} ^{+24}_{-13} {\rm (syst)}$\\
Ridge & $456 \pm 4 {\rm (stat)} ^{+73}_{-38} {\rm (syst)}$ & $444 \pm 12 {\rm (stat)} ^{+48}_{-40} {\rm (syst)}$&&\\
$\pi$-region $-$ Ridge && $226 \pm 14 {\rm (stat)}$ && $249 \pm 13 {\rm (stat)}$\\\hline
Inclusive charged hadron & \multicolumn{4}{c}{$256$}\\

\end{tabular}
\end{ruledtabular}
\end{table*}

There is much other experimental evidence suggesting that the ridge and the jet-like component may be physically unrelated despite the apparent correlation between the ridge and the high $\pt$ trigger particle. For example, three-particle correlations suggest that the production of the jet-like component and the production of the ridge are uncorrelated~\cite{Abelev:2009jv}. The particle composition of the ridge has been found to be similar to that of the bulk medium~\cite{Nattrass:2008pn}. The ridge magnitude has been observed to be rather independent of the trigger particle $\ptt$, persisting to very large $\ptt$~\cite{Abelev:2009af} where jets are almost the sole source of those large $\ptt$ trigger particles. The parent parton energies triggered by the wide range $\ptt$ trigger particles vary greatly, and yet the ridge is independent of $\ptt$. This, again, suggests that the ridge and the jet-like component may be unrelated.

It has been suggested that the ridge may be generated by fluctuations of color flux tubes stretched between the colliding nuclei at the initial time of contact~\cite{Dumitru:2008wn,Dusling:2009ar,Gavin:2008ev,Takahashi:2009na}. The ridge particles from the color flux tubes near the surface of the collision zone are boosted radially by the medium expansion, becoming correlated in relative azimuth. If the ridge is indeed due to color flux tube fluctuations, i.e. entirely from the medium without connection to high $\pt$ trigger particles, then the meaning of ``near side'' as defined by the high $\pt$ trigger particle bears no significance to the ridge. In such a case, there ought to exist a ridge partner on the away side due to symmetry, i.e. a back-to-back ridge. In addition, it is conceivable that the ridge would be stronger along the reaction plane direction where both the flux tube strength and the medium flow are stronger. This would naturally explain our observation that the ridge decreases from in-plane to out-of-plane and the ridge amplitude and the $\pi$-region amplitude trace other other. 
The trigger particles in the in-plane direction happen to have ridge particles associated within a narrow $\dphi$ region (near-side), while those trigger particles out-of-plane cannot accidentally pick up ridge particles to be within a narrow $\dphi$ azimuth. 
In fact, the above mechanism, where the ridge particles are aligned with the trigger particle in azimuth, is similar to the ridge formation mechanism proposed in CEM~\cite{Chiu:2008ht}; however, the underlying physics is quite different.

Examining the $\pt$ dependences of the different correlation components can give further insights into the physics mechanisms responsible for their formation. We show in Fig.~\ref{fig:ampl_pt} the $\pta$ spectra of the average correlation amplitudes from various $\dphi$ regions, $\frac{dN}{\pt d\pt}$. The upper panels show results for in-plane trigger particles, $0<\phis<\pi/4$. Four $\dphi$ regions are shown: the $\pi$-region, the double-peak region, the jet, and the ridge. The lower panels show results for out-of-plane particles, $\pi/4<\phis<\pi/2$. The ridge, which is consistent with zero, is not shown in the lower panel for clarity. Both $\ptt$ selections are shown, $3<\ptt<4$~\gev\ in the left-hand panels and $4<\ptt<6$~\gev\ in the right-hand panels. Note that the upper panels and the lower panels have the same order of magnitude span in their coordinates, so the spectral shapes can be readily compared. 

To investigate the spectral shapes quantitatively, we fit an exponential function to the spectra. The inverse slopes from the fits are tabulated in Table~\ref{tab:slope}. The systematic uncertainties of the fitted inverse slope parameters have already taken into account the correlations in the various sources of systematic uncertainties of the spectra. As expected, the jet-like spectra are harder for the higher trigger $\ptt$ range. The difference in the jet-like spectra between in-plane and out-of-plane is small. 
On the other hand, the double-peak hadron spectra do not seem to depend on trigger particle $\ptt$, nor trigger particle orientation relative to the reaction plane. The double-peak region appears to be universal. In addition, the double-peak hadron spectra are significantly softer than the jet-like spectra, suggesting different production mechanisms for the near-side jet-like hadrons and the away-side double-peak correlated hadrons. Yet, the hadron yields in the jet-like correlation region and in the double-peak region appear to trace each other. This would be a natural consequence if the away-side parton, in rough energy balance with the near-side jet, loses most of its energy to form the double-peak structure~\cite{CasalderreySolana:2004qm}. 

It is interesting to note that the $\pi$-region hadrons are similar to the double-peak hadrons in the in-plane direction; however, in the out-of-plane region they are softer. In fact, the out-of-plane hadrons in the $\pi$-region are not much different from the inclusive hadrons in their $\pt$ distributions. In the scenario of only jet-quenching, the in-plane away-side partons do not have enough medium to interact in 20-60\% Au+Au collisions to completely wash out their identity. On the other hand, the out-of-plane away-side partons have a longer path-length and the lost energy appears to have equilibrated with the medium, a result found in the inclusive dihadron correlation in central collisions~\cite{Adams:2005ph}.

Surprisingly, the ridge particles are relatively hard, not much softer than the jet-like particle spectra (see Fig.~\ref{fig:ampl_pt} and Table~\ref{tab:slope}). Yet, the $\phis$ dependence of the ridge yield is completely different from that of the jet-like yield. We note that the ridge spectrum measured at larger $\pta>2$~\gev\ is significantly softer than the jet-like hadron spectrum also at large $\pta$~\cite{Abelev:2009af}, suggesting that the ridge might be related to the medium. If the ridge comes from the medium, then our result implies that it is not a simple uniform share of medium particles at our measured relatively low $\pta$, because the ridge particles are harder than the bulk medium particles. 

For the associated particle $\pt$ range of $1<\pta<2$~\gev\ shown in Fig.~\ref{fig:ampl_phis}, the $\pi$-region amplitude is slightly smaller than the ridge amplitude for $3<\ptt<4$~\gev. 
For the higher trigger $\pt$ range of $4<\ptt<6$~\gev, there appears an excess of particles in the away-side $\pi$-region over those in the near-side ridge for all $\pta$ bins. The excess appears to be insensitive to $\phis$. Experimentally, it is interesting to examine the $\pt$ dependence of this excess by taking the difference between the away-side $\pi$-region and the near-side ridge. This difference is rather robust because all the systematic uncertainties cancel. The difference (excess of particles in the away-side $\pi$-region over the near-side ridge) for the trigger particle $4<\ptt<6$~\gev\ range is shown as diamonds in Fig.~\ref{fig:ampl_pt}(b,d). It is remarkable to note that the $\pt$ spectra of those excess particles are rather similar for in-plane and out-of-plane in terms of their inverse slopes. Direct comparison is made in the lower panels, where the excess particle spectra in the upper panels are superimposed in the corresponding lower panel as dotted lines. 
The agreement is excellent. Those excess particles have rather soft $\pt$'s, similar to the inclusive charged hadrons. This is already evident from the ridge and $\pi$-region spectra--the away-side $\pi$-region spectra are softer than the ridge spectra. If the ridge is generated by fluctuating color flux tubes and is back-to-back~\cite{Dumitru:2008wn,Dusling:2009ar,Gavin:2008ev,Takahashi:2009na}, then the excess particles in the away-side $\pi$-region must come from other physics mechanisms. One such mechanism is punch-through jets. However, it is counterintuitive to have a much softer spectrum for punch-through jet-like particles, as well as the agreement between in-plane and out-of-plane directions. Another mechanism is statistical global momentum conservation to balance the extra momentum carried by the ridge particles (because they are harder than the particles in the $\pi$-region). However, one may expect a somewhat harder spectrum for the recoil from statistical global momentum conservation than the inclusive spectrum~\cite{Borghini:2006yk}.

\subsection{Properties of the Correlation Peaks\label{sec:results:peaks}}

\begin{figure*}[hbt]
\centerline{\includegraphics[width=\textwidth]{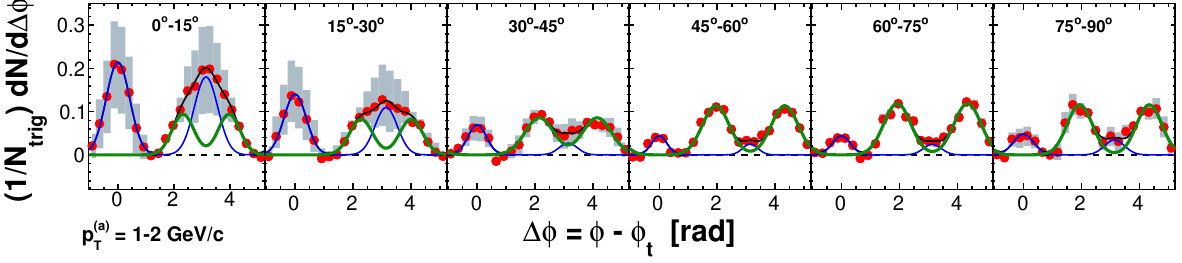}}
\caption{(Color online) Fit to the background-subtracted $\dphi$ correlation functions at $|\deta|>0.7$ in six slices of $\phis=|\phit-\psiEP|$ by four Gaussians (curves): a near-side Gaussian at $\dphi=0$ corresponding to the ridge, an away-side Gaussian at $\dphi=\pi$ with identical width to the near-side Gaussian but with varying amplitude, and two identical away-side Gaussians symmetric about $\dphi=\pi$. Data are from 20-60\% Au+Au collisions, the same as those in the upper panel of Fig.~\ref{fig:jetRidge}. The trigger and associated particle $\pt$ ranges are $3<\ptt<4$~\gev\ and $1<\pta<2$~\gev, respectively. Flow background is subtracted by Eq.~(\ref{eq:bkgd}) using measurements in Table~\ref{tab:v2} and the parameterization in Eq.~(\ref{eq:v4}). The systematic uncertainties due to elliptic flow are shown by the shaded areas; those due to \zyam\ normalization are not shown.}
\label{fig:fit4Gaus}
\centerline{
\includegraphics[width=0.4\textwidth]{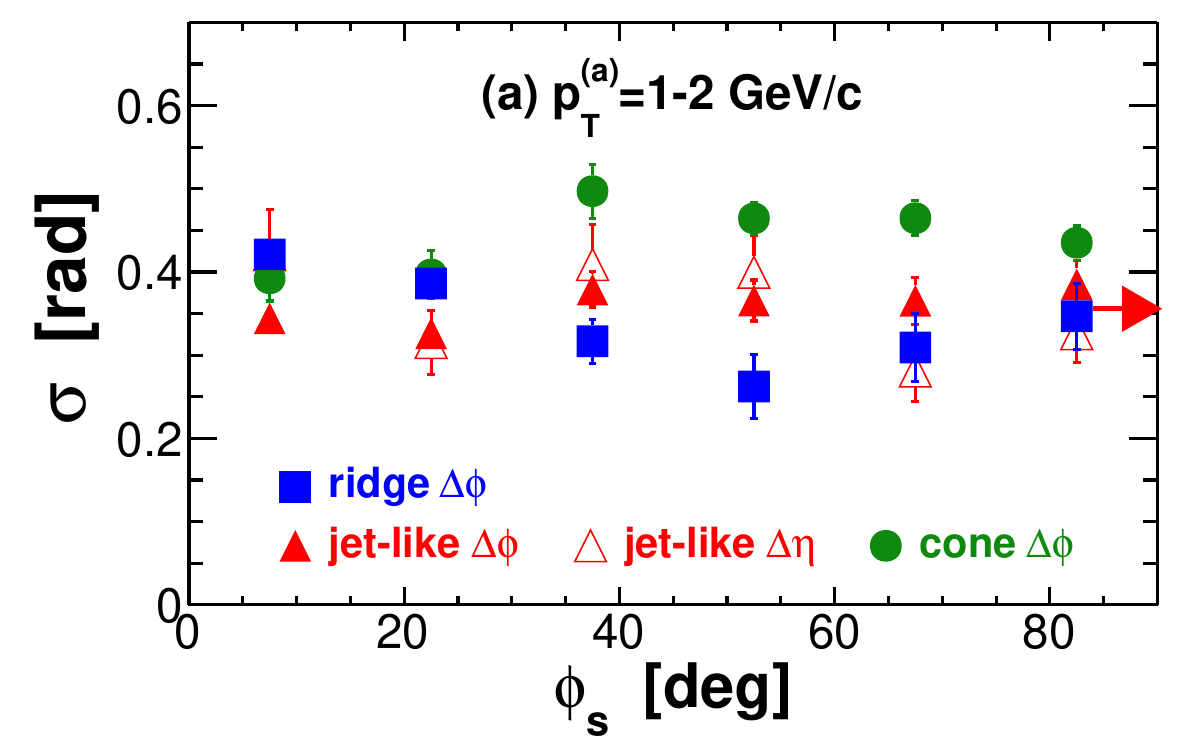}
\includegraphics[width=0.4\textwidth]{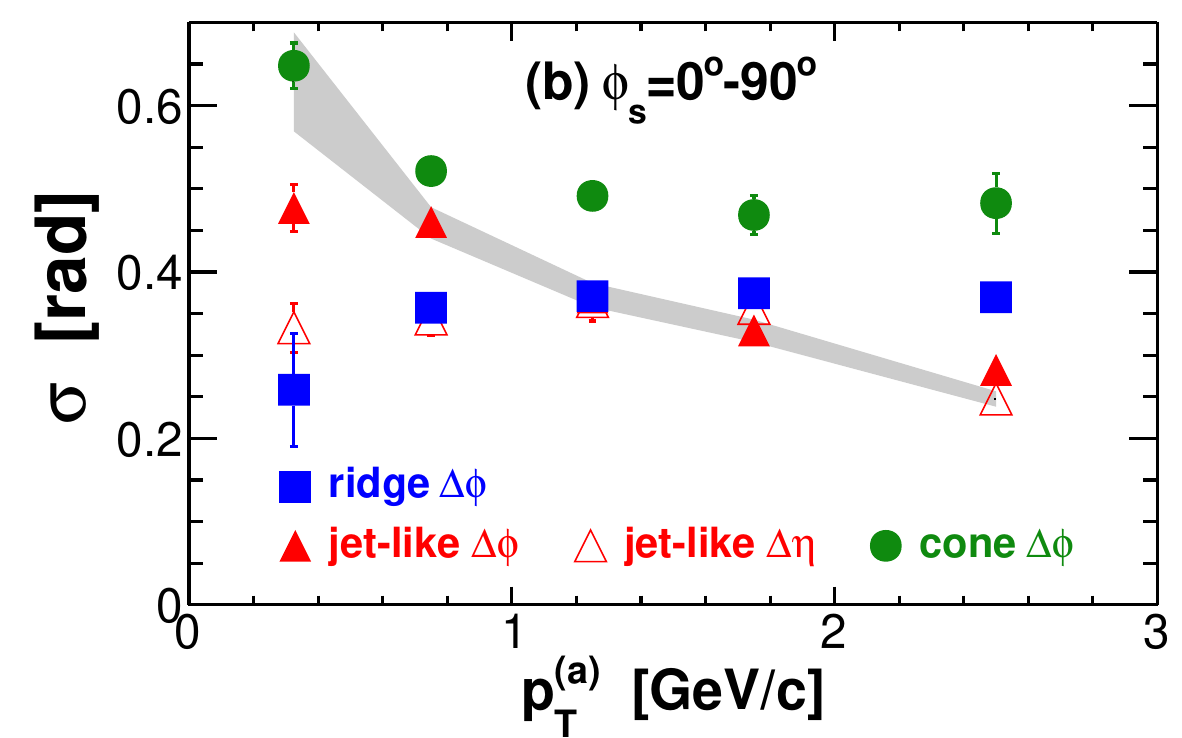}
}
\caption{(Color online) Fit Gaussian widths of the near-side jet-like correlation, the near-side ridge, and the away-side double-peak: (a) as a function of $\phis=|\phit-\psiEP|$ for $1<\pta<2$~\gev, and (b) as a function of $\pta$ for the $\phis$-integrated correlation. The $\deta$ Gaussian width for the jet-like correlation is also shown. Data are from 20-60\% Au+Au collisions. The trigger $\ptt$ range is $3<\ptt<4$~\gev. Error bars are statistical only. The near-side jet-like $\dphi$ correlation Gaussion width from the minimum bias \dAu\ data is indicated by the arrow in (a) and by the shaded area in (b); the width of the shaded area indicates the statistical uncertainty.}
\label{fig:sigma_4gaus}
\centerline{
\includegraphics[width=0.4\textwidth]{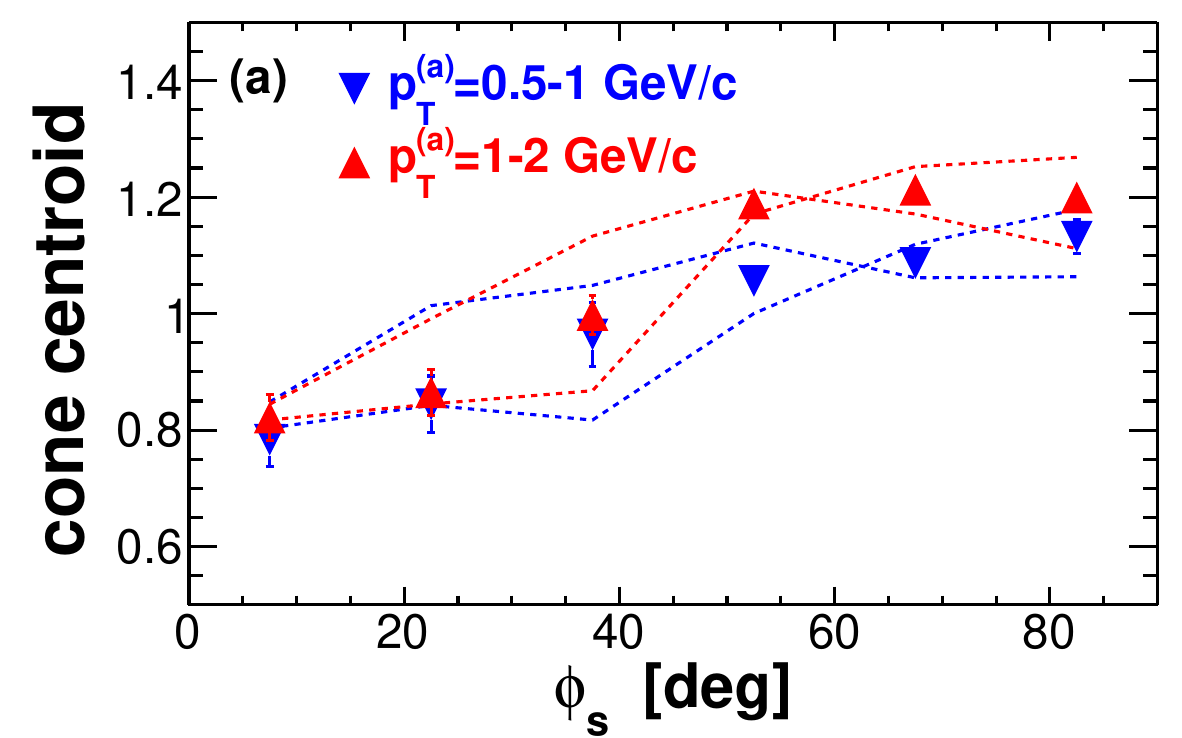}
\includegraphics[width=0.4\textwidth]{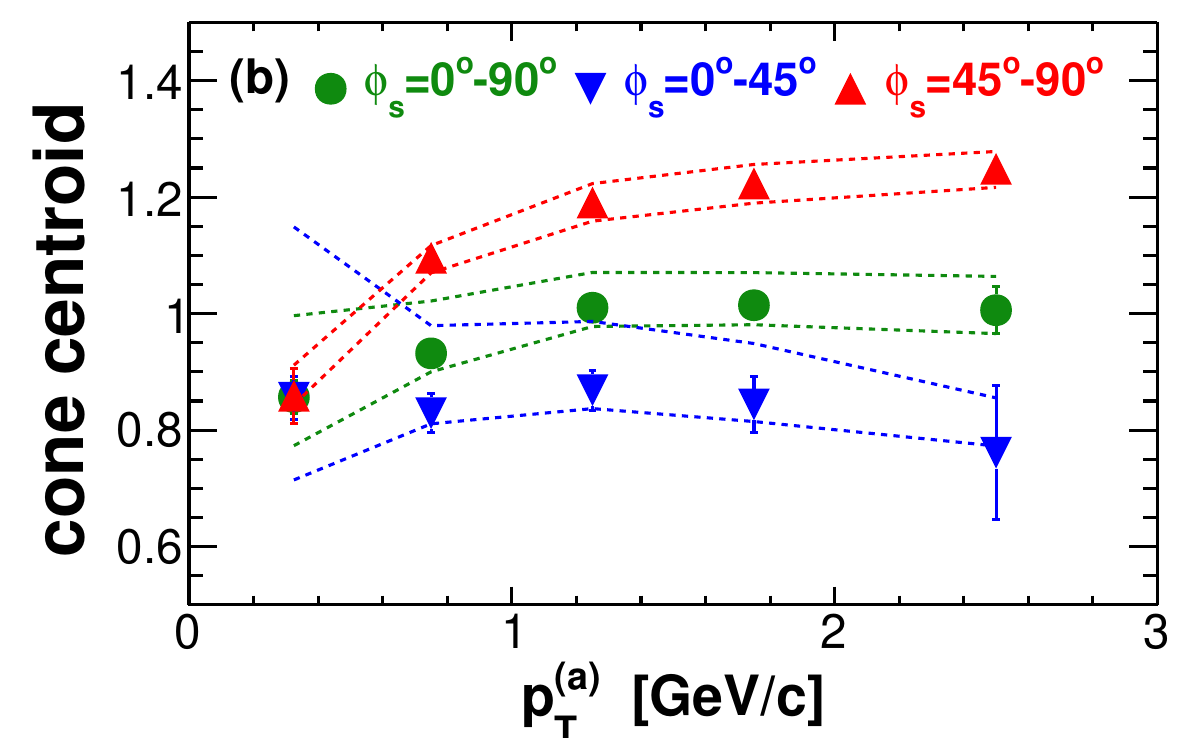}
}
\caption{(Color online) Away-side double-peak position relative to $\dphi=\pi$ from four-Gaussian fits to $\dphi$ correlations at $|\deta|>0.7$: (a) as a function of $\phis=|\phit-\psiEP|$ for two associated particle $\pta$ bins, and (b) as a function of $\pta$ for $\phis$-integrated as well as in-plane and out-of-plane correlations. Data are from 20-60\% Au+Au collisions. The trigger $\ptt$ range is $3<\ptt<4$~\gev. Error bars are statistical. The systematic uncertainties due to elliptic flow are indicated by the dashed lines.}
\label{fig:cone_4gaus}
\end{figure*}

To characterize the structure of the correlation functions, we fit the large-$\deta$ azimuthal correlations with two away-side Gaussian peaks symmetric about $\dphi=\pi$ and two ridge Gaussians (at $\dphi=0$ and $\pi$) with identical widths. We allow the ridge Gaussian magnitudes to vary independently because physics mechanisms other than the back-to-back ridge can also contribute to the $\pi$-region, as discussed earlier.
The fit results are shown by the curves in Fig.~\ref{fig:fit4Gaus} for $3<\ptt<4$~\gev\ and $1<\pta<2$~\gev\ as an example. 
The away-side to near-side ridge ratio for $3<\ptt<4$~\gev\ is generally larger than unity at low $\pta$ but becomes smaller than unity at large $\pta$. For $4<\ptt<6$~\gev\ the fit error is too large to draw a firm conclusion.

We study the peak positions and the Gaussian widths of the various components in the dihadron correlation obtained from our four-Gaussian fit. The Gaussian widths are shown in Fig.~\ref{fig:sigma_4gaus}(a) as a function of $\phis$ for the associated particle $\pta$ range of $1<\pta<2$~\gev. Also shown are the jet-like peak widths in $\deta$ fit to the near-side $\deta$ correlation functions ($|\dphi|<1$), as in Fig.~\ref{fig:deta}. The jet-like peak widths as a function of $\phis$ are flat and are consistent between $\dphi$ and $\deta$ for this associated $1<\pta<2$~\gev\ bin. This again indicates that the jet-like correlation component is independent of the orientation of the trigger particle. 
The widths for the double peaks and the ridge seem to have some dependence on $\phis$, being most different in the $\phis\sim\pi/3$ region.

Figure~\ref{fig:sigma_4gaus}(b) shows the peak Gaussian widths as a function of $\pta$ for integrated $\phis$. The jet-like width in $\dphi$ decreases with increasing associated particle $\pta$, consistent with expectations for jet fragmentation. The $\deta$ width of the jet-like component generally agrees with the $\dphi$ width for $\pta>1$~\gev, but appears narrower than the $\dphi$ width for lower $\pta$. The double-peak width does not vary significantly with $\pta$ except for a drop at low $\pta$, and is wider than the near-side jet-like peak. The ridge peak width does not seem to vary with $\pta$ except perhaps for an increase at low $\pta$. This is in contrast to the $\dphi$ width of the jet-like peak. This may be taken as a confirmation that the ridge and the jet-like component, both on the near side of the trigger particle, may come from rather different physics mechanisms.

For comparison, we fit the near-side jet-like $\dphi$ correlation in minimum bias \dAu\ collisions (which was also obtained by the difference between small and large $\deta$ correlations by Eq.~(\ref{eq:jet})) with a single Gaussian centered at $\dphi=0$. The fitted Gaussian widths are shown in Fig.~\ref{fig:sigma_4gaus} by the shaded area, whose vertical breadth indicates the statistical uncertainty. The $\pt$-integrated correlation Gaussian width is shown by the arrow in the left panel. As seen from the figure, the $\dphi$ widths of the near-side jet-like correlations in Au+Au collisions are consistent with those from \dAu\ collisions at the corresponding $\pt$. 
In addition, as shown in Fig.~\ref{fig:jetRidgeYield}, the near-side jet-like yields are the same for Au+Au and \dAu. In fact, the near-side jet-like correlations in Au+Au collisions of all $\phis$ bins are consistent with the minimum bias \dAu\ data for all $\ptt$ and $\pta$ bins, as shown in Figs.~\ref{figApp:corr34jet} and~\ref{figApp:corr46jet}. This strongly suggests that the near-side jet-like correlations in Au+Au collisions are result of in-vacuum jet fragmentation, just as in \dAu\ collisions.

Figure~\ref{fig:cone_4gaus}(a) shows the fitted double-peak angle as a function of $\phis$ for two associated particle $\pta$ bins. The peak angle increases with increasing $\phis$, and becomes somewhat different for low and high associated particle $\pta$. The larger double-peak angle for out-of-plane trigger particles may be due to a more significant influence from medium flow. For the in-plane orientation, the away-side double-peak hadrons are likely aligned with the medium flow, receiving only a small deflection to their $\pt$. Moreover, the overlap collision zone is thinner in the in-plane direction, thus the away-side correlated hadrons can escape the collision zone more easily. For the out-of-plane orientation, on the other hand, the away-side double-peak hadrons move more or less perpendicularly to the medium flow direction because of the long path-length they have to traverse. They receive a large side-kick from the medium flow, broadening their final emission angle. 

Figure~\ref{fig:cone_4gaus}(b) shows the double-peak angle as a function of $\pta$ for in-plane, out-of-plane, and all trigger particle orientations. The peak angle is relatively independent of the associated particle $\pta$ for in-plane trigger particles. They may more closely reflect the average emission angle of correlated away-side hadrons because the medium flow effect is expected to be small, as discussed above. 

On the other hand, the double-peak angle for the out-of-plane orientation is larger, consistent with a larger deflection from the medium flow. However, the angle position increases with $\pta$, which is naively not expected if those particles are pushed by media with the same flow velocity. 
We note that the medium flow can either broaden or shrink the double-peak angle, depending on the relative orientations of the double-peak hadron direction and the direction of the flow. Investigations of medium flow effects on the correlated hadron emission require realistic dynamical modeling which is outside the scope of this paper.

It is also worth noting that the peak positions reported here are from fits to dihadron correlations. They are different from those obtained from three-particle correlations~\cite{Abelev:2008ac}, where the away-side correlated hadron angle was found to be independent of the associated particle $\pta$. The angle obtained from the three-particle correlation fit is cleaner because the peaks are more cleanly separated in the two-dimensional angular space, while the fit to dihadron correlations is more affected by other physics effects. One such effect is jet deflection~\cite{Armesto:2004pt,Chiu:2006pu}, which was found to be present by three-particle correlations where the diagonal peaks are stronger than the off-diagonal peaks~\cite{Abelev:2008ac}.

\subsection{Effect of Higher Order Harmonics\label{sec:v3}}

It has been suggested by the NeXSPheRIO model~\cite{Takahashi:2009na,Hama:2009vu,Andrade:2009em,Qin:2010pf,Qian:2012qn,Wen:2018agh} that initial energy density fluctuations (hot spots) with subsequent hydro evolution may generate a near-side ridge and a double-peak correlation on the away side. The physics mechanism appears to be side-splashes of particles by the hot spot on the surface resulting in two peaks in the single particle azimuthal distribution event-by-event separated by about two radians~\cite{Hama:2009vu,Andrade:2009em}. These two-peaked single particle distributions produce two-particle correlations of a near-side ridge and an away-side double peak. The near-side ridge and the away-side double peak are due to the same physics, and the near-side ridge amplitude should be larger (by a factor of two) than each of the two away-side peaks. This relative amplitude is a unique feature of the NeXSPheRIO model because of the topology of particle distributions in the model~\cite{Hama:2009vu,Andrade:2009em}. This feature is not observed in the out-of-plane large $\deta$ correlations in data.

It has also been shown based on the AMPT, UrQMD and other models that incorporate Glauber initial geometry~\cite{Alver:2010gr,Lacey:2010hw,Xu:2010du,Xu:2011fe,Petersen:2010cw,Qian:2013nba} that there can be large triangularity in the initial collision geometry event-by-event and those initial geometry fluctuations could produce a triangular anisotropy (triangular flow) in the final momentum space. Such triangular flow would result in three peaks at $\dphi=0$, $2\pi/3$, and $4\pi/3$ in the two-particle correlation, which appear to be qualitively consistent with the inclusive dihadron correlation data integrated over all reaction plane directions~\cite{Adare:2008ae,Aggarwal:2010rf}. 

Hydrodynamic calculations~\cite{Song:2010mg,Schenke:2010rr,Werner:2010aa,Andrade:2010sd,Qiu:2011iv,Schenke:2011bn,Gardim:2011xv,Qiu:2011hf,Schenke:2012wb,Gardim:2012yp,Gale:2013da}, incorporating event-by-event fluctuations in the initial collision geometry, all confirm the existence of $v_3$ and higher-order harmonics in the azimuthal distributions of final-state hadrons. The hydrodynamic evolutions translate the initial configuration space anisotropy (and fluctuations) into final-state momentum anisotropy. The triangular anisotropy $v_3$ has been measured at RHIC~\cite{Adare:2011tg,Adamczyk:2013waa,Adare:2014kci,Adare:2015cpn,Adamczyk:2016exq,Adamczyk:2017ird} with the event-plane as well as the two-particle cumulant method. Event-by-event hydrodynamic calculations are able to reproduce the measurements qualitatively, and in some cases even quantitatively~\cite{Heinz:2013th}.

Because the minor axis direction of the initial fluctuating triangular geometry is random with respect to the reaction plane or the participant plane~\cite{Lacey:2010av,Teaney:2013dta}, the three-peak structure in two-particle correlation from triangular flow should be independent of $\phis$. 
However, the near-side peak of our dihadron correlation data decreases with increasing $\phis$ and is consistent with zero at large $\deta$ with trigger particles out-of-plane, as shown in the upper panel of Fig.~\ref{fig:jetRidge} as well as in Figs.~\ref{figApp:corr34ridge} and~\ref{figApp:corr46ridge}. 
The $\phis$ dependence of the observed correlation structures suggests that the ridge-like correlation is unlikely to be due solely to the possible triangular flow.

\subsubsection{Subtraction of $v_3$}

In order to make a quantitative estimate of $v_3$ effect on our dihadron correlations, we measure the $v_3$ using STAR data and apply flow background subtraction including $v_3$ by Eq.~(\ref{eq:bkgd_v3}). We obtain the $v_3$ of trigger and associated particles using the two-particle cumulant method with a reference particle of $0.15<\pt<2$~\gev\ by Eq.~(\ref{eq:vn}). An $\eta$-gap of 0.7 is applied between the particle of interest and the reference particle, similar to the $\flow{2}$ described in Sec.~\ref{sec:v2}. The $\veta{3}{0.7}$ values are listed in Table~\ref{tab:v3}. Also listed are the $\veta{4}{0.7}$ values, which will be discussed later.

\begin{table}[hbt]
\caption{Triangular and quadrangular anisotropies, $\ff{3}{2}$ and $\ff{4}{2}$, measured by the two-particle cumulant method (with a reference particle) as a function of $\pt$ in 20-60\% minimum-bias Au+Au collisions. An $\etagap=0.7$ is applied. The errors are statistical.} 
\label{tab:v3}
\begin{ruledtabular}
\begin{tabular}{c|cccc}
$\pt$ (\gev) & $\veta{3}{0.7}$ & $\veta{4}{0.7}$ \\ \hline
0.15 - 0.5 & 0.0079 $\pm$ 0.0002 & 0.0019 $\pm$ 0.0004 \\
0.5 - 1 & 0.0246 $\pm$ 0.0002 & 0.0080 $\pm$ 0.0006 \\
1 - 1.5 & 0.0482 $\pm$ 0.0004 & 0.0236 $\pm$ 0.0010 \\
1.5 - 2 & 0.0688 $\pm$ 0.0007 & 0.0376 $\pm$ 0.0018 \\
2 - 3 & 0.0858 $\pm$ 0.0012 & 0.0558 $\pm$ 0.0028 \\
3 - 4 & 0.0905 $\pm$ 0.0038 & 0.0648 $\pm$ 0.0088 \\
4 - 6 & 0.0748 $\pm$ 0.0092 & 0.0687 $\pm$ 0.0214 \\

\end{tabular}
\end{ruledtabular}
\end{table}

\begin{figure*}[hbt]
\centerline{\includegraphics[width=1.03\textwidth]{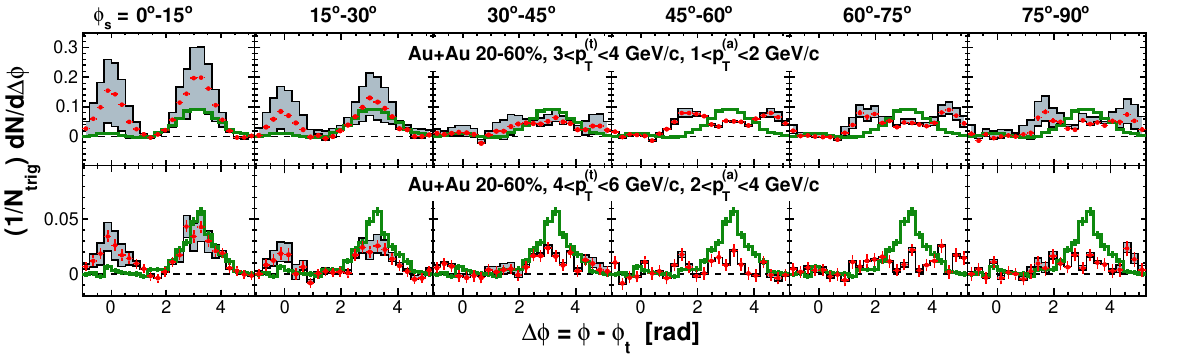}}
\caption{(Color online) Background-subtracted dihadron correlations with trigger particles in six slices of azimuthal angle relative to the event plane, $\phis=|\phit-\psiEP|$, with a cut on the trigger-associated pseudo-rapidity difference of $|\deta|>0.7$. The triangle two-particle $\deta$ acceptance is not corrected. The trigger and associated particle $\pt$ ranges are $3<\ptt<4$~\gev\ and $1<\pta<2$~\gev\ (upper panel), and $4<\ptt<6$~\gev\ and $2<\pta<4$~\gev\ (lower panel), respectively. The data points are from minimum-bias 20-60\% Au+Au collisions. Flow background is subtracted by Eq.~(\ref{eq:bkgd_v3}) using $v_2$ measurements in Table~\ref{tab:v2} and $v_3$ in Table~\ref{tab:v3} and the parameterization in Eq.~(\ref{eq:v4}). Systematic uncertainties due to flow subtraction are shown in the thin histograms embracing the shaded area; those due to \zyam\ normalization are not shown. Error bars are statistical. For comparison, the inclusive dihadron correlations from \dAu\ collisions are superimposed as the thick (green) histograms (only statistical errors are depicted).}
\label{fig:ridge_v3}
\end{figure*}

Figure~\ref{fig:ridge_v3} shows the dihadron correlation functions for $|\deta|>0.7$ and $3<\ptt<4$~\gev\ obtained with $\ff{3}{2}$ included in the background subtraction. The change from Fig.~\ref{fig:jetRidge} is the additional subtraction of the $\ff{3}{2}$ contribution. As seen from Fig.~\ref{fig:ridge_v3}, the qualitative features of the correlation functions are unchanged from those in the upper panels of Fig.~\ref{fig:jetRidge}. The away-side double-peak structure out-of-plane remains prominent. The decreasing trend of the ridge magnitude from in-plane to out-of-plane is unaffected because a constant $v_3$ contribution over $\phis$ is subtracted. This demonstrates that the main features of the measured near-side ridge and away-side double peak in the dihadron correlations with high $\pt$ trigger particles, whether or not integrated over $\phis$, are unlikely to be due solely to the possible triangular flow contributions, but also to other physics mechanisms.

\subsubsection{Subtraction of uncorrelated $v_4$\label{sec:v4uc}}

We have so far subtracted the $\vvPsi$ background correlated with the second harmonic plane, $\psiPP$, by Eq.~(\ref{eq:bkgd}) and Eq.~(\ref{eq:bkgd_v3}). We have used the parameterization of Eq.~(\ref{eq:v4}) to the previous $\vvPsi$ measurement~\cite{Adams:2004bi}. 
There is an additional contribution to $v_4$ that is uncorrelated with $\psiPP$ and arises from fluctuations.
The uncorrelated component can be obtained by
\begin{equation}
\VVuc=\fff{4}{t}{2}\fff{4}{a}{2}-\fff{4}{t}{\psiPP}\fff{4}{a}{\psiPP}\,.\label{eq:v4uc}
\end{equation}
where $\ff{4}{2}$ is the two-particle cumulant $v_4$ with $\etagap=0.7$ given in Table~\ref{tab:v3}.

\begin{figure*}[hbt]
\centerline{\includegraphics[width=1.03\textwidth]{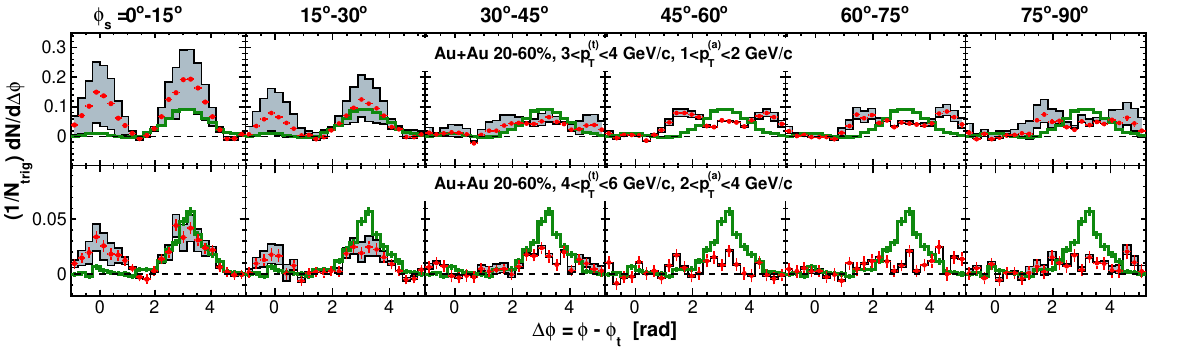}}
\caption{(Color online) The same as Fig.~\ref{fig:ridge_v3}, except with subtraction of an additional background of $\VVuc$ uncorrelated with the second harmonic event plane ($\psi_2$). The data in the upper panels have been published in Ref.~\cite{Agakishiev:2014ada}.}
\label{fig:ridge_v4uc}
\end{figure*}

The flow background including the uncorrelated $\VVuc$ is given by Eq.~(\ref{eq:bkgd_v4}). Figure~\ref{fig:ridge_v4uc} shows the dihadron correlation functions for $|\deta|>0.7$ and $3<\ptt<4$~\gev\ obtained with additional $\VVuc$ included in the background subtraction. As can be seen, the correlation results are effectively as same as those shown in Fig.~\ref{fig:ridge_v3}. This is because $\VVuc$ is small for the 20-60\% centrality and its effect on dihadron correlation is negligible. 

\subsubsection{Subtraction of $\phis$-Dependent $v_2$}

One can always attribute all azimuthal dependence to Fourier harmonics. In fact, Luzum~\cite{Luzum:2010sp} argued that our $|\deta|>0.7$ correlation data can be fitted by Fourier harmonics up to the $4^{\rm th}$ order and the fitted coefficients are consistent with features expected from anisotropic flows. This is not surprising because nonflow effects, which must be contained in the fitted Fourier coefficients, are relatively small compared to the flow contributions in our kinematic regions. If the observed $\phis$-dependent ridge is due to anisotropic flow, then the harmonic flows must be $\phis$-dependent. This may not be impossible because the requirement of trigger particles in a particular $\phis$ bin from the event plane reconstructed from particles in $0.15<\pt<2$~\gev\ could preferentially select events with associated particle $v_2$ displaced from the average. In the following, we analyze the two-particle cumulant $v_n$ in events of different $\phis$ values separately, and subtract them from the dihadron correlations. 

\begin{figure}[hbt]
\centerline{\includegraphics[width=0.45\textwidth]{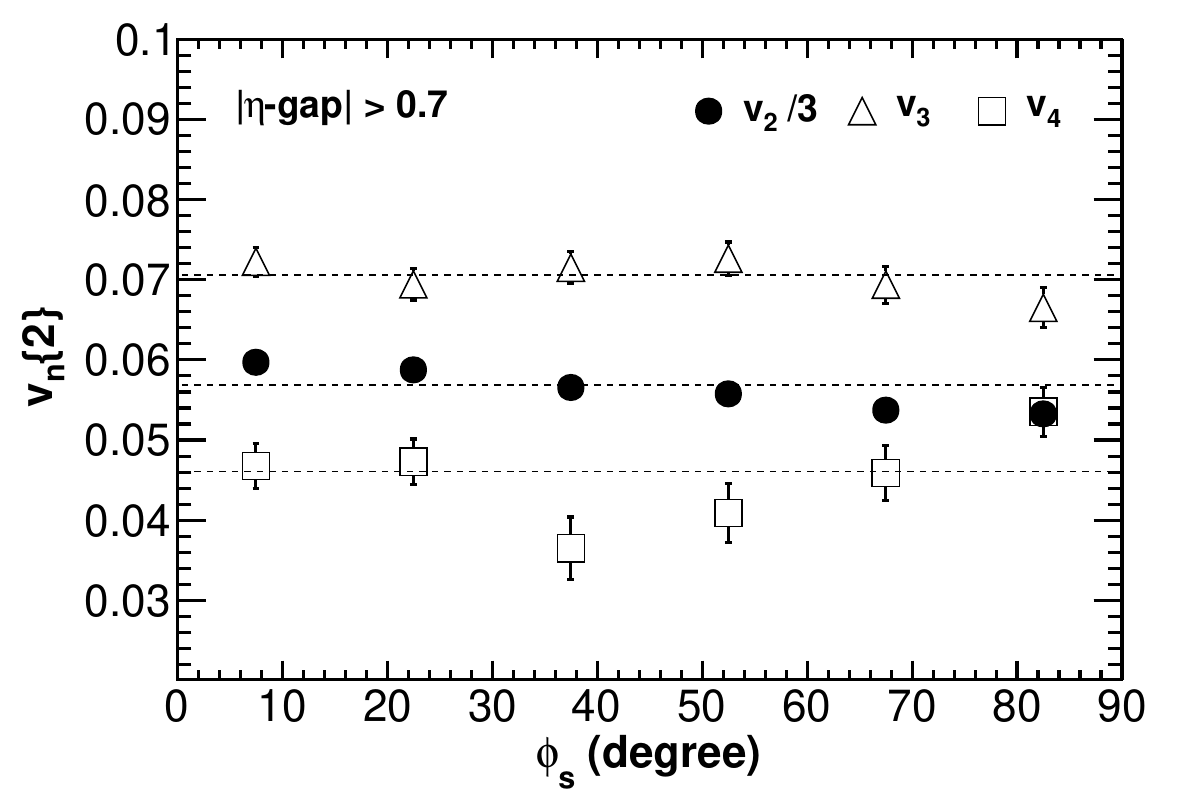}}
\caption{Harmonic $\vpt{n}$ of associated particles of $1.5<\pta<2$~\gev\ as a function of $\phis$ of trigger particles of $3<\ptt<4$~\gev. Note that $\vpt{2}$ is scaled down by a factor of 3 to fit into the plot coordinate range. The $\vpt{n}$ is measured by the two-particle cumulant method with particle pairs from the same associated $\pta$ bin and with $\etagap=0.7$. The data are from minimum-bias 20-60\% Au+Au collisions. Error bars are statistical. The horizontal lines are to guide the eye.}
\label{fig:vn_phisDep}
\end{figure}

Since reference particles are used to reconstruct the EP to determine the $\phis$, one cannot calculate $v_n$ from the cumulant of the associated particle and a reference particle in event sample selected according to $\phis$. Instead, we form a two-particle cumulant from particles in a given associated $\pta$ bin, applying an $\eta$-gap of 0.7. The $v_n$ of the associated particles is simply the square root of the cumulants:
\begin{equation}
\vpt{n}(\phis)=\sqrt{\Vpteta{n}{0.7}(\phis)}\,.\label{eq:vn_phisDep}
\end{equation}
Here $\Vpteta{n}{0.7}$ indicates the two-particle cumulant with particle pairs from the same $\pt$ bin. We use $\vpteta{n}{0.7}$ or simply $\vpt{n}$ to stand for the resultant anisotropy measurement. Figure~\ref{fig:vn_phisDep} shows the obtained $\vpt{n}$ of $1.5<\pt<2$~\gev\ as a function of $\phis$ of trigger particles of $3<\ptt<4$~\gev. The $\vpt{2}$ decreases with $\phis$. The decrease is a consequence of the decreasing ridge with increasing $\phis$. The $\vpt{4}$ is found to be smallest with $\phis=45^\circ$ and largest with $\phis=0^\circ$ and $90^\circ$. On the other hand, the $\vpt{3}$ is independent of $\phis$, consistent with the expectation that the third and second harmonic planes are uncorrelated at mid-rapidity. The $\vpt{3}$ from the cumulant of same-$\pt$ bin pairs is consistent with that obtained from the cumulant with a reference particle, $\ff{3}{2}$, given in Table~\ref{tab:v3}. The $\vpt{2}$ values are listed in Table~\ref{tab:v2_phisDep} as a function of $\pt$ and $\phis$. Results for two $\etagap$ values are listed, $\vpteta{2}{0.7}$ and $\vpteta{2}{1.2}$, to estimate the range of $\vpt{2}$.

\begin{table*}[hbt]
\caption{Elliptic flow anisotropy, $\ff{2}{\pt\mbox{-}\pt}$, measured by the two-particle cumulant method using pairs from the same $\pt$ bin, as a function of $\pt$ and $\phis$ in 20-60\% minimum-bias Au+Au collisions. Two $\etagap$ values (0.7 and 1.2) are used. Errors are statistical.}
\label{tab:v2_phisDep}
\begin{ruledtabular}
\begin{tabular}{c|cccccc}
$\pta$ (\gev)	& $0 - \pi/12$ & $\pi/12 - \pi/6$ & $\pi/6 - \pi/4$ & $\pi/4- \pi/3$ & $\pi/3 - 5\pi/12$ & $5\pi/12 - \pi/2$ \\ \hline
 & \multicolumn{6}{c}{$\vpteta{2}{0.7}$} \\
0.15 - 0.5 & 0.0432 $\pm$ 0.0002 & 0.0421 $\pm$ 0.0003 & 0.0416 $\pm$ 0.0003 & 0.0403 $\pm$ 0.0003 & 0.0393 $\pm$ 0.0003 & 0.0379 $\pm$ 0.0004 \\
0.5 - 1 & 0.0923 $\pm$ 0.0002 & 0.0916 $\pm$ 0.0002 & 0.0903 $\pm$ 0.0002 & 0.0878 $\pm$ 0.0002 & 0.0858 $\pm$ 0.0002 & 0.0854 $\pm$ 0.0002 \\
1 - 1.5 & 0.1427 $\pm$ 0.0003 & 0.1399 $\pm$ 0.0003 & 0.1371 $\pm$ 0.0004 & 0.1347 $\pm$ 0.0004 & 0.1301 $\pm$ 0.0004 & 0.1296 $\pm$ 0.0004 \\
1.5 - 2 & 0.1791 $\pm$ 0.0007 & 0.1763 $\pm$ 0.0008 & 0.1697 $\pm$ 0.0008 & 0.1673 $\pm$ 0.0009 & 0.1612 $\pm$ 0.0010 & 0.1598 $\pm$ 0.0010 \\
2 - 3 & 0.2108 $\pm$ 0.0015 & 0.2081 $\pm$ 0.0016 & 0.1976 $\pm$ 0.0018 & 0.1905 $\pm$ 0.0020 & 0.1860 $\pm$ 0.0021 & 0.1885 $\pm$ 0.0022 \\
 & \multicolumn{6}{c}{$\vpteta{2}{1.2}$} \\
0.15 - 0.5 & 0.0435 $\pm$ 0.0004 & 0.0422 $\pm$ 0.0005 & 0.0417 $\pm$ 0.0005 & 0.0403 $\pm$ 0.0006 & 0.0397 $\pm$ 0.0006 & 0.0366 $\pm$ 0.0007 \\
0.5 - 1 & 0.0914 $\pm$ 0.0003 & 0.0903 $\pm$ 0.0004 & 0.0891 $\pm$ 0.0004 & 0.0861 $\pm$ 0.0004 & 0.0850 $\pm$ 0.0004 & 0.0843 $\pm$ 0.0005 \\
1 - 1.5 & 0.1409 $\pm$ 0.0006 & 0.1401 $\pm$ 0.0006 & 0.1356 $\pm$ 0.0007 & 0.1324 $\pm$ 0.0007 & 0.1297 $\pm$ 0.0008 & 0.1274 $\pm$ 0.0008 \\
1.5 - 2 & 0.1752 $\pm$ 0.0013 & 0.1755 $\pm$ 0.0014 & 0.1673 $\pm$ 0.0015 & 0.1649 $\pm$ 0.0017 & 0.1593 $\pm$ 0.0018 & 0.1553 $\pm$ 0.0019 \\
2 - 3 & 0.2136 $\pm$ 0.0027 & 0.2037 $\pm$ 0.0030 & 0.1963 $\pm$ 0.0032 & 0.1959 $\pm$ 0.0035 & 0.1773 $\pm$ 0.0040 & 0.1863 $\pm$ 0.0040 \\

\end{tabular}
\end{ruledtabular}
\end{table*}

Although the measured $\vpt{4}$ is $\phis$-dependent, the contribution of the $\psiPP$-uncorrelated $v_4$ to flow background is negligibly small, as discussed in Sec.~\ref{sec:v4uc}. We therefore use the $\phis$-independent $\ff{4}{2}$ measured by the two-particle cumulant with a reference particle, as in Sec.~\ref{sec:v4uc}. We have checked our results using the $\phis$-dependent $\vpt{4}$ and found no observable difference. 

\begin{figure*}[hbt]
\centerline{\includegraphics[width=1.03\textwidth]{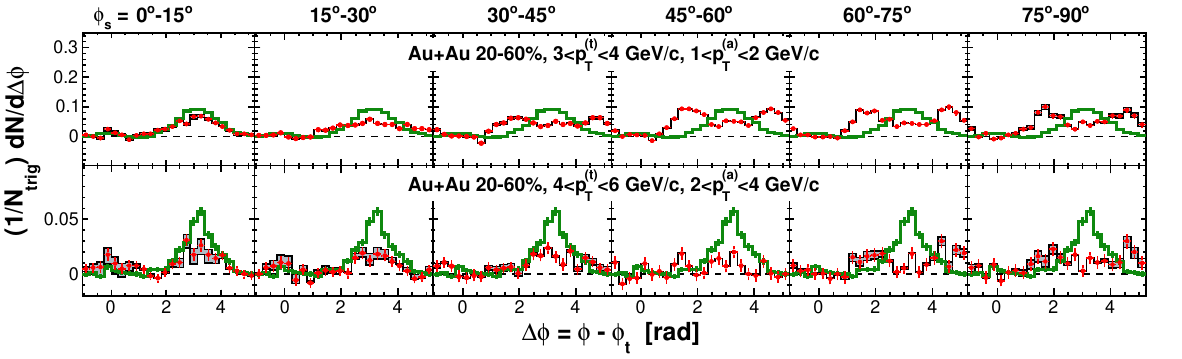}}
\caption{(Color online) Background-subtracted dihadron correlations with trigger particles in six slices of azimuthal angle relative to the event plane, $\phis=|\phit-\psiEP|$, with a cut on the trigger-associated pseudo-rapidity difference of $|\deta|>0.7$. The triangle two-particle $\deta$ acceptance is not corrected. The trigger and associated particle $\pt$ ranges are $3<\ptt<4$~\gev\ and $1<\pta<2$~\gev\ (upper panel), and $4<\ptt<6$~\gev\ and $2<\pta<4$~\gev\ (lower panel), respectively. The data points are from minimum-bias 20-60\% Au+Au collisions. Flow background is subtracted by Eq.~(\ref{eq:bkgd_v4}). The $\phis$-dependent $\vpt{2}$ measured by two-particle cumulants with $\etagap=0.7$ and 1.2 in Table~\ref{tab:v2_phisDep} are used (the thin histograms embracing the shaded area), with their average shown in the data points. The subtracted $\ff{3}{2}$ is given in Table~\ref{tab:v3}. The subtracted $\vvPsi$ is parameterized by Eq.~(\ref{eq:v4}), and the $\VVuc$ is given by Eq.~(\ref{eq:v4uc}). Error bars are statistical; systematic uncertainties are not shown. The shaded areas show the range of the results using $\vpt{2}$ values from two $\eta$-gaps of 0.7 and 1.2. For comparison, the inclusive dihadron correlations from \dAu\ collisions are superimposed as the thick (green) histograms.}
\label{fig:ridge_phisDep}
\end{figure*}

In the following we subtract flow background using the $\phis$-dependent $\vpt{2}(\phis)$, as discussed above and tabulated in Table~\ref{tab:v2_phisDep}. The trigger particle $v_2$ is still given by the two-particle cumulant flow obtained with a reference particle from Table~\ref{tab:v2} as in Sec.~\ref{sec:v2}. This is because the trigger $v_2$ is the second harmonic modulation of trigger particles, which determines the $\phis$. The flow background is given by Eq.~(\ref{eq:bkgd_v4}) and is normalized by \zyam. 
Figure~\ref{fig:ridge_phisDep} shows the dihadron correlation results for $3<\ptt<4$~\gev\ and $|\deta|>0.7$ with subtraction of $\phis$-dependent $\vpt{2}$. The change from the lower systematic bound in Fig.~\ref{fig:ridge_v3} is the subtraction of the $\phis$-dependent $v_2$ in place of the $\phis$-independent one.

As seen from Fig.~\ref{fig:ridge_phisDep}, the near-side ridge is diminished, maybe as expected, because the large $\deta$ ridge is presumably included in the subtracted $v_n$. However, it is important to point out that it is not automatically guaranteed that the ridge will be gone just because the $v_n$'s are measured by two-particle cumulant either with a reference particle or with a particle from the same $\pt$ region. This is because they are not simply measured by the trigger-associated particle pair at $|\deta|>0.7$. If they were, then the correlation would be strictly zero everywhere, both on the near side and on the away side. This would be similar to the fitting method in Ref.~\cite{Sharma:2015qra}, where the large $\deta$ region on the near side is fitted and treated as background Fourier coefficients. Such Fourier coefficients would be inevitably $\phis$-dependent, and the near-side ridge would by definition be zero.

It is interesting to note that, despite the diminished near-side ridge, the away-side correlation is not diminished. It still evolves from a single peak with in-plane trigger particles to a double peak with out-of-plane trigger particles. The observation of the away-side double-peak structure for out-of-plane triggers seems robust against the wide range of flow background subtraction.

It is worth noting that, if $v_2$ depends on $\phis$, then the factorization of $\vt$ and $\va$ in inclusive dihadron correlation analysis is no longer valid and the flow background there may be underestimated. This is discussed in Appendix~\ref{app:inclusive}.

As noted in Sec.~\ref{sec:v3bkgd}, we have neglected the effect of dipole fluctuations (rapidity-even $v_1$) in flow background subtraction. STAR measurements~\cite{Pandit:2012hp,Star:2018zpt} indicate that the dipole fluctuation effect changes sign at $\pt\approx1$~\gev, negative at lower $\pt$ and positive at higher $\pt$. For $\pta=1$-2~\gev\ shown in Fig.~\ref{fig:ridge_phisDep}, the dipole fluctuation effect is approximately zero and can be neglected. The qualitative conclusions on the near-side and away-side correlations are therefore unaffected by the potential dipole fluctuations. 

Figure~\ref{figApp:corr34ridge_phisDep} shows results similar to Fig.~\ref{fig:ridge_phisDep} but for other associated $\pta$ bins. Figure~\ref{figApp:corr46ridge_phisDep} shows the results for $4<\ptt<6$~\gev. For all kinematic cuts studied, the near-side ridges all seem to vanish after the subtraction of the $\phis$-dependent $\vpt{2}$, the two-particle cumulant $\Flow{3}{2}$, and the $\psiPP$-correlated $\ff{4}{\psiPP}$ and uncorrelated $\vuc$. The evolution of the away-side correlation function from in-plane to out-of-plane appears different for high and low associated $\pta$. At relatively high $\pta$, the away-side correlation is single-peaked for in-plane triggers and double-peaked for out-of-plane triggers, as already noted earlier in Fig.~\ref{fig:ridge_phisDep}. At low $\pta$, however, the trend is opposite--the away-side correlation is double-peaked in-plane and single-peaked out-of-plane. As noted above, we have neglected the effect of dipole fluctuations in flow background subtraction. The effect of dipole fluctuations is negative at low $\pta$. This may be responsible for the concave shape of the near-side correlation. However, the away-side correlation shape would be more strongly double-peaked after the subtraction of a negative dipole background. Thus the qualitative conclusion of the double-peaked away-side correlations at low $\pta$ for in-plane triggers seems robust.

Since the ridge is diminished after subtraction of $\phis$-dependent $\vf$ from the two-particle cumulant at large $\deta$, can we conclude that the physics origin of the ridge is hydrodynamic $v_n$ flow? The answer is no, because any non-hydrodynamic origin of $v_n$ is also included in the two-particle $v_n$ measurements. In other words, any ridge signal (whatever its physics origin might be) is included in $v_n$, and the ridge would be subtracted after subtraction of $v_n$. However, one also cannot rule out the ridge being part of hydrodynamic flow. This is because it is still possible that hydrodynamic flow of the underlying event is biased by the selection of the trigger particle orientation, and all the long-range $\deta$ correlation may indeed be due to flow. 

\section{Conclusions\label{conclusion}}

Dihadron azimuthal correlations in non-central 20-60\% Au+Au collisions are reported by the STAR experiment as a function of trigger particle azimuthal angle relative to the event plane ($\phis=|\phit-\psiEP|$) in six equal-size slices. 
The correlations have first been studied with subtraction of the even harmonic elliptic and quadrangular flow backgrounds. 
The $\phis$ dependence of the dihadron correlation signal, as well as the trigger and associated particle transverse momentum ($\pt$) dependences, have been studied. Minimum-bias \dAu\ collision data have been presented for comparison. The correlation functions have also been obtained for small and large pseudo-rapidity separations ($|\deta|$) independently in order to isolate the jet-like and ridge (long range $\deta$ correlation) contributions. The resulting jet-like and ridge components have been studied as a function of $\phis$, trigger particle $\ptt$ and associated particle $\pta$.

The \zyam\ background subtraction method has been described in detail. The flow subtraction has been carried out to the order of $\vf\vv$. The systematic uncertainties in the background subtraction have been discussed extensively. 
The effect of the triangular flow harmonic was not subtracted in the results quantitatively characterizing the main features of the correlation function in the jet-like versus ridge-like regions.
However, the effects of triangular flow fluctuations, as well as $\phis$-dependent elliptic flow, on these main features have been investigated and discussed.

The dihadron correlations are strongly modified in Au+Au collisions with respect to minimum-bias \dAu\ collisions. The modifications strongly depend on the trigger particle orientation relative to the event plane and evolve with associated particle $\pta$. No significant changes are observed between trigger particle $\ptt$ ranges of $3<\ptt<4$~\gev\ and $4<\ptt<6$~\gev. The $\phis$ and $\pta$ dependences of the correlation functions are quantitatively similar in the two trigger particle $\ptt$ ranges.

The away-side dihadron correlation broadens from in-plane to out-of-plane. 
The away-side correlation for $\phis<\pi/6$ is single-peaked, independent of $\pta$, and not much wider than in \dAu, while the amplitude is larger than the \dAu\ data. 
For $\phis>\pi/6$, the away-side double-peak structure starts to develop and becomes stronger for increasing $\phis$ and increasing $\pta$. The strongest double-peak structure is found at large $\pta$ in the out-of-plane direction.

The away-side dihadron correlation amplitude at $\dphi=\pi$ drops from in-plane to out-of-plane, while that in the double-peak region remains approximately constant over $\phis$.
For in-plane $\phis$, the amplitude ratio in the $\pi$-region to the double-peak region increases with $\pta$, consistent with \dAu\ and qualitatively consistent with punch-through jets or away-side jets not interacting with the medium. However, the individual amplitudes in these two regions are both higher than in \dAu, suggesting other physics mechanisms are at work. For out-of-plane $\phis$, the amplitude ratio decreases strongly with $\pta$, opposite to what would be expected from punch-through jets.

The near-side dihadron correlation amplitude decreases with increasing $\phis$. The decrease comes entirely from the decrease in the ridge. The ridge is extracted from correlations at $|\deta|>0.7$. Its amplitude is found to decrease with increasing $\phis$ significantly in the 20-60\% centrality. 
This feature is present for all associated particle $\pta$, and appears to be independent of $\pta$.

The jet-like contribution to the near-side correlation has been extracted from the difference between small- and large-$\deta$ azimuthal correlations, subject to small experimental systematic uncertainties. 
The jet-like contribution is invariant from in-plane to out-of-plane within our systematic uncertainties, and is found to be the same as in \dAu\ collisions.

The different behaviors of the jet-like component and the ridge with respect to $\phis$ suggest that their production mechanisms are different. The jet-like component is insensitive to the reaction plane and appears to be universal, suggesting in-vacuum jet-fragmentation of partons whose production is biased towards the surface of the collision zone by requiring the high $\pt$ trigger particles. The strong dependence of the ridge on the reaction plane suggests its origin to be connected to the medium, not to the jet.

There might be strong connections between the near- and aways-side of the dihadron correlations. It is found that the near-side jet-like yield and the away-side double-peak yield both have little dependence on $\phis$. The jet-like spectral shape and the double-peak hadron spectral shape do not change with $\phis$, and the double-peak region spectra are not much softer than the jet-like spectra. The jet-like spectrum for $4<\ptt<6$~\gev\ is somewhat harder than that for $3<\ptt<4$~\gev, while the double-peak hadron spectra remain the same for the two trigger $\ptt$ ranges. 

On the other hand, the near-side ridge and the away-side $\pi$-region appear to trace each other as a function of $\phis$, and also approximately as a function of $\pta$, suggesting the possibility of a back-to-back ridge. This would be consistent with the recent suggestion that the ridge may be generated by fluctuations in the initial color flux tubes focused by transverse radial flow. Such a picture would also explain the decreasing ridge from in-plane to out-of-plane because both the color flux tubes and the radial flow are the strongest along the in-plane direction. However, it remains unclear why the ridge particles are much harder than inclusive hadrons in our measured $\pta$ region. 

The dihadron correlation structure has been fitted with a four-Gaussian model representing a back-to-back ridge and an away-side double peak. The fitted away-side double-peak angle increases from in-plane to out-of-plane. 
For in-plane trigger particles, the fitted double-peak angle is approximately constant over the associated particle $\pta$. For out-of-plane trigger particles, it increases with $\pta$. 
Whether and how much the medium flow influences the emission directions of the away-side correlated particles warrants further investigation.

The dihadron correlations have been further studied with subtraction of triangular anisotropy ($v_3$) independent of $\phis$. The $v_3$ was measured by the two-particle cumulant method with a $\eta$-gap ($\etagap$) of 0.7. The triangular anisotropy with larger $\etagap$ is significantly smaller. The qualitative feature of the correlation data seems unchanged. The ridge magnitude is reduced, but seems still to be present for in-plane trigger particles, decreases from in-plane to out-of-plane, and vanishes for out-of-plane trigger particles.

Finally, we have considered the effect of a $v_n$ that is dependent on the trigger-particle $\phis$. We analyzed the two-particle cumulants $\vpt{n}$ in events with different trigger particle $\phis$ separately. The second harmonic $\vpt{2}$ is found to decrease with increasing $\phis$. This is synonymous to the decreasing ridge magnitude with $\phis$. The fourth harmonic $\vpt{4}$ is found to also depend on $\phis$, but its effect on dihadron correlation is negligible. The third harmonic $\vpt{3}$ is found to be independent of $\phis$. The dihadron correlations have been studied relative to the event plane with the subtraction of the two-particle cumulants, $\veta{3}{0.7}$, $\veta{4}{0.7}$ and the $\phis$-dependent $\vpteta{2}{0.7}(\phis)$. With this exploratory subtraction of the $v_n$ values, the ridge is found to be eliminated. However, this result does not enlighten us as to the origin of the ridge because the measured $\phis$-dependent $v_2$ has likely already included the ridge; whether the ridge is due to flow or nonflow is undetermined. On the other hand, the away-side double-peak structure for out-of-plane triggers remains robust even with the subtraction of $\phis$-dependent $\vpteta{n}{0.7}$. This indicates a medium effect on the away-side jet propagation, and the effect depends on the pathlength the away-side jet traverses. 

To summarize our main findings, high $\pt$ triggered particles are biased towards surface emission, and the near-side jet fragmentation is hardly modified by the medium. 
Away-side partner jets interact maximally with the medium in the direction perpendicular to the reaction plane. These interactions may be responsible for the double-peak correlation structure remaining on the away side even after $v_3$ subtraction. 
The near-side jet-like component is accompanied by the ridge in the reaction-plane direction. The ridge magnitude drops rapidly with increasing $\phis$ and largely disappears out-of-plane in mid-central 20-60\% Au+Au collisions. 
The most natural explanation for our results seems to be the combination of a near-side in-vacuum jet-fragmentation, a near-side ridge and away-side double-peak structure significantly contributed by triangular flow, and remaining double-peak correlations on the away side for out-of-plane trigger particles. 

\appendix

\section{Effect of Possible Biases in Event-Plane Reconstruction\label{app:EP}}

\begin{figure*}[hbt]
\centerline{\includegraphics[width=1.03\textwidth]{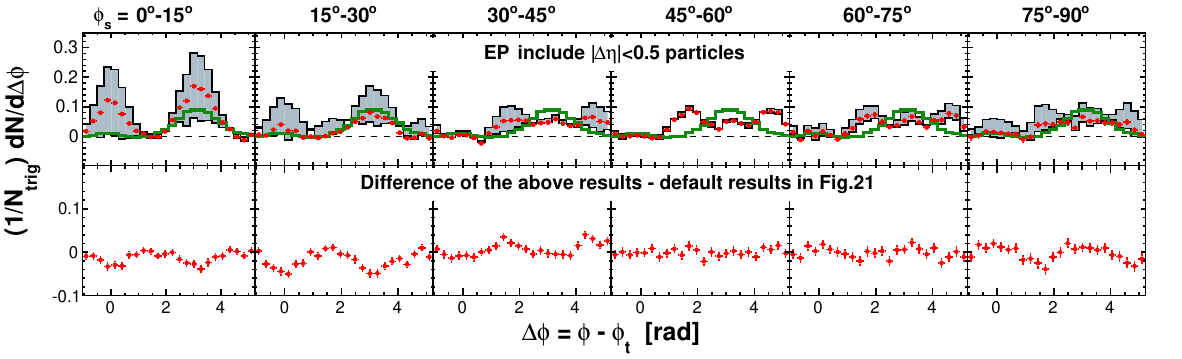}}
\caption{(Color online) Upper panels: As same as Fig.~\ref{fig:ridge_v3} upper panels but relative to event plane reconstructed without excluding particles within $|\deta|<0.5$ of the trigger. The $v_2$, $v_4(\psi_2)$, and $v_3$ backgrounds are subtracted with resolutions corresponding to the new EP. Lower panels: the difference between the upper panel results minus the default results in Fig.~\ref{fig:ridge_v3}.}
\label{fig:EP_bias_check}
\end{figure*}

In our analysis, the event plane is reconstructed by particles excluding those within $|\deta|<0.5$ of the trigger particle. The question remains of how large the effect is of possible biases in the reconstructed event plane from particles correlated to the trigger, especially on the away side. One way to estimate this possible effect is to analyze dihadron correlations relative to the event plane reconstructed from particles without excluding those within $|\deta|<0.5$ of the trigger, thereby maximizing the biases from jet-correlations. These results (subtracted by $v_2$, $v_4(\psi_2)$, and $v_3$ backgrounds with resolutions corresponding to the new EP) are shown in the upper panels of Fig.~\ref{fig:EP_bias_check}. The differences between these results and our default results in Fig.~\ref{fig:ridge_v3} are shown in the lower panels of Fig.~\ref{fig:EP_bias_check}. By including in the EP those particles close to the trigger in $\eta$, the correlated yield at $\dphi=0$ for in-plane triggers is smaller, and for out-of-plane triggers, larger. The correlated yield at $\dphi=0$ is not larger for in-plane triggers, as one would naively expect from a more aligned EP. This is because the associated $\pt$ bin is always excluded from EP reconstruction. We have verified that if the associated $\pt$ bin is included in EP, the associated yield at $\dphi=0$ for in-plane triggers is significantly enhanced, as expected. 

As seen from Fig.~\ref{fig:EP_bias_check}, introducing a stronger bias in EP reconstruction causes a relatively small change in the correlation signals. This suggests that possible EP biases in our default results in Fig.~\ref{fig:ridge_v3} may be also relatively small.

\section{Implications of Possible $\phis$-Dependent $v_2$ on Inclusive Dihadron Correlations\label{app:inclusive}}

If $v_2$ depends on $\phis$, then there is an important implication for the inclusive dihadron correlation (i.e.~without cutting on $\phis$). For inclusive dihadron correlation, a flow background $\mean{\fff{2}{t}{2}}\cdot\mean{\fff{2}{a}{2}}$ has been used so far for $\mean{\fff{2}{t}{2}\cdot\fff{2}{a}{2}}$. (Note, for clarity, we have omitted the $\mean{...}$ notation throughout the paper except here.) This is correct because fluctuations are already included in the two-particle cumulant flow measurement of $\mean{\ff{2}{2}}$. However, if $v_2$ depends on trigger particle orientation $\phis$, then the equality $\mean{\fff{2}{t}{2}(\phis)\cdot\fff{2}{a}{2}(\phis)}=\mean{\fff{2}{t}{2}(\phis)}\cdot\mean{\fff{2}{a}{2}(\phis)}$ is no longer valid. The left-hand side is always larger than the right-hand side. This means that the inclusive dihadron flow background is underestimated by $\mean{\fff{2}{t}{2}}\cdot\mean{\fff{2}{a}{2}}$. In fact, because $\fff{2}{t}{2}(\phis)$ is positive for $\phis\sim0$ and negative for $\phis\sim\pi/2$, the true background magnitude for inclusive dihadron correlation is even larger than that for the $\phis=0$ dihadron correlation, which has the largest background magnitude of all $\phis$ bins. Namely, for all $\phis$,
\begin{widetext}
\begin{equation}
\mean{\fff{2}{t}{2}(\phis)\cdot\fff{2}{a}{2}(\phis)} >
\mean{\fff{2}{t}{2}(\phis)}\cdot\mean{\fff{2}{a}{2}(\phis)} >
\mean{\fff{2}{t}{2}(\phis=0)}\cdot\mean{\fff{2}{a}{2}(\phis=0)}\,.
\end{equation}
\end{widetext}

\begin{figure}[hbt]
\centerline{\includegraphics[width=0.35\textwidth]{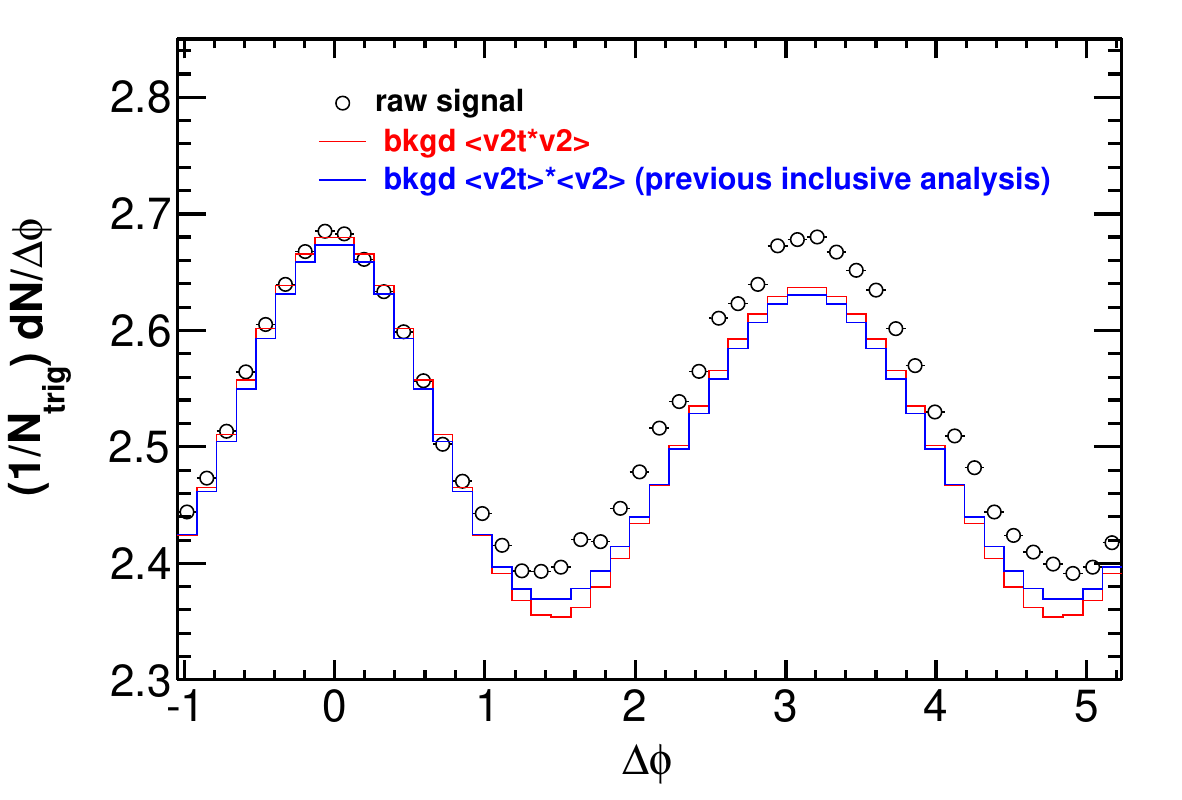}}
\centerline{\includegraphics[width=0.35\textwidth]{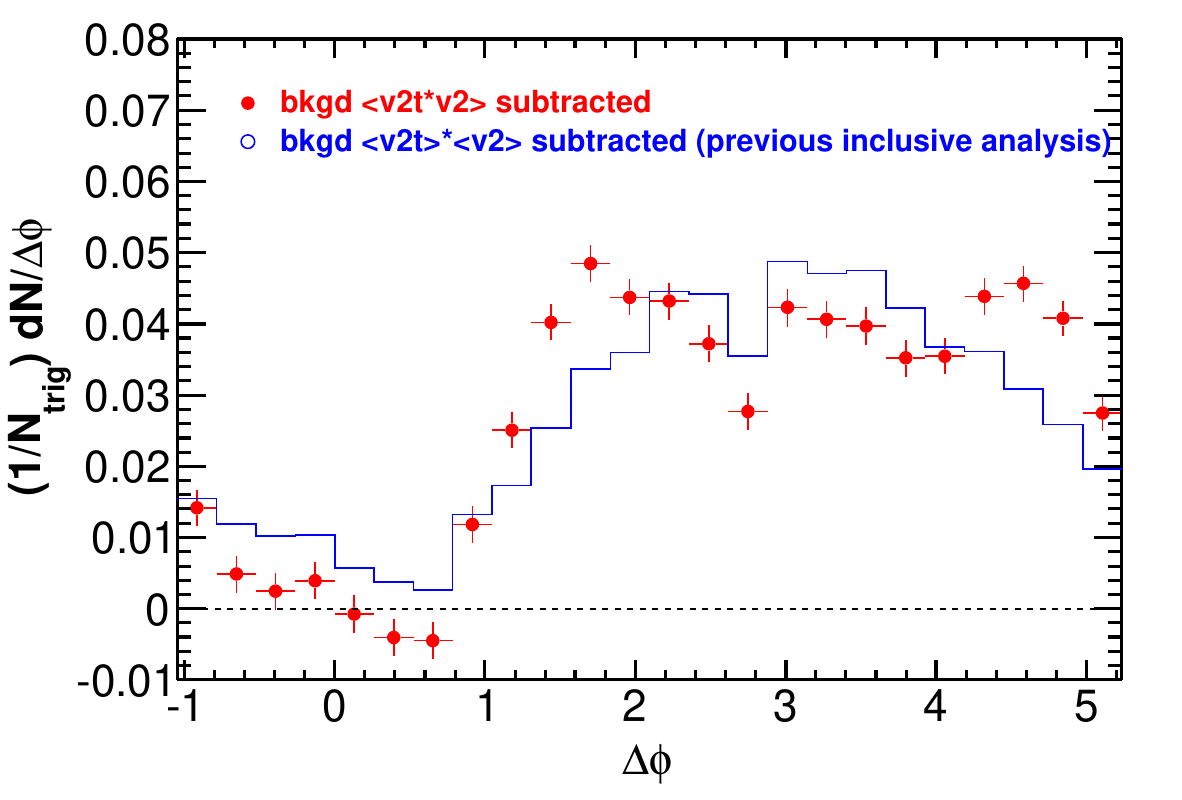}}
\caption{(Color online) Effect of possible $\phis$-dependent elliptic flow anisotropy on inclusive dihadron correlations. Upper panel: raw $\dphi$ correlation together with flow background obtained by two different ways, one by the average of the product of the trigger $\vtR$ and the associated particle $\va(\phis)$ as from this analysis (red histogram), and the other by the product of the average trigger and associated $v_2$ as from the standard inclusive dihadron correlation analysis (blue histogram). Lower panel: the correlation signals subtracted by the background from this analysis (red points) and by the standard background from inclusive dihadron correlation analysis (blue histogram). The data are from minimum-bias 20-60\% Au+Au collisions. The trigger and associated particle $\pt$ ranges are $3<\ptt<4$~\gev\ and $1<\pta<1.5$~\gev, respectively. A $|\deta|>0.7$ cut is applied to the trigger-associated pairs. Error bars are statistical.}
\label{fig:inclusive_dihadron}
\end{figure}

Fig.~\ref{fig:inclusive_dihadron} illustrates the effect. The upper panel shows the raw dihadron correlation for $3<\ptt<4$~\gev\ and $1<\pta<1.5$~\gev\ together with two flow background curves, both \zyam-normalized. The blue histogram is from a traditional inclusive dihadron correlation analysis with the $v_2$ modulation calculated from $\mean{\fff{2}{t}{2}}\cdot\mean{\fff{2}{a}{2}}$. 
The red histogram is that calculated from the $\phis$-dependent $\ff{2}{2}(\phis)$ by $\mean{\fff{2}{t}{2}(\phis)\cdot\fff{2}{a}{2}(\phis)}$ which is the correct flow background provided $v_2(\phis)$ is the real flow. (The $v_3$ and $v_4$ contributions are included in both flow background histograms). 
As seen from Fig.~\ref{fig:inclusive_dihadron}, the traditional flow background is underestimated. The lower panel of Fig.~\ref{fig:inclusive_dihadron} shows the dihadron correlation signals after subtraction of the traditional background, shown by the histogram, and of the correct flow background, shown by the data points. The signal from the traditional average flow background subtraction is less double-peaked. This means, if the ridge is entirely due to flow that must be $\phis$-dependent, then all the inclusive dihadron correlation analyses have under-subtracted the flow background, resulting in a more peaked away-side correlation signal.

\section{Dihadron Correlation Functions\label{app}}

\begin{figure*}[hbt]
\centerline{\includegraphics[width=\textwidth]{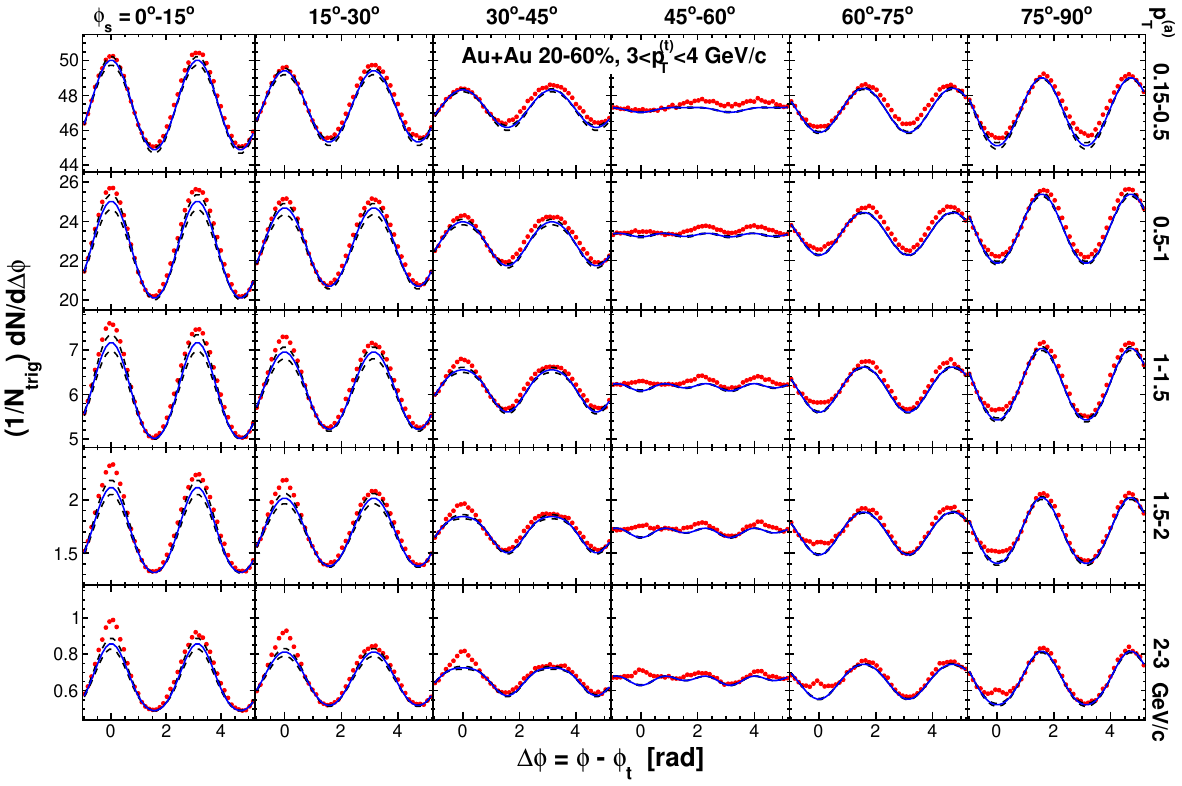}}
\caption{(Color online) Raw dihadron $\dphi$ correlations with trigger particles in six slices of azimuthal angle relative to the event plane, $\phis=|\phit-\psiEP|$. The data are from minimum-bias 20-60\% Au+Au collisions. The trigger $\pt$ range is $3<\ptt<4$~\gev. Five associated particle $\pta$ bins are shown. Both the trigger and associated particles are restricted within $|\eta|<1$. The triangle two-particle $\deta$ acceptance is not corrected. Statistical errors are smaller than the symbol size.  The curves are flow modulated \zyam\ background including $\vf$ and $\vvPsi$ by Eq.~(\ref{eq:bkgd}). The used $v_2$ values are given in Table~\ref{tab:v2} from four-particle $\flow{4}$ and two-particle $\veta{2}{0.7}$ (dashed curves) and the average $\vf$ from the two methods (solid curve). The $\vvPsi$ is taken from the parameterization in Eq.~(\ref{eq:v4}).}
\label{figApp:raw34}
\end{figure*}

\begin{figure*}[hbt]
\centerline{\includegraphics[width=\textwidth]{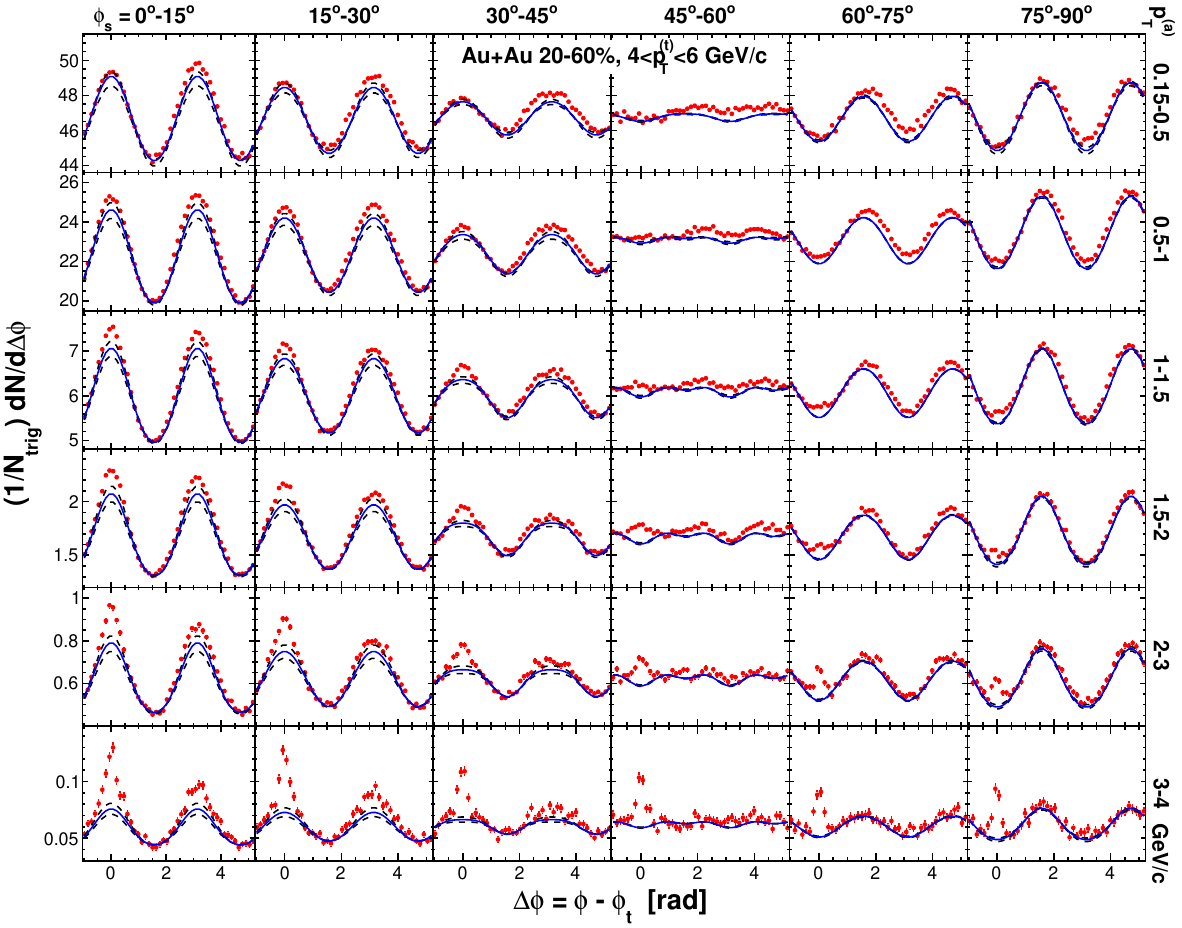}}
\caption{(Color online) Same as in Fig.~\ref{figApp:raw34} but for trigger particle $4<\ptt<6$~\gev\ and six bins in associated particle $\pta$.}
\label{figApp:raw46}
\end{figure*}

\begin{figure*}[hbt]
\centerline{\includegraphics[width=\textwidth]{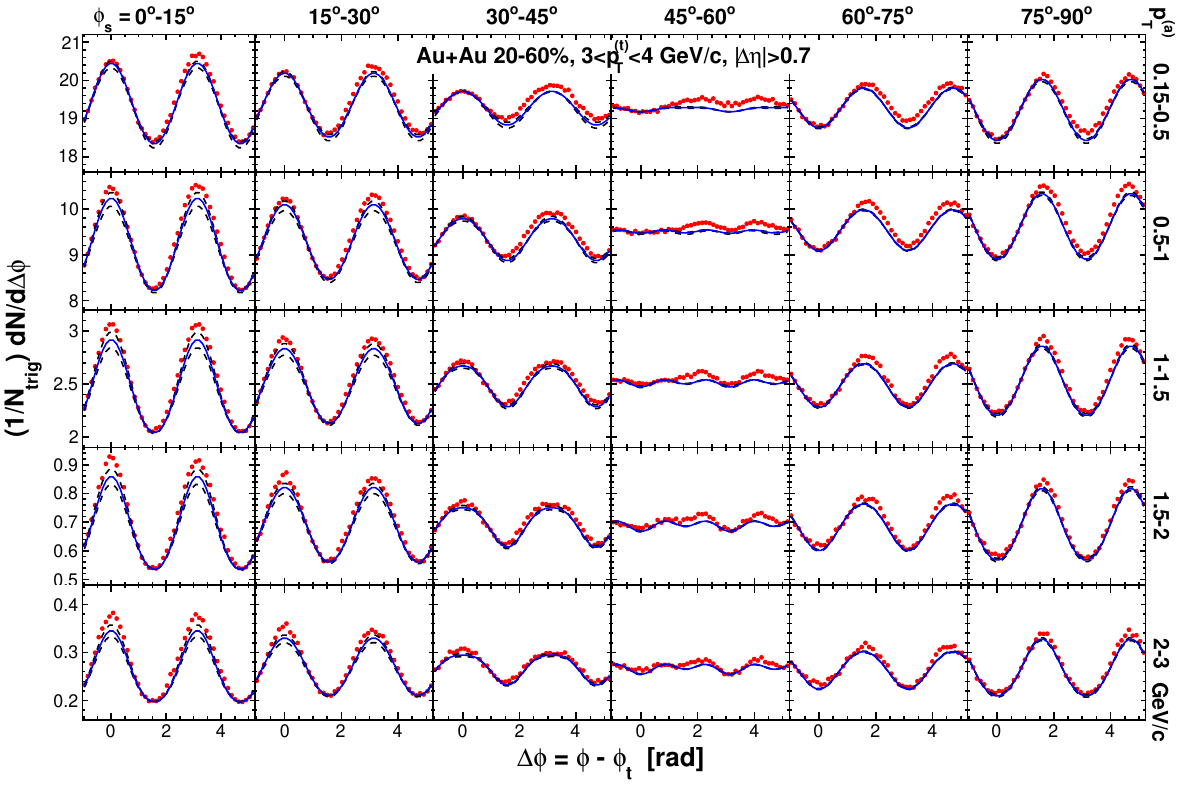}}
\caption{(Color online) Same as in Fig.~\ref{figApp:raw34} but for $|\deta|>0.7$.}
\label{figApp:raw34ridge}
\end{figure*}

\begin{figure*}[hbt]
\centerline{\includegraphics[width=\textwidth]{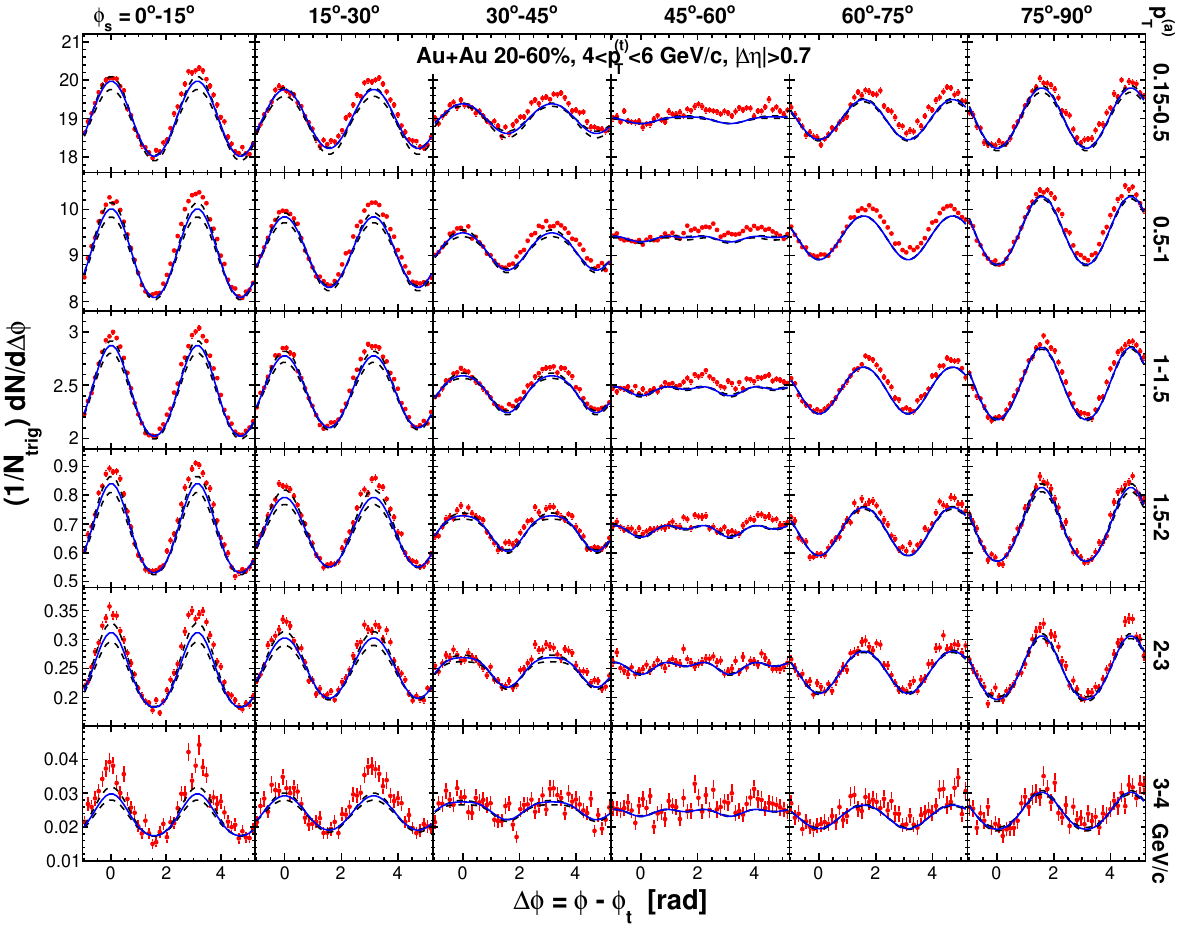}}
\caption{(Color online) Same as in Fig.~\ref{figApp:raw46} but for $|\deta|>0.7$.}
\label{figApp:raw46ridge}
\end{figure*}

\begin{figure*}[hbt]
\centerline{\includegraphics[width=\textwidth]{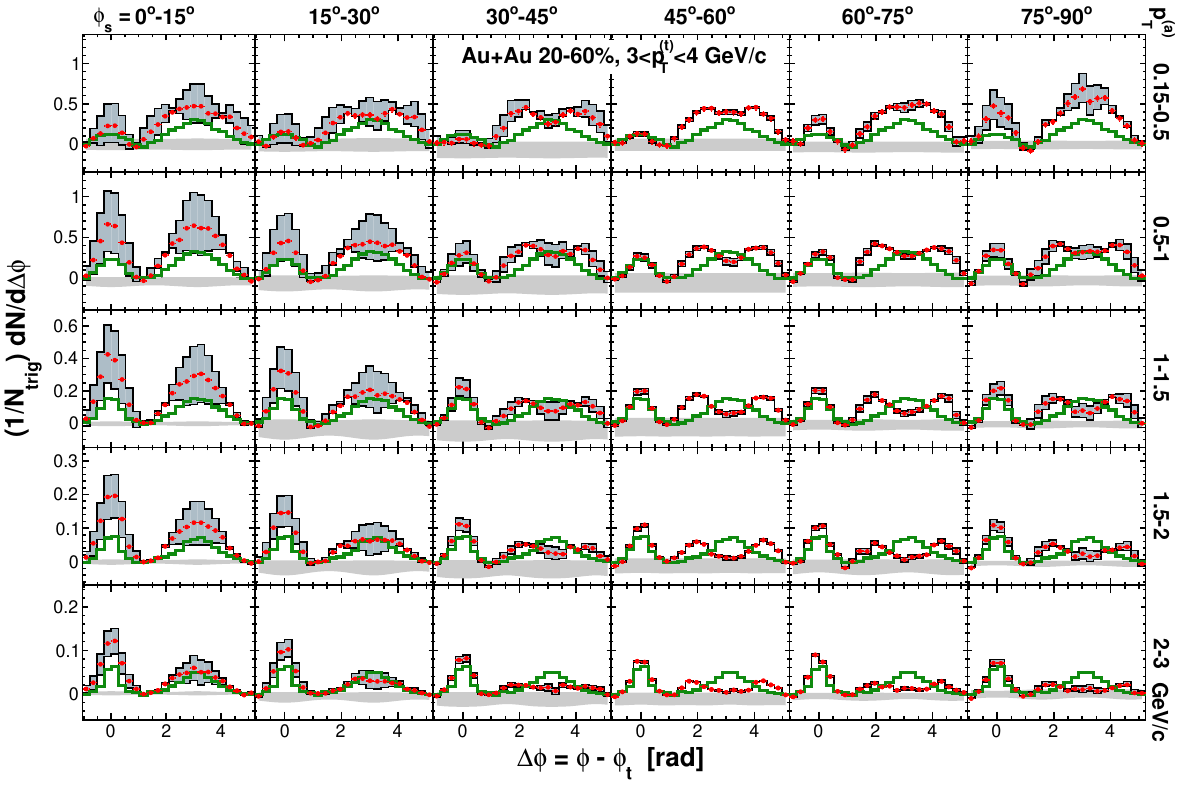}}
\caption{(Color online) Background-subtracted dihadron correlations with trigger particle in six slices of azimuthal angle relative to the event plane, $\phis=|\phit-\psiEP|$. The trigger $\pt$ range is $3<\ptt<4$~\gev. Five associated particle $\pta$ bins are shown. Both the trigger and associated particles are restricted to be within $|\eta|<1$. The triangle two-particle $\deta$ acceptance is not corrected. The figure corresponds to the raw correlations in Fig.~\ref{figApp:raw34}. The data points are from minimum-bias 20-60\% Au+Au collisions. Flow background is subtracted by Eq.~(\ref{eq:bkgd}) using measurements in Table~\ref{tab:v2} and the parameterization in Eq.~(\ref{eq:v4}). Systematic uncertainties are shown in the thin histograms embracing the shaded area due to flow subtraction and in the horizontal shaded band around zero due to \zyam\ background normalization. Statistical errors are mostly smaller than symbol size. For comparison, the inclusive dihadron correlations from \dAu\ collisions are superimposed as the thick (green) histograms (only statistical errors are depicted).}
\label{figApp:corr34}
\end{figure*}

\begin{figure*}[hbt]
\centerline{\includegraphics[width=\textwidth]{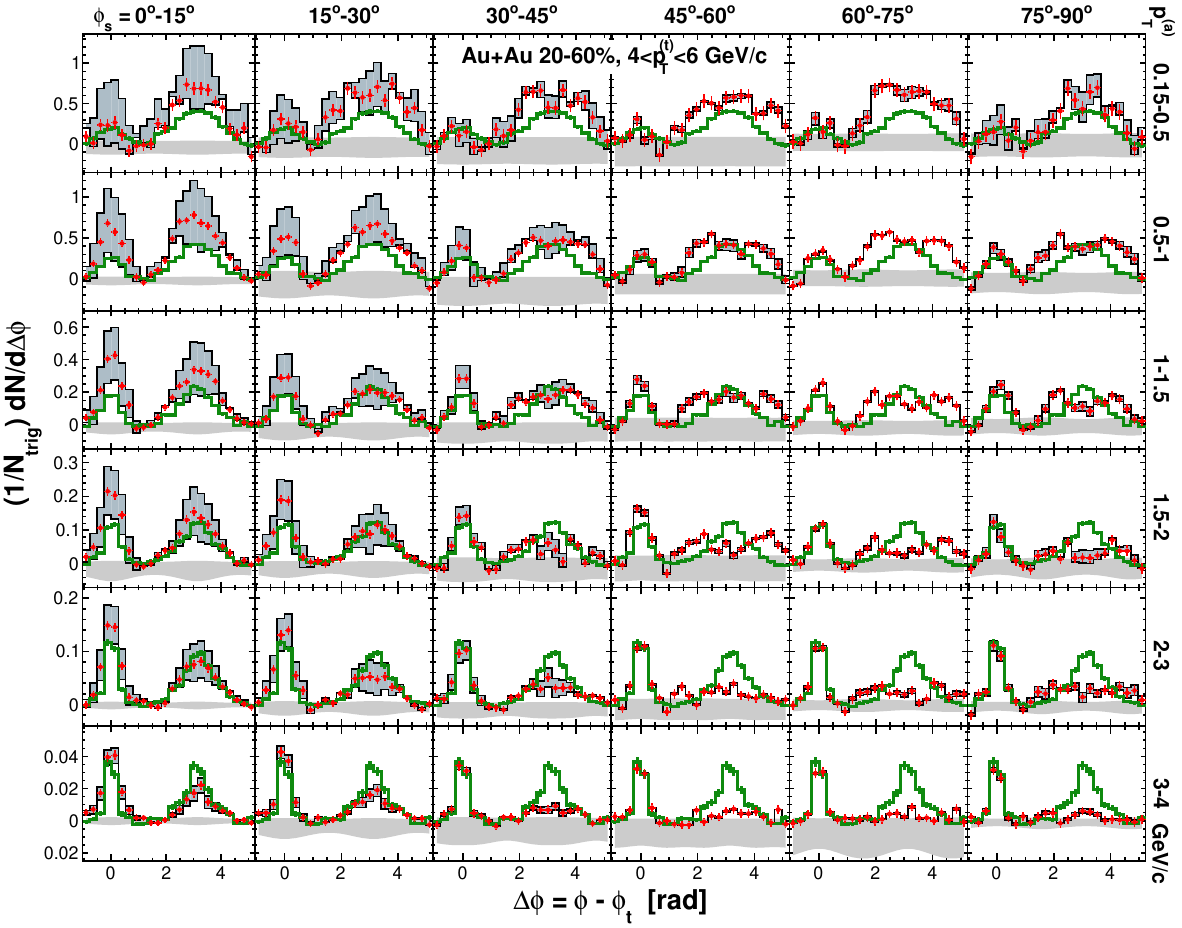}}
\caption{(Color online) Same as in Fig.~\ref{figApp:corr34} but for trigger particle $4<\ptt<6$~\gev\ and six bins in associated particle $\pta$. The figure corresponds to raw correlations in Fig.~\ref{figApp:raw46}.}
\label{figApp:corr46}
\end{figure*}

\begin{figure*}[hbt]
\centerline{\includegraphics[width=\textwidth]{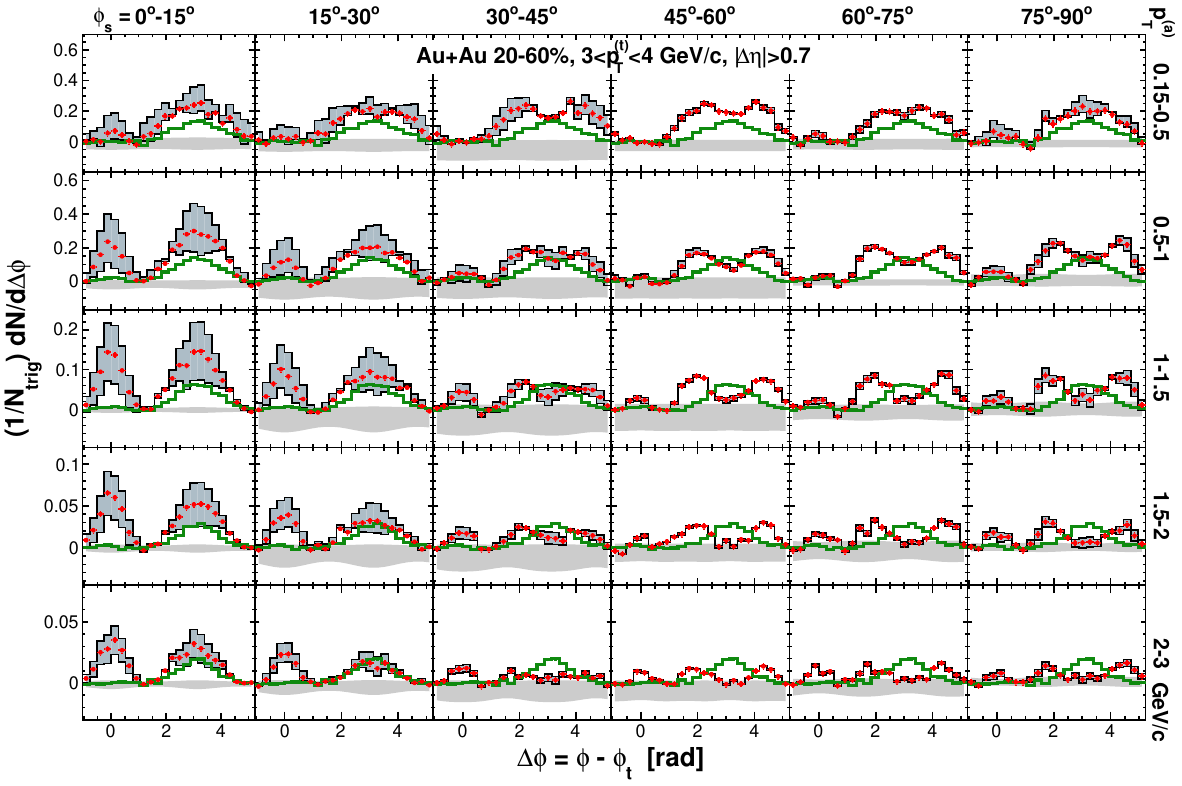}}
\caption{(Color online) Same as in Fig.~\ref{figApp:corr34} but for $|\deta|>0.7$. The figure corresponds to raw correlations in Fig.~\ref{figApp:raw34ridge}.}
\label{figApp:corr34ridge}
\end{figure*}

\begin{figure*}[hbt]
\centerline{\includegraphics[width=\textwidth]{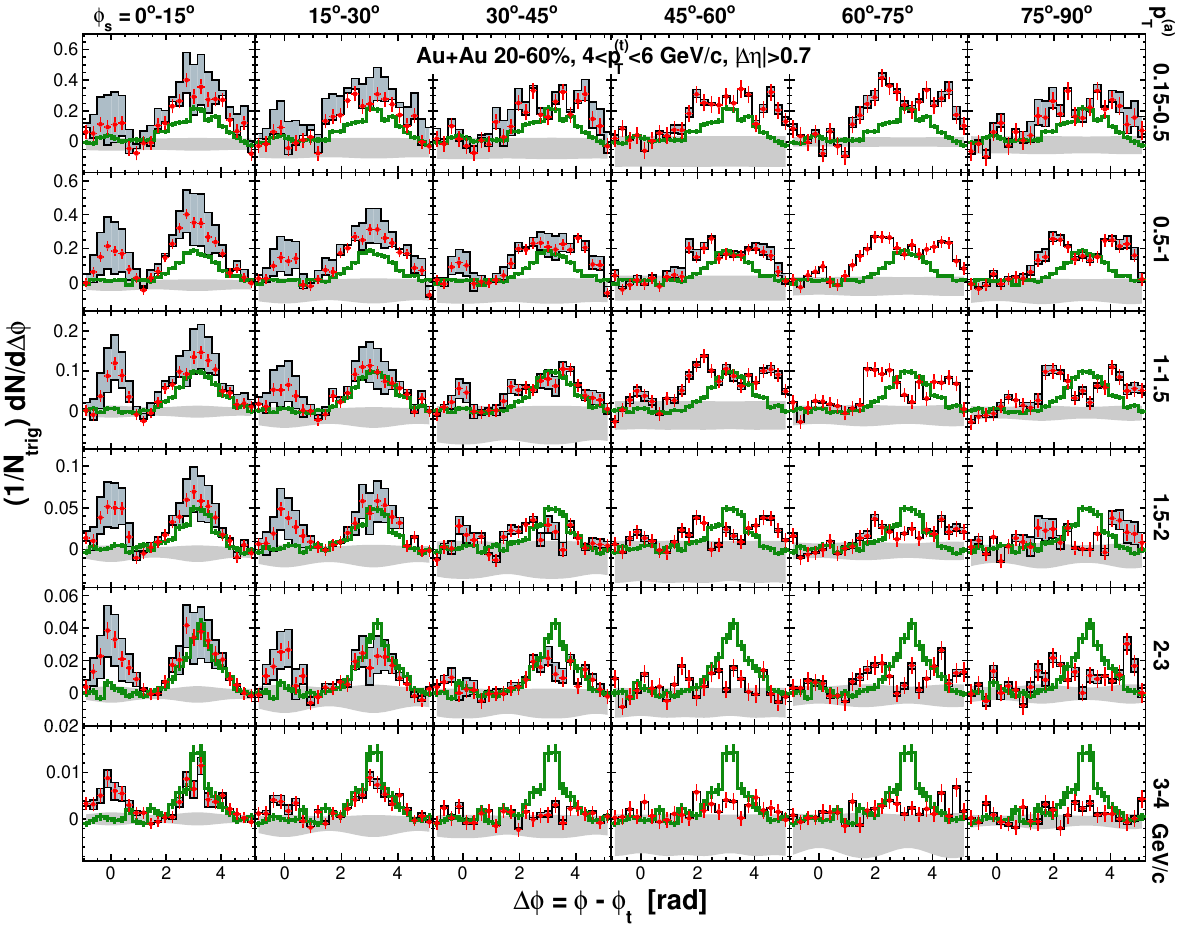}}
\caption{(Color online) Same as in Fig.~\ref{figApp:corr46} but for $|\deta|>0.7$. The figure corresponds to raw correlations in Fig.~\ref{figApp:raw46ridge}.}
\label{figApp:corr46ridge}
\end{figure*}

\begin{figure*}[hbt]
\centerline{\includegraphics[width=\textwidth]{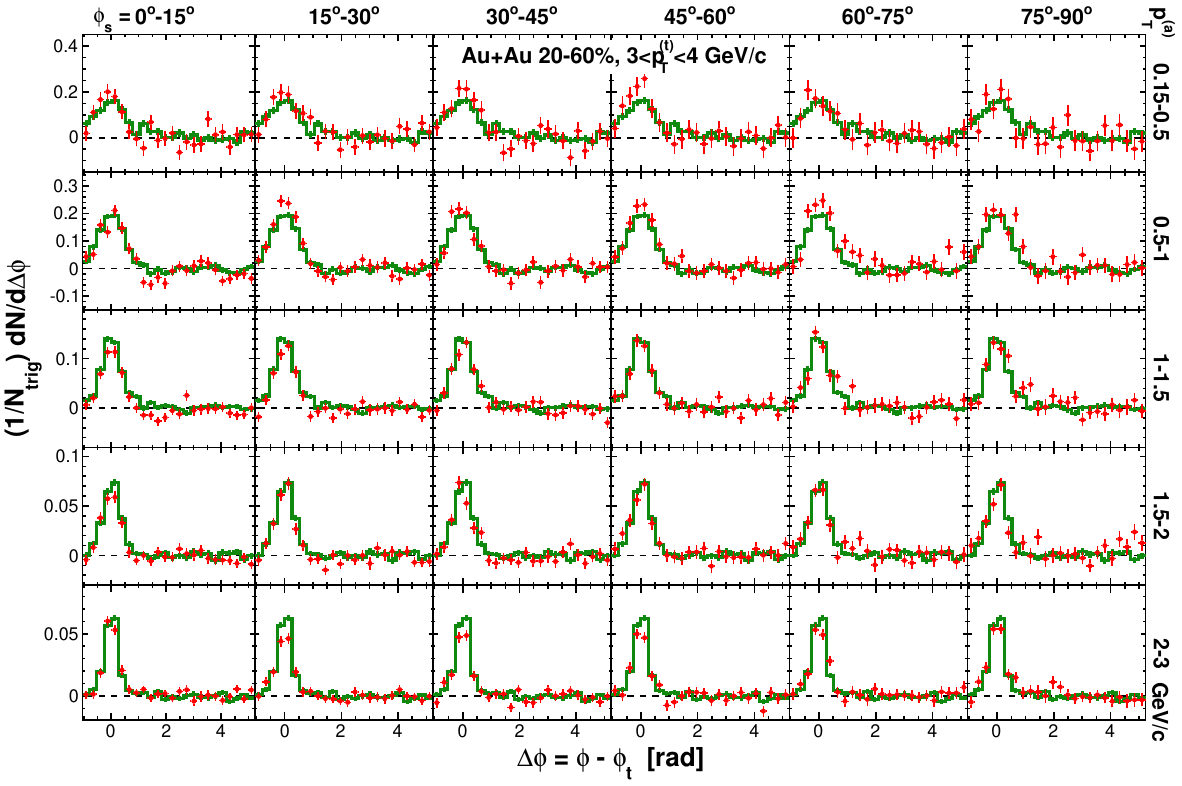}}
\caption{(Color online) Jet-like dihadron correlations with trigger particle in six slices of azimuth relative to the event plane, $\phis=|\phit-\psiEP|$. The jet-like dihadron correlations are obtained from the difference between $|\deta|<0.7$ and (acceptance weighted) $|\deta|>0.7$ correlations. The triangle two-particle $\deta$ acceptance is not corrected. The trigger $\pt$ range is $3<\ptt<4$~\gev. Five associated particle $\pta$ bins are shown. Both the trigger and associated particles are restricted to be within $|\eta|<1$. The data points are from minimum-bias 20-60\% Au+Au collisions. Superimposed for comparison in the thick histograms are the inclusive jet-like dihadron correlation from \dAu\ collisions. Errors bars are statistical; Systematic uncertainties are small.}
\label{figApp:corr34jet}
\end{figure*}

\begin{figure*}[hbt]
\centerline{\includegraphics[width=\textwidth]{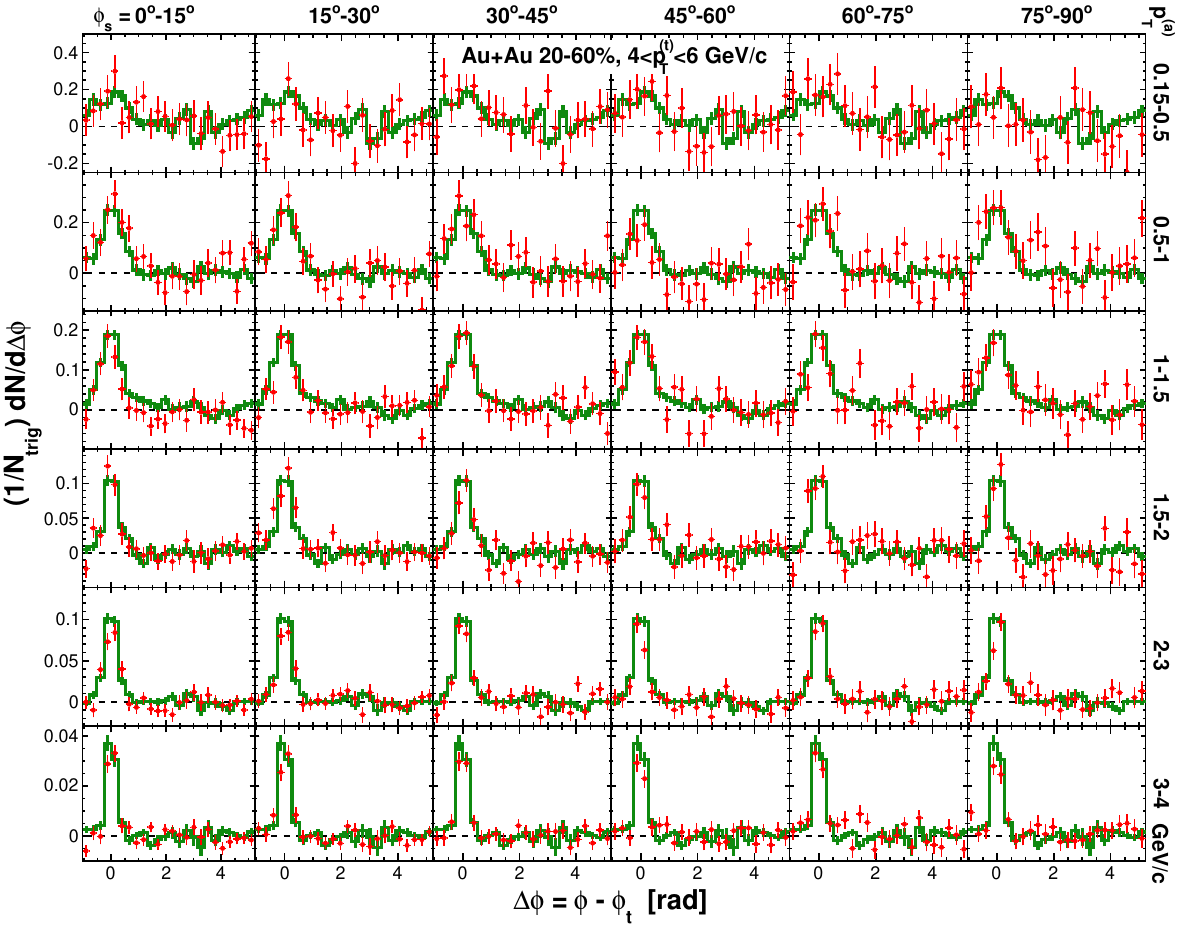}}
\caption{(Color online) Same as in Fig.~\ref{figApp:corr34jet} but for trigger particle $4<\ptt<6$~\gev\ and six bins in associated particle $\pta$.}
\label{figApp:corr46jet}
\end{figure*}

\begin{figure*}[hbt]
\centerline{\includegraphics[width=\textwidth]{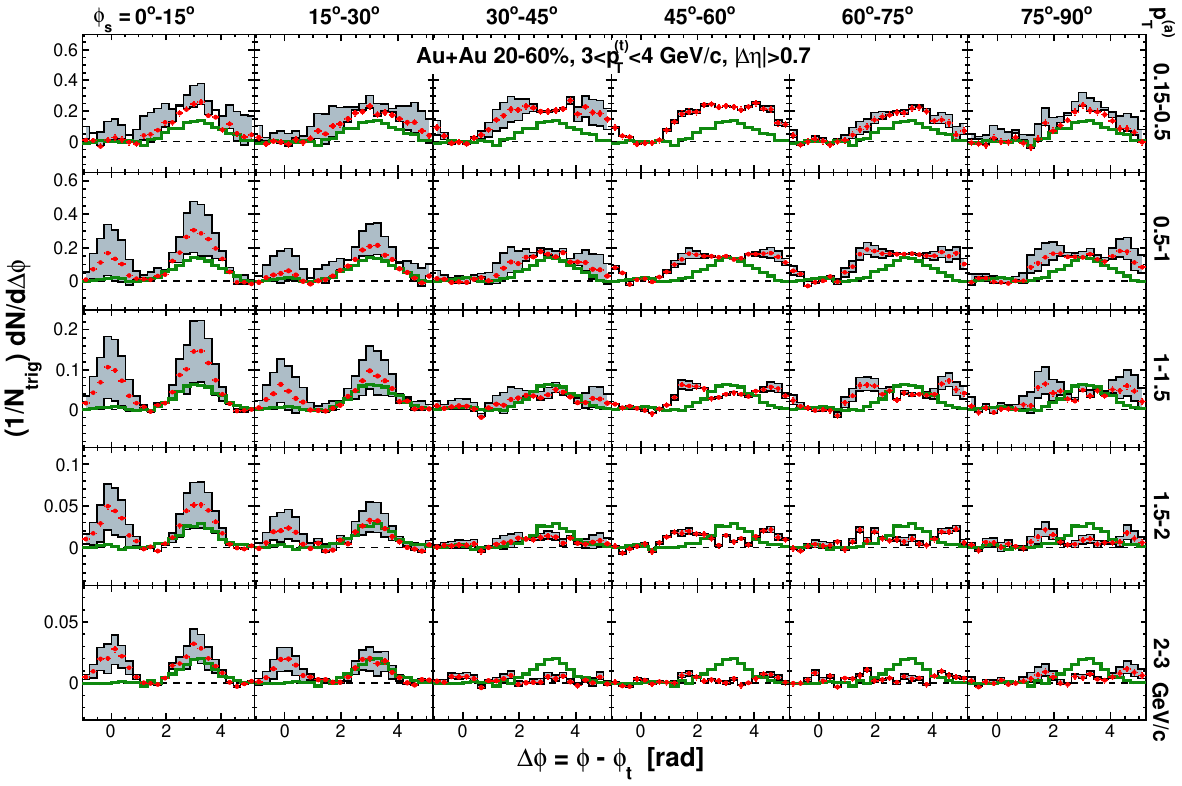}}
\caption{(Color online) Background-subtracted dihadron correlations with trigger particle in six slices of azimuthal angle relative to the event plane, $\phis=|\phit-\psiEP|$, with a cut on the trigger-associated pseudo-rapidity difference of $|\deta|>0.7$. The triangle two-particle $\deta$ acceptance is not corrected. The trigger $\pt$ range is $3<\ptt<4$~\gev. Five associated particle $\pta$ bins are shown. Both the trigger and associated particles are restricted to be within $|\eta|<1$. The figure corresponds to the raw correlations in Fig.~\ref{figApp:raw34ridge}. The data points are from minimum-bias 20-60\% Au+Au collisions. Flow background is subtracted by Eq.~(\ref{eq:bkgd_v3}) using $v_2$ measurements in Table~\ref{tab:v2} and $v_3$ in Table~\ref{tab:v3} and the parameterization in Eq.~(\ref{eq:v4}). Systematic uncertainties due to flow subtraction are shown in the thin histograms embracing the shaded area; those due to \zyam\ background normalization are not shown. Error bars are statistical. For comparison, the inclusive dihadron correlations from \dAu\ collisions are superimposed as the thick (green) histograms.}
\label{figApp:corr34ridge_v3}
\end{figure*}

\begin{figure*}[hbt]
\centerline{\includegraphics[width=\textwidth]{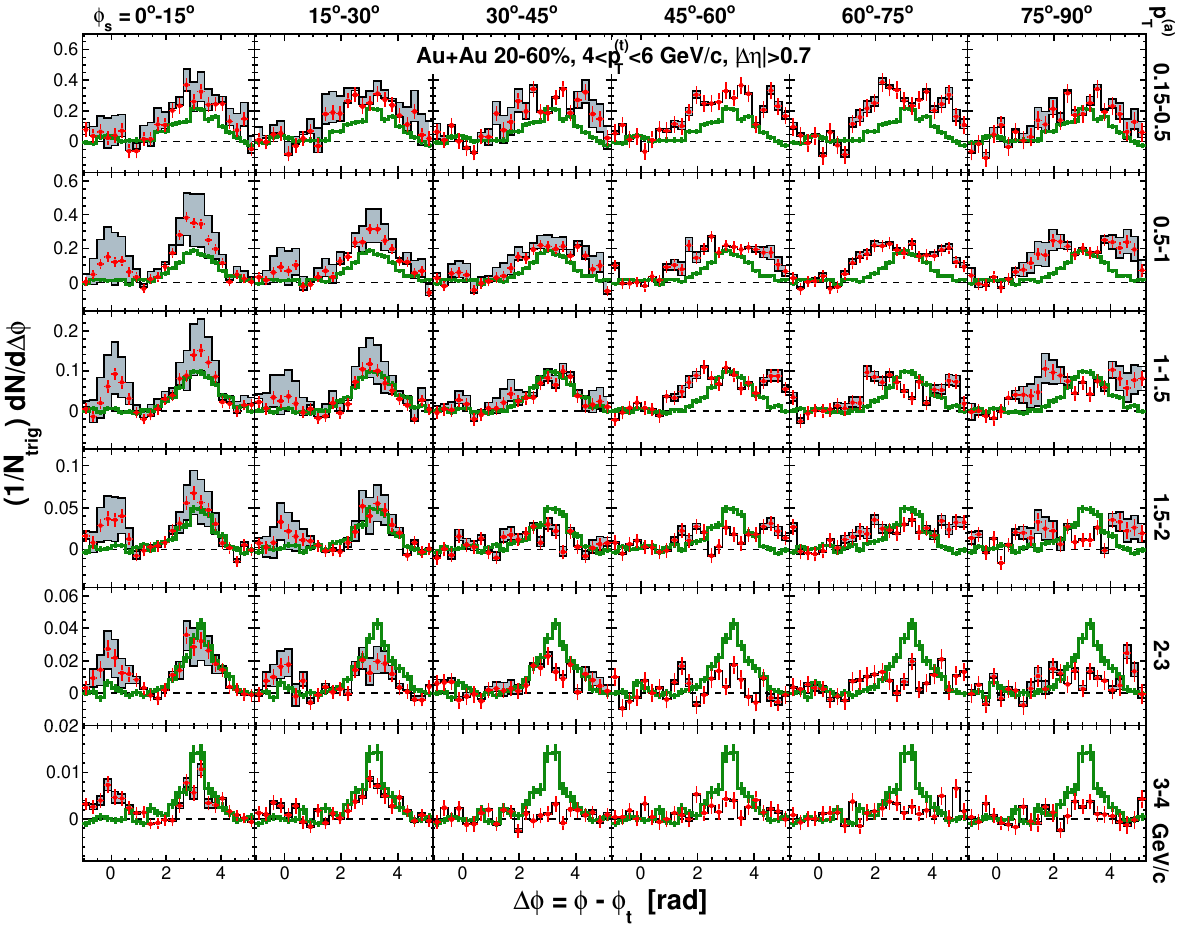}}
\caption{(Color online) Same as in Fig.~\ref{figApp:corr34ridge_v3} but but for trigger particle $4<\ptt<6$~\gev\ and six bins in associated particle $\pta$.}
\label{figApp:corr46ridge_v3}
\end{figure*}

\begin{figure*}[hbt]
\centerline{\includegraphics[width=\textwidth]{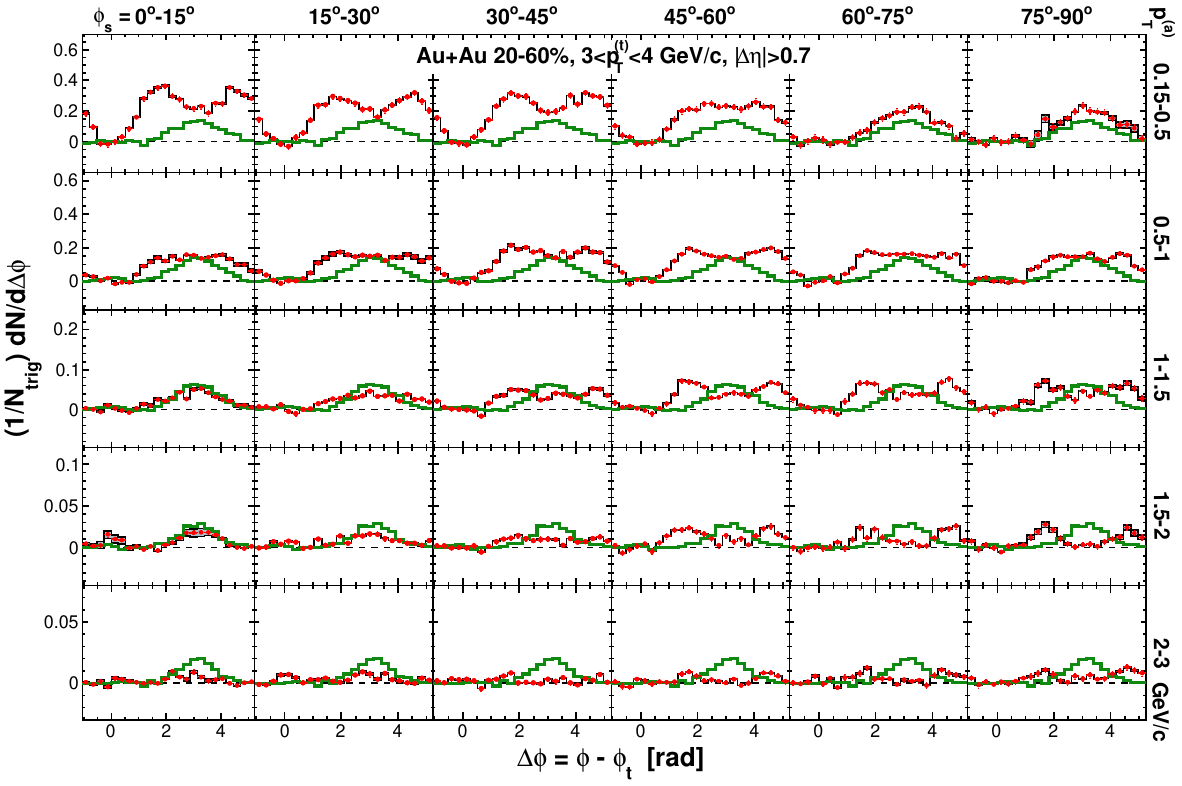}}
\caption{(Color online) Background-subtracted dihadron correlations with trigger particle in six slices of azimuthal angle relative to the event plane, $\phis=|\phit-\psiEP|$, with a cut on the trigger-associated pseudo-rapidity difference of $|\deta|>0.7$. The triangle two-particle $\deta$ acceptance is not corrected. The trigger $\pt$ range is $3<\ptt<4$~\gev. Five associated particle $\pta$ bins are shown. Both the trigger and associated particles are restricted to be within $|\eta|<1$. The figure corresponds to the raw correlations in Fig.~\ref{figApp:raw34ridge}. The data points are from minimum-bias 20-60\% Au+Au collisions. Flow background is subtracted by Eq.~(\ref{eq:bkgd_v4}). The $\phis$-dependent $\vpt{2}$ measured by two-particle cumulants with $\etagap=0.7$ and 1.2 in Table~\ref{tab:v2_phisDep} are used (the thin histograms embracing the shaded area), with their average shown in the data points. The subtracted $\ff{3}{2}$ is given in Table~\ref{tab:v3}. The subtracted $\vvPsi$ is parameterized by Eq.~(\ref{eq:v4}), and the $\VVuc$ is given by Eq.~(\ref{eq:v4uc}). Error bars are statistical; systematic uncertainties are not shown. For comparison, the inclusive dihadron correlations from \dAu\ collisions are superimposed as the thick (green) histograms.}
\label{figApp:corr34ridge_phisDep}
\end{figure*}

\begin{figure*}[hbt]
\centerline{\includegraphics[width=\textwidth]{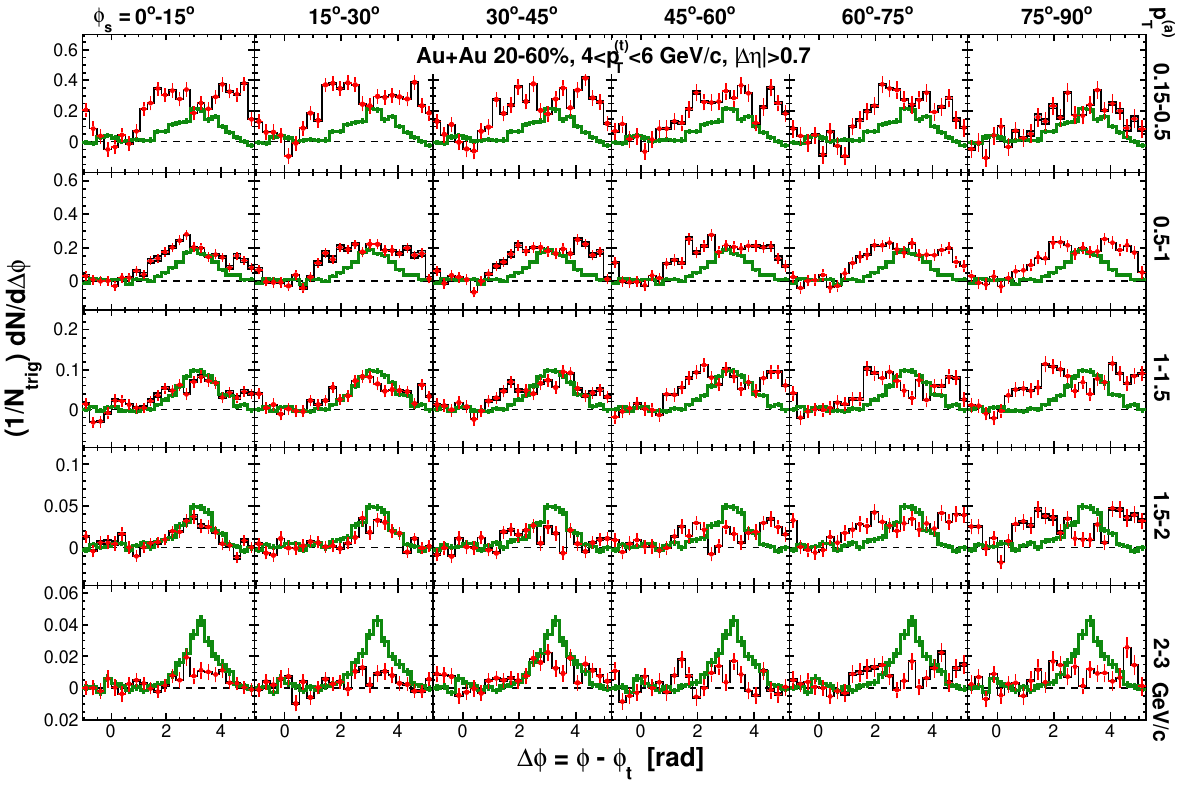}}
\caption{(Color online) Same as in Fig.~\ref{figApp:corr34ridge_phisDep} but but for trigger particle $4<\ptt<6$~\gev.}
\label{figApp:corr46ridge_phisDep}
\end{figure*}

This appendix presents dihadron correlation functions. 
Figures~\ref{figApp:raw34}, \ref{figApp:raw46}, \ref{figApp:raw34ridge}, and \ref{figApp:raw46ridge} show the raw correlation functions. Figures~\ref{figApp:corr34}, \ref{figApp:corr46}, \ref{figApp:corr34ridge}, and \ref{figApp:corr46ridge} show the $v_2$, $v_4$ and \zyam\ background-subtracted correlation functions. Figures~\ref{figApp:corr34jet} and \ref{figApp:corr46jet} show the near-side jet-like correlation functions. Figures~\ref{figApp:corr34ridge_v3}, \ref{figApp:corr46ridge_v3}, \ref{figApp:corr34ridge_phisDep}, and \ref{figApp:corr46ridge_phisDep} show the $v_2$, $v_3$, $v_4$ and \zyam\ background-subtracted correlation functions. The data for the correlation functions and all other figures in the paper are published online at http://www.star.bnl.gov/central/publications/.

\section*{Acknowledgments}

We thank the RHIC Operations Group and RCF at BNL, the NERSC Center at LBNL and the Open Science Grid consortium for providing resources and support. This work was supported in part by the Offices of NP and HEP within the U.S.~DOE Office of Science, the U.S.~NSF, the Sloan Foundation, the DFG cluster of excellence `Origin and Structure of the Universe' of Germany, CNRS/IN2P3, STFC and EPSRC of the United Kingdom, FAPESP CNPq of Brazil, Ministry of Ed.~and Sci.~of the Russian Federation, NNSFC, CAS, MoST, and MoE of China, GA and MSMT of the Czech Republic, FOM and NWO of the Netherlands, DAE, DST, and CSIR of India, Polish Ministry of Sci.~and Higher Ed., Korea Research Foundation, Ministry of Sci., Ed.~and Sports of the Rep.~Of Croatia, Russian Ministry of Sci.~and Tech, and RosAtom of Russia.

\bibliographystyle{unsrt}
\bibliography{../rc/ref}

\pagebreak

\end{document}